\documentclass{ptephy_v1}
\usepackage{bm}
\usepackage{amsmath}
\usepackage{mathrsfs}

\preprintnumber{1611-07295} 

\begin{document}

\title{Kinetic theory of discontinuous rheological phase transition for a dilute inertial suspension}


\author{Hisao Hayakawa}
\affil{Yukawa Institute for Theoretical Physics, Kyoto University, Kitashirakawa Oiwakecho, Sakyo-ku, Kyoto 606-8502, Japan \email{hisao@yukawa.kyoto-u.ac.jp}}

\author[2,3,4]{Satoshi Takada}
\affil{Department of Physics, Kyoto University, Kitashirakawa Oiwakecho, Sakyo-ku, Kyoto 606-8502, Japan}
\affil[3]{Earthquake Research Institute, The University of Tokyo, 1-1-1, Yayoi, Bunkyo-ku, Tokyo 113-0032, Japan}
\affil[4]{Institute of Engineering, Tokyo University of Agriculture and Technology, 2-24-16, Naka-cho, Koganei, Tokyo 184-8588, Japan\footnote{present address}}


\begin{abstract}%
A kinetic theory for a dilute inertial suspension under a simple shear is developed. 
With the aid of the corresponding Boltzmann equation, it is found that the flow curves (the relations between the stress and the strain rate) exhibit the crossovers from the Newtonian to the Bagnoldian for a granular suspension and from the Newtonian to a fluid having a viscosity proportional to the square of the shear rate for a suspension consisting of elastic particles, respectively.
The existence of the negative slope in the flow curve directly leads to a discontinuous shear thickening (DST).
This DST corresponds to the discontinuous transition of the kinetic temperature between a quenched state and an ignited state. 
The results of the event-driven Langevin simulation of hard spheres perfectly agree with the theoretical results without any fitting parameter.
The introduction of an attractive interaction between particles is also another source of the DST in dilute suspensions.
Namely, there are two discontinuous jumps in the flow curve if the suspension particles have the attractive interaction. 
\end{abstract}

\subjectindex{J44}

\maketitle

\section{Introduction}

Shear thickening is a drastic rheological process in which the viscosity increases as the shear rate increases. 
The shear thickening fluid typically behaves as a liquid at rest or weakly stirred situation, but its resistance becomes large as if it is a solid above a critical shear rate.
The shear thickening can be a continuous shear thickening (CST) or a discontinuous shear thickening (DST) depending on its situation.
The DST can be easily observed in densely packed suspensions of cornstarch in water.
There are many industrial applications of the DST such as a body armor and a traction control.

The DST attracts much attention even from  physicists~\cite{Barnes89,Mewis11,Brown14,Lootens05,Cwalina14,Brown10,Waitukaitis12}. 
Typical suspensions exhibiting DSTs have some common features. 
The first one is that the discontinuous jump can be only observed below the jamming point~\cite{Brown09,Mari14}.
The second one is that the normal stress difference becomes large when the DST takes place~\cite{Cwalina14,Lootens05}. 
The third one is
that the mutual frictions between grains play important roles in the DSTs for dense suspensions and dry granular materials~\cite{Lootens05,Otsuki11,Seto13,Mari14,Bi11,Pica11,Haussinger13,Grob14}.
There are some phenomenologies to understand the mechanism of the DST for dense suspensions and dry granular materials~\cite{Nakanishi11,Nakanishi12,Nagahiro13,Grob16,Wyart14}.

So far, the most of studies on the DST focus on the rheological behavior of dense suspensions.
However, we can look for the possibility of the occurrence of 
the DST-like processes even in dilute inertial suspensions.
 Such suspensions are usually discussed in the context of fluidized beds of gas-solid mixtures~\cite{Gidaspow94,Jackson00} as a typical example of inertial suspensions~\cite{Koch01}, though the uniform flows are often unstable~\cite{Batchelor88,Batchelor93,Sasa92,Komatsu93,Ichiki95}.
Nevertheless, there is a homogeneous phase if we control the rate of the injected gas flow from the bottom of the container.
Moreover, the gravity effect, the origin of the instability, seems to be not important for aerosols which are suspensions of solid particles in the air, though attractive interactions between aerosol grains might not be ignored~\cite{Friedlander}.
When the gravity effect is negligible, Tsao and Koch~\cite{Tsao95} demonstrated the existence of a discontinuous phase transition for the kinetic temperature of an inertial suspension under a simple shear between a quenched state (a low temperature state) and an ignited state (a high temperature state)~\cite{Koch01} in terms of the analysis of the Boltzmann equation. 
They also illustrated the existence of a rapid increment of the shear viscosity as the shear rate increases, though it is not clear whether it is the DST or the CST~\cite{comment_Tsao1}.   
Sangani {\it et al.} extended the analysis of Ref.~\cite{Tsao95} to the finite density and found that the discontinuous transition of the kinetic temperature for dilute suspensions becomes continuous at relatively low density~\cite{Sangani96}.
Recently, Saha and Alam have revised the theoretical analysis of quenched-ignited phase transition for dilute gas-solid suspensions~\cite{Saha17}.
Santos {\it et al.} also 
demonstrated the existence of a CST in moderately dense hard-core gases by using the revised Enskog theory~\cite{Santos98}.
See also Refs.~\cite{Garzo12,Hayakawa-Takada-Garzo} for the theoretical analysis of granular gas-solid suspensions in terms of the kinetic theory.

As another example of an inertial suspension, we can consider shaken granular materials. 
Because the exact theoretical modeling of shaking processes is difficult,
we usually introduce a thermostat as a stochastic model to express shaking effects~\cite{Williams96}. 
So far we have used various thermostats such as the white noise thermostat~\cite{vanNoije98,Henrique00,Barrat02,Dahl02}, the Gaussian thermostat~\cite{Montanero00} and the Fokker-Planck thermostat~\cite{Hayakawa03,Sarracino10}.
In this paper we adopt the Fokker-Planck thermostat to describe suspensions, because it can be relaxed to an equilibrium state if we do not add any external force nor ignore mutual collisions between particles. 
Note that the white noise thermostat~\cite{vanNoije98} is used for the explanation for an experiment of shaken granular gases in a microgravity environment~\cite{Tatsumi09}.

One paper suggests the existence of a DST for relatively dilute suspensions in terms of the analysis of a phenomenological BGK equation~\cite{BGK2016}.
Their paper predicted that the DST becomes the CST as the density increases~\cite{BGK2016} as Sangani {\it et al.} indicated~\cite{Sangani96}
(see also Refs.~\cite{Saha17,Hayakawa-Takada-Garzo}).
Note that the asymptotic form of this model in high shear limit agrees with those in Refs.~\cite{Tsao95,Kawasaki14}.
Nevertheless, the previous theory is only a qualitative one but not a quantitative one, because the BGK equation is only applicable to elastic particles and the relaxation time cannot be determined within the theory.   
We also need to explain the relationship between this paper and Ref.~\cite{Hayakawa-Takada-Garzo}. 
Although Ref.~\cite{Hayakawa-Takada-Garzo} has already extended the kinetic theory of dilute inertial suspensions to the moderately dense suspensions,
the detailed calculation of that paper relies on unpublished Ref. [28] of Ref.~\cite{Hayakawa-Takada-Garzo}. 
This paper is nothing but the published version of Ref. [28] of Ref.~\cite{Hayakawa-Takada-Garzo} with adding some new aspects.\footnote{ 
Ref. [28] of Ref.~\cite{Hayakawa-Takada-Garzo} (arXiv:1611.07295) was submitted prior to Ref.~\cite{Hayakawa-Takada-Garzo} but it has not been accepted for publication in any scientific journals so far.
}

As explained, a typical example of dilute inertial suspensions is the aerosol which has the attractive interaction between particles to form clusters~\cite{Friedlander}.
The attractive interaction is also important for wet granular materials~\cite{Castellanos05,Mitarai06}.
Among several studies on the effects of attractive interactions on granular rheology~\cite{Rahbari10,Gu14, Irani14,Takada14,Saitoh15,Takada16,Takada18,Irani18}, 
the finding by Irani et al.~\cite{Irani18} on the existence of a hysteresis loop in the flow curve is remarkable, which is similar to that observed in the DST.
Because its effect on suspension rheology is not well understood, we need to clarify the role of the attractive interaction in the inertial suspensions. 

The purpose of this paper is to extract the mechanism of the DST-like phenomena in terms of the Boltzmann equation for dilute inertial suspensions 
by identifying the discontinuous transition between the quenched state and the ignited state as the DST.
We examine the quantitative validity of our theory from the comparison of results between our theory and the event-driven Langevin simulation of hard spheres (EDLSHS)~\cite{Scala12}.
Then, we demonstrate that (i) the mutual friction between grains is not always necessary for the DST-like phenomena and 
(ii) the DST-like phenomena can take place even in the dilute inertia suspensions.
We also illustrate the role of an attractive interaction in the flow curve, which can cause another DST-like process.

The organization of this paper is as follows.
In the next section, we present a microscopic basic equation, the Boltzmann equation under the influence of the background fluid.
We also derive a set of equations for the kinetic temperature $T$, the anisotropy of the temperature $\Delta T$ which is proportional to the normal stress difference and the shear stress $P_{xy}$ when we adopt Grad's approximation.
In Sec. 3, we obtain $T$, $\Delta T$, the stress ratio $\mu\equiv -P_{xy}/P$ and the second normal stress difference, where $P$ is the pressure.
From these results, we demonstrate the existence of the DST-like process in the inertial suspension.
In Sec. 4, we perform the simulation (EDLSHS) for the corresponding Langevin equation to verify the quantitative validity of our theoretical predictions.
In Sec. 5, we illustrate that an attractive interaction between grains can cause another discontinuous jump in the flow curve. 
In the final section, we discuss and summarize our obtained results.
We have seven appendices to describe the detailed calculation in our paper.


\section{ The framework of our kinetic theory for the dilute inertial suspension}

Let us consider a collection of smooth mono-disperse spherical grains (the diameter $\sigma$, the mass $m$ and the restitution coefficient $e$ which is ranged $0<e\le 1$) distributed in a $d-$dimensional space, which are suspended in the background fluid.
We assume that the macroscopic velocity field $\bm{u}=(u_x,\bm{u}_\perp)$ satisfies the simple shear flow
\begin{equation}
u_x=\dot\gamma y, \quad \bm{u}_\perp=\bm{0} ,
\end{equation}
where $\dot\gamma$ is the shear rate.
Because we are interested in the homogeneous phase in the inertial suspension, 
we assume that the effects of the gravity for the motion of particles are negligible.
Introducing the peculiar momentum of $i-$th particle as $\bm{p}_i\equiv m(\bm{v}_{i}-\dot\gamma y \bm{e}_x)$ with the unit vector $\bm{e}_x$ parallel to $x$-direction with the velocity of $i-$th particle $\bm{v}_i$, 
a reasonable starting point for the motion of grains at low Reynolds number flows is the Langevin equation
\begin{equation}\label{Langevin_eq}
\frac{d{\bm{p}}_i}{dt}=-\zeta \bm{p}_i + \bm{F}_i^{({\rm imp})}+ m\bm{\xi}_i,
\end{equation} 
where we have introduced the impulsive force $\bm{F}_i^{(\rm imp)}$ to express collisions and the 
noise $\bm{\xi}_i(t)=\xi_{i,\alpha}(t)\bm{e}_\alpha$ satisfying 
\begin{equation}\label{noise}
\langle \bm{\xi}_i(t)\rangle=0, \quad
\langle \xi_{i,\alpha}(t)\xi_{j,\beta}(t')\rangle = 2\zeta T_{\rm ex} \delta_{ij}\delta_{\alpha\beta}\delta(t-t').
\end{equation}
Here, we have introduced $\zeta$ and $T_{\rm ex}$ to characterize the drag from the background fluid and the environmental temperature under the unit of the Boltzmann constant $k_{\rm B}=1$, respectively.
The bracket $\langle \cdot \rangle$ stands for the average over the noise distribution.
This model is essentially a dilute version of the model used by Kawasaki {\it et al.}~\cite{Kawasaki14}.
We, here, assume that the inertia of the particles is important but the inertia of the fluid (or the gas) is negligible. 
Typically these conditions hold for the particles with the diameters in the range 1-70 $\mu$m~\cite{Koch01}. 
Even when we consider such gas-solid suspensions,
the realistic drag coefficient $\zeta$ must be a resistance matrix  
which strongly depend on the configuration of particles~\cite{Ichiki95}.
For simplicity, however, we regard $\zeta$ as a scalar constant which is independent of the configuration of particles as in Refs.~\cite{Tsao95,Sangani96}.
This treatment might be justified if we consider cases in the dilute limit or the dense limit.\footnote{
There are numerous papers to use a constant drag for systems of dense grains.
This simplification can be justified as follows: the dense grains near the jamming points are almost at contact and the gap between grains can be replaced by the cutoff length of the resistance matrix. 
Indeed, the behavior of dense suspensions embedded in a fluid is almost identical to that observed in dense granular systems as shown in a recent paper
~\cite{Pradipto}.  
}

In this paper, we also assume that $\zeta$ is proportional to $\sqrt{T_{\rm ex}}$, because the drag coefficient is proportional to the viscosity of the solvent $\eta_0$ which is proportional to $\sqrt{T_{\rm ex}}$ if the solvent consists of hard-core molecules.
\footnote{
If we regard the gas as a dilute hard-core gas for $d=3$, 
the drag coefficient is given by
$\zeta=3\pi \eta_0 \sigma/m$ where $\eta_0$ and $\sigma_0$ are $\eta_0=(5/16\sigma_0^2)\sqrt{m_0T_{\rm ex}/\pi}$ with the mass of the molecule $m_0$ and the diameter of the molecule, respectively.
}

We emphasize that Eq.~\eqref{Langevin_eq} contains both the collision and the thermal noise. 
Although some previous papers~\cite{Tsao95,Sangani96} ignored the thermal noise,
the existence of the thermal noise is crucially important because
(i) thanks to this term, the system can reach a thermal equilibrium state characterized by $T_{\rm ex}$,
(ii) the background viscosity $\eta_0$ and the drag $\zeta$ become zero if we take the limit $T_{\rm ex}\to 0$,
(iii) relatively small suspensions, which are the target of our study, are affected by the thermal noise if there is no external shear 
and 
(iv) the Newtonian rheology cannot be recovered at $T_{\rm ex}=0$ in zero shear limit, as will be shown. 
It is obvious that aerosol particles can diffuse because of the thermal fluctuations~\cite{Friedlander}.
Of course, the thermal noise is not important for high shear regime because the inertial collisions between grains play dominant roles. 
We can use the Stokes number ${\rm St}=\rho_p\sigma^2\dot\gamma/\eta_0$ where $\rho_p$ is the mass density of a suspended particle to characterize whether the inertial or the noise term is dominant.\footnote{
The Stokes number can be interpreted as the ratio of the kinetic energy $m C_0^2/2$ to the work $\eta_0 \sigma^2 C_0$ due to the drag force proportional to $\eta_0 \sigma$, where $C_0=\sigma \dot\gamma$ is the characteristic speed for collisions of particles.
In other words, the Stokes number is expressed as ${\rm St}=\dot\gamma/\zeta$. Therefore, small (large) ${\rm St}$ corresponds to collisionless (collisional) regime.
We also note that ${\rm St}$ is proportional to the P\'{e}clet number ${\rm Pe}=3\pi \eta_0\dot\gamma \sigma^3/(4T_{\rm ex}) \propto \dot\gamma/\sqrt{T_{\rm ex}}\sim {\rm St}$. 
Thus, the rheology in the low shear regime is governed by the thermal motion. }
Therefore, both contributions can coexist only for the transient regime between two limiting (low shear and high shear) cases.

It is well known that an arbitrary stochastic equation can be converted into the corresponding stochastic Liouville equation.
Then, if we consider the Langevin equation activated by the white-Gaussian noise as in Eq.~\eqref{Langevin_eq}, the Langevin equation can be mapped onto the Liouville equation for the $N-$body distribution function $f^{(N)}(\{\bm{r}_i,\bm{V}_i\},t)$ with $\bm{V}_i=\bm{p}_i/m=\bm{v}_i-\bm{u}$, which is represented by a sum of the Fokker-Planck type equation and the collision term.
If we are interested in a dilute suspension,
the equation of $N-$body distribution function is reduced to that of the one-body velocity distribution function $f(\bm{V},t)$ under the simple shear in the Boltzmann-Grad limit as~\cite{Tsao95,Sangani96,Hayakawa03,Chamorro15}
\begin{equation}
\left(\frac{\partial}{\partial t}-\dot\gamma
V_{y}\frac{\partial}{\partial V_{x}}
\right)
f(\bm{V},t) 
=
\zeta\frac{\partial}{\partial \bm{V}} \cdot \left(
\left\{ \bm{V}+ \frac{T_{\rm ex}}{m} \frac{\partial}{\partial \bm{V}} \right\} 
 f(\bm{V},t) \right)
+
 J(\bm{V}|f),
\label{boltzmann} 
\end{equation}
where we have ignored the spatial fluctuating term in Eq.~\eqref{boltzmann} and introduced the peculiar velocity $\bm{V}\equiv \bm{v}-\bm{u}$, 
because the uniform shear flow is stable as long as we have checked.
Note that Eq.~\eqref{boltzmann} is also used for moderately dense gases as in Refs.~\cite{Sangani96,Hayakawa-Takada-Garzo,Garzo13} with the aid of the Enskog equation, though the validity of the Enskog equation is not mathematically justified.
The collisional integral $J(\bm{V}|f)$ for a dilute suspension characterized by hard-core collisions is given by 
\begin{equation}\label{J(V|f)}
J(\bm{V}_1|f)
=\sigma^{d-1}\int d\bm{v}_2\int d\hat{\bm{\sigma}}\Theta(\bm{v}_{12}\cdot\hat{\bm{\sigma}})
|\bm{v}_{12}\cdot\hat{\bm{\sigma}}| 
\left\{
\frac{f(\bm{V}_1^{**})f(\bm{V}_2^{**})}{e^2}-f(\bm{V}_1)f(\bm{V}_2) 
\right\} ,
\end{equation}
where $\hat{\bm{\sigma}}$ is the normal unit vector at contact, $\Theta(x)=1$ for $x\ge 0$ and $\Theta(x)=0$ otherwise and
$\bm{V}_i^{**}=\bm{v}_i^{**}-\bm{u}$ for $i=1,2$ is the pre-collisional velocity of $\bm{V}_i$ defined through the pre-collisional velocities $\bm{v}_i^{**}$:
\begin{equation}\label{collision_rule}
\bm{v}_1^{**}=\bm{v}_1-\frac{1+e}{2e}(\bm{v}_{12}^{}\cdot\hat{\bm{\sigma}})\hat{\bm{\sigma}}, \quad
\bm{v}_2^{**}=\bm{v}_2+\frac{1+e}{2e}(\bm{v}_{12}^{}\cdot\hat{\bm{\sigma}})\hat{\bm{\sigma}} 
\end{equation}
with $\bm{v}_{12}=\bm{v}_1-\bm{v}_2$.
Once we adopt Eq.~\eqref{boltzmann} associated with Eqs.~\eqref{J(V|f)} and \eqref{collision_rule} instead of Eq.~\eqref{Langevin_eq}, we can construct a theory describing the shear thickening.

One of the most important quantities to characterize the rheology of the dilute suspension is the pressure tensor
\begin{equation}
P_{\alpha\beta}=m \int d\bm{v} V_\alpha V_\beta f(\bm{V},t) .
\label{pressure_tensor}
\end{equation}
This is related to the pressure as $P\equiv P_{\alpha\alpha}/d$, 
where we adopt Einstein's notation for the sum rule {\it i. e.} $P_{\alpha\alpha}=\sum_{\alpha=1}^d P_{\alpha\alpha}$.

Multiplying Eq.~\eqref{boltzmann} by $m V_\alpha V_\beta$ and integrate it over 
$\bm{v}$, we obtain
\begin{equation}\label{stress_eq}
\frac{d}{d t}P_{\alpha\beta}+\dot\gamma(\delta_{\alpha x}P_{y\beta}+\delta_{\beta x}P_{y\alpha})
=-\Lambda_{\alpha\beta}+
2\zeta (n T_{\rm ex} \delta_{\alpha\beta}-P_{\alpha\beta} ),
\end{equation}
where we have introduced the number density $n\equiv \int d\bm{V} f(\bm{V},t)$
 and
\begin{equation}
\Lambda_{\alpha\beta}\equiv -m \int d\bm{v} V_\alpha V_\beta J(\bm{V}|f) .
\label{Lambda}
\end{equation}
Because Eqs.~\eqref{stress_eq} and \eqref{Lambda} are not closed equations,
we adopt Grad's approximation~\cite{BGK2016,Chamorro15,Grad49,Jenkins85a,Jenkins85b,Garzo02,Santos04,Garzo13A}
\begin{equation}\label{Grad}
f(\bm{V})=f_{\rm eq}(\bm{V})\left[1+\frac{m}{2T}\left(\frac{P_{\alpha\beta}}{nT}-\delta_{\alpha\beta} \right)V_\alpha V_\beta \right] 
\end{equation}
with 
\begin{equation}\label{Maxwell}
f_{\rm eq}(\bm{V})=
n\left(\frac{m}{2nT}\right)^{d/2}\exp\left(-\frac{m V^2}{2T} \right) ,
\end{equation}
where we have introduced the kinetic temperature $T$ defined by $T\equiv \int d\bm{v} m(\bm{v}-\bm{u})^2f(\bm{V})/(dn)$.
Note that the pressure satisfies the equation of state for an ideal gas $P=nT$ in our model.
Grad's approximation or Grad's 13 moments method for $d=3$ is the well established method to describe the slow motion of nonequilibrium gases~\cite{BGK2016,Chamorro15,Grad49,Jenkins85a,Jenkins85b,Garzo02,Santos04,Garzo13A}.
In fact, Grad's 13 moment method is a natural extension of the Chapman-Enskog expansion~\cite{Chapman} which can be regarded as Grad's $5$ moments method in terms of $d+2$ collisional invariance (the number of particles, the components of momentum and the energy).
Grad's 13 moment method consists of $d+2$ collisional invariants 
 plus the heat flux and the stress tensor.~\footnote{
Note that the number of independent components of the stress tensor for Grad's expansion is $(d-1)(d+2)/2\to 5$ for $d=3$, because the stress tensor is symmetric and the trace of the stress tensor is proportional to the kinetic energy. }   
We note that Grad's approximation is compatible with the Green-Kubo formula within the BGK approximation (see Appendix \ref{G-K_formula}).
It should be noted that the contribution from the Chapman-Enskog expansion is irrelevant in the present analysis, because its contribution disappears if the system is spatially uniform.
We also note that the heat flux is irrelevant for our problem.
Therefore, Eq.~\eqref{Grad} is a natural assumption to describe the nonequilibrium fluid under the shear.
The quantitative justification of Eq.~\eqref{Grad} will be examined through the comparison between the theoretical results in terms of Eq.~\eqref{Grad} and the results of simulation of Eqs.~\eqref{Langevin_eq} and \eqref{noise} in Sec. 4.

When we adopt Eq.~\eqref{Grad}, it is straightforward to show the relation 
\begin{equation}
\Lambda_{\alpha\beta}=\Lambda_{\alpha\beta}^L+\Lambda_{\alpha\beta}^{\rm NL} .
\end{equation}
Here the main contribution of $\Lambda_{\alpha\beta}$ is $\Lambda_{\alpha\beta}^L$ which is given by 
\begin{equation}\label{Lambda_maintext}
\Lambda_{\alpha\beta}^L=\nu (P_{\alpha\beta}-nT\delta_{\alpha\beta})+\lambda  nT \delta_{\alpha\beta} ,
\end{equation}
where 
$\nu$ and $\lambda$ are, respectively, given by~\cite{Santos04}
 (see Eq.~\eqref{eq:Lambda} in Appendix \ref{details_app} for details)
\begin{eqnarray}\label{nu}
\nu&=&n \sqrt{T}\nu_0
; \quad 
\nu_0 =\frac{2\pi^{(d-1)/2}   \sigma^{d-1} (1+e)(2d+3-3e) }{d(d+2)\Gamma(d/2)\sqrt{m}} ,
\\
\lambda&=&(1-e^2)n\sqrt{T} \lambda_0;
\quad
\lambda_0=
\frac{2\pi^{(d-1)/2}  \sigma^{d-1} }{d\Gamma(d/2)\sqrt{m}} 
,
\label{nu'}
\end{eqnarray}
where $\Gamma(x)\equiv \int_0^\infty dt t^{x-1} e^{-t}$.
Here, we have introduced constants $\nu_0$ and $\lambda_0$ whose details are unimportant for later discussion.

As will be shown, the nonlinear correction $\Lambda_{\alpha\beta}^{\rm NL}$
is expressed as
\begin{equation}\label{Lambda^NL_main}
\Lambda_{\alpha\beta}^{\rm NL}={\cal L}_{\alpha\beta\gamma\mu\nu\rho} (P_{\gamma\mu}-nT \delta_{\gamma\mu})(P_{\nu\rho}-nT\delta_{\nu\rho})
\end{equation}
where the explicit form of the six-order tensor
${\cal L}_{\alpha\beta\gamma\mu\nu\rho}$ for general dimension $d$ is not important.
Instead, we will use $\Lambda_{\alpha\beta}^{\rm NL}$ to evaluate the second normal stress difference for $d=3$. 
This term is almost invisible for most of the cases except for the second normal stress difference $N_{p,2}\equiv (P_{yy}-P_{zz})/P$.

With the aid of the linearization, i. e. $\Lambda_{\alpha\beta}\approx \Lambda_{\alpha\beta}^L$, and thus
from Eqs.~\eqref{Lambda_maintext}-\eqref{nu'}, 
Eq.~\eqref{stress_eq} can be rewritten as
 three coupled equations:
\begin{eqnarray}\label{d_tT}
\frac{d T}{d t} &=& -\frac{2\dot\gamma}{d n}P_{xy}-\lambda T+2\zeta (T_{\rm ex}-T) , \\
\label{d_tDT}
\frac{d \Delta T}{dt} &=&-\frac{2}{n}\dot\gamma P_{xy}-(\nu+2\zeta) \Delta T , \\
\label{d_tP_{xy}}
 \frac{d P_{xy}}{dt} &=& \dot\gamma n\left(\frac{\Delta T}{d}-T\right)
-(\nu+2\zeta)P_{xy} ,
\end{eqnarray} 
where we have introduced $\Delta T\equiv(P_{xx}-P_{yy})/n$ and used $P_{yy}=P_{zz}$ for $d=3$.
It should be noted that $P_{yy}$ is not always equal to $P_{zz}$ in general~\cite{Chamorro15,Sangani96}, but the equality $P_{yy}=P_{zz}$ is held when we ignore nonlinear contributions, i. e. $\Lambda_{\alpha\beta}^{\rm NL}$. 

Because the set of linearized equations~\eqref{d_tT}--\eqref{d_tP_{xy}} gives the precise results for $P_{\alpha\beta}$ except for $N_{p,2}$, 
we only include the effects of $\Lambda_{\alpha\beta}^{\rm NL}$ for the evaluation of the second normal stress difference $N_{p,2}$ in terms of the perturbation as
\begin{equation}\label{eq_N2}
N_{p,2}=-\frac{\Lambda_{yy}-\Lambda_{zz}}{2\zeta P} 
\approx
-\frac{\Lambda_{yy}^{\rm NL}-\Lambda_{zz}^{\rm NL}}{(\nu+2\zeta)P} .
\end{equation}

We stress that Eqs.~\eqref{d_tT}--\eqref{d_tP_{xy}} are coupled equations for the pressure, the shear stress and the normal stress difference.
Thanks to this set of coupled equations once the normal stress difference becomes large, the shear stress and the pressure can be large.
Therefore, we expect that the DST can take place if the discontinuous transition of the kinetic temperature between a quenched state and an ignited state exists or the sudden increment of the normal stress difference exists.

\section{Rheology of hard-core suspensions}

Now, let us derive a relation between the shear rate $\dot\gamma$ and the viscosity $\eta\equiv -P_{xy}/\dot\gamma$  for hard-core dilute inertial suspensions from
Eqs.~\eqref{d_tT}-\eqref{d_tP_{xy}} as well as the relations of $T$ and $\Delta T$ against $\dot\gamma$. 
Here, we introduce the following dimensionless quantities:
\begin{equation}\label{nu**}
\nu^{*}=\frac{\nu}{\sqrt{\theta}\zeta}, \quad
\lambda^{*}=\frac{\lambda}{\sqrt{\theta}\zeta}, \quad
\dot\gamma^{*}=\frac{\dot\gamma}{\zeta} 
\end{equation}
where we have introduced $\theta\equiv T/T_{\rm ex}$.
Note that $\dot\gamma^*$ corresponds to the Stokes number ${\rm St}=\rho_p\sigma^2\dot\gamma/\eta_0=18\dot\gamma^*$ in Refs.~\cite{Tsao95,Sangani96}.
With the aid of Eqs. \eqref{nu} and \eqref{nu'} the explicit expressions of $\nu^*$ and $\lambda^*$ are, respectively, given by 
\begin{equation}
\nu^*=\frac{2^d(1+e)(2d+3-3e)}{(d+2)\sqrt{\pi}}\xi_{\rm ex} \varphi, 
\quad \lambda^*=\frac{2^d}{\sqrt{\pi}}(1-e^2)\xi_{\rm ex}\varphi ,
\end{equation}
where we have introduced the dimensionless quantity $\xi_{\rm ex}\equiv \sqrt{T_{\rm ex}/m}/(\sigma \zeta)$ and the volume fraction $\varphi$. 
Note that $\xi_{\rm ex}$ is expected to be independent of $T_{\rm ex}$ because of $\zeta\propto \eta_0\propto \sqrt{T_{\rm ex}}$ with the viscosity of the solvent $\eta_0$. 

In a steady state, Eqs.~\eqref{d_tT} and \eqref{d_tDT} are reduced to
\begin{equation}\label{DT/T**}
\frac{\Delta \theta}{\theta}=
\frac{d\{\lambda^{*}\sqrt{\theta}+2(1-\theta^{-1}) \}}{\nu^{*}\sqrt{\theta}+2} ,
\end{equation}
where $\Delta\theta\equiv \Delta T/T_{\rm ex}$.
Substituting this into Eq.~\eqref{d_tDT} we obtain the equation for $P_{xy}^{*}=P_{xy}/(nT_{\rm ex})$ :
\begin{equation}\label{P_{xy}**}
P_{xy}^{*}=
-\frac{d\theta}
{2\dot\gamma^{*}}
\left\{ 
\lambda^{*}\sqrt{\theta}+2(1-\theta^{-1})
\right\} .
\end{equation}
Then, substituting Eqs.~\eqref{DT/T**} and \eqref{P_{xy}**} into the steady equation of Eq.~\eqref{d_tP_{xy}} we obtain
\begin{equation}\label{dgamma**}
\dot\gamma^{*}=
(\nu^{*}\sqrt{\theta}+2)
\displaystyle\sqrt{
\frac{
d[\lambda^{*}\sqrt{\theta}+2(1-\theta^{-1})]
}
{
2[(\nu^{*}-\lambda^{*})\sqrt{\theta}+2\theta^{-1}]
}
} .
\end{equation}
Therefore, the dimensionless viscosity $\eta^*=-P_{xy}^*/\dot\gamma^*$ is given by
\begin{equation}\label{eta**}
\eta^*= \frac{\theta\{(\nu^{*}-\lambda^{*})\sqrt{\theta}+2\theta^{-1}\}}
{(\nu^{*}\sqrt{\theta}+2)^2} .
\end{equation} 

Unfortunately, we cannot express $\eta^*$ in Eq.~\eqref{eta**} as a function of  $\dot\gamma^*$ in Eq.~\eqref{dgamma**} explicitly, but
$\eta^*$ and $\dot\gamma^*$ can be parametrically expressed as functions of $\theta$.
We can also give the explicit asymptotic forms in the limits $\dot\gamma^*\to 0$ and $\dot\gamma^*\to \infty$. 
This model exhibits the crossover from the Newtonian regime for $\dot\gamma\to 0$: 
\begin{equation}\label{Newtonian_regime}
\eta^{*}\to 
\frac{\theta_{\rm min}}
{ \nu^*\sqrt{\theta_{\rm min}}+2}
\end{equation}
at $\theta=\theta_{\rm min}$, where $\theta_{\rm min}$ is a real solution of $\theta=(1+\lambda^*\sqrt{\theta}/2)^{-1}$ ($\theta_{\rm min}=1$ for $e=1$),
 to the Bagnoldian viscosity 
\begin{equation}\label{Bagnold_viscosity}
\eta^{*}\to \displaystyle\sqrt{\frac{2}{d}} \frac{(\nu^*-\lambda^*)^{3/2}\dot\gamma^*}{\sqrt{\lambda^*}\nu^{*3}
}, 
\quad
\theta\to \frac{2(\nu^*-\lambda^*)\dot\gamma^{*2}}{d\nu^{*2}\lambda^*}
\end{equation}
in the high shear rate limit  for $e<1$.
Note that the asymptotic forms 
\begin{equation}\label{asymptotic_e=1}
\eta^* \to \frac{\dot\gamma^{*2}}{d\nu^{*2}}
, \quad
\theta\to \frac{\dot\gamma^{*4}}{(d\nu^*)^2} 
\end{equation}
 in the high shear limit 
for the elastic case $e=1$ are quite different from those for $e<1$.
The asymptotic behavior of $\eta^*$ corresponding to Eq.~\eqref{asymptotic_e=1} is consistent with the observation in Ref.~\cite{Kawasaki14} and theoretical prediction in Ref.~\cite{BGK2016}.
We also note that $\nu^*$ and $\lambda^*$ depend on the restitution coefficient $e$ as shown in Eqs. \eqref{nu} and \eqref{nu'}, where $\lambda^*$ becomes zero at $e=1$.

Equations \eqref{Newtonian_regime}--\eqref{asymptotic_e=1} can be rewritten as
\begin{eqnarray}\label{Newton_phys}
 \eta&\to &\frac{n T_{\rm min}}{2\zeta}  , \\
\eta &\to& 
\frac{\sqrt{2}(\nu_0-(1-e^2)\lambda_0)^{3/2}}{\sqrt{(1-e^2)d\lambda_0}\nu_0^3n}\dot\gamma
 ,
\quad 
T\to 
\frac{2(\nu_0-(1-e^2)\lambda_0)}{(1-e^2)d\nu_0^2\lambda_0 n^2}\dot\gamma^2 , 
\label{bagnold_dim}
\\
\eta &\to& 
\frac{1}{\zeta d\nu_0^2 n}\dot\gamma^2 ,
\quad
T\to \frac{1}{\zeta^2d\nu_0^2 n^2}\dot\gamma^4 ,
\label{high_shear_dim}
\end{eqnarray}
respectively, in dimensional forms, where $T_{\rm min}=\theta_{\rm min} T_{\rm ex}$ reduces to $T_{\rm ex}$ in the elastic limit $e\to 1$.
To derive Eq.~\eqref{Newton_phys} we have used $\nu^*\sqrt{\theta_{\rm min}}\ll 1$ in the dilute limit because of $\nu^*\propto \varphi$ and $\theta_{\rm min}\approx 1$ for $e \lesssim 1$.
The apparent viscosity in Eq.~\eqref{Newton_phys} is proportional to $\eta_0$ for $e=1$ because of $T_{\rm min}\to T_{\rm ex}$ and
 $\zeta\propto \eta_0\propto \sqrt{T_{\rm ex}}$.
Equations \eqref{Newton_phys}-\eqref{high_shear_dim} have interesting characteristics.
In the low shear regime in Eq.~\eqref{Newton_phys}, the viscosity is Newtonian which approaches zero if $T_{\rm ex}\to 0$.
Thus, our theory rescues the drawback of the previous analysis in Ref.~\cite{Tsao95} without adding the Newtonian viscosity by hand~\cite{comment_Tsao2}.
Note that both $\eta$ and $T$ diverge in the dilute limit $n\to 0$ and the high shear limit $\dot\gamma\to \infty$ (see Eqs.~\eqref{bagnold_dim} and \eqref{high_shear_dim}).
Therefore, there exist the large contrasts in both the viscosity and the kinetic temperature between the low shear (quenched) regime and the high shear (ignited) regime for the dilute gas.
We also note that Eqs.~\eqref{bagnold_dim} and \eqref{high_shear_dim} satisfy
$\eta=\{1-(1-e^2)\lambda_0/\nu_0\}\nu_0^{-1}\sqrt{T}$ and $\eta=\sqrt{T}/(\sqrt{d}\nu_0)$, respectively, which are known results~\cite{Santos04}.

Figure \ref{fig:fig1} (a) corresponding to Eq.~\eqref{dgamma**}  plots the behavior of the dimensionless kinetic temperature $\theta$ against $\dot\gamma^*$, which exhibits a discontinuous transition between the quenched state for low shear regime and the ignited state for high shear regime.
Note that we have adopted the set of parameters for this plot
$d=3$, $n\sigma^3=0.01$ corresponding to $\varphi=(\pi/6)n\sigma^3 \approx 0.0052$, $\xi_{\rm ex}=1.0$, which will be commonly used for the later discussions including our simulation in the next section.
This result corresponds to that obtained by Tsao and Koch~\cite{Tsao95}, but
the behavior for $\dot\gamma^*\to 0$ is different because of the different asymptotic viscosities in the low shear regime.   
Indeed, the theoretical prediction of $T$ in Ref.~\cite{Tsao95} is $T\sim {\rm St}^3\propto \dot\gamma^3$ in the limit $\dot\gamma^*\to 0$ but our result becomes $T\to T_{\rm min}$ for $\dot\gamma^*\to 0$~\cite{comment_Tsao2}.
Needless to say, it is natural that the temperature of suspension is identical to the environmental temperature {\it i. e.} $T=T_{\rm ex}$ in the absence of the shear in the elastic limit $e\to 1$. 
Note that there is a linearly unstable region for the steady solution which is indicated by the thick line in Fig.~\ref{fig:fig1} (a). 
See Appendix \ref{app:linear} for the details of the linear stability analysis. 

As shown in Fig.~\ref{fig:fig1} (b) corresponding to Eq.~\eqref{eta**}
 (the solid line for $e=0.9$ and the dashed line for $e=1$), 
we have also confirmed the existence of DSTs, in which
the flow curves have S-shapes. 
Then, if we gradually increase/decrease $\dot\gamma^*$, 
the viscosity $\eta^*$ discontinuously increases/decreases at a certain value of the shear rate. 
Therefore, this S-shape flow curve directly leads to the existence of the DST.
Note that the viscosity also has the linearly unstable region of the steady solution indicated by the thick line in Fig.~\ref{fig:fig1} (b).


We also plot the theoretical prediction of $\Delta \theta$ against $\dot\gamma^*$ (see Fig.~\ref{fig:fig2}) corresponding to Eq.~\eqref{DT/T**}.
It is remarkable that $\Delta \theta$ is insensitive to $e$. 
This is because the asymptotic forms,
 $\Delta \theta \to
\theta_{\rm min}^4\dot\gamma^{*2}\{(\nu^*-\lambda^*)\sqrt{\theta_{\rm min}}+2\theta_{\rm min}^{-1})\}/4
$
 in the limit $\theta\to \theta_{\rm min}$ and $\Delta \theta\to 2(\nu^*-\lambda^*)\dot\gamma^{*2}/\nu^{*3}$ in the limit $\dot\gamma^*\to \infty$,
 do not have any singularity at $e=1$.

The stress ratio $\mu\equiv -P_{xy}/P$ is also an important quantity to characterize the rheology~\cite{GDRMidi}.
With the aid of Eq.~\eqref{P_{xy}**} and the equation of state $P=nT$, we obtain
\begin{equation}
\mu=-\frac{P_{xy}^*}{\theta}=
\frac{d}
{2\dot\gamma^{*}}
\left\{ 
\lambda^{*}\sqrt{\theta}+2(1-\theta^{-1})
\right\} .
\end{equation}
Because we do not control $P$, we plot $\mu$ against $\dot\gamma^*$ (see Fig.~\ref{fig:fig3}).
The stress ratio satisfies $\mu\to \dot\gamma^*/(\nu^*\sqrt{\theta_{\rm min}}+2)$ in the limit $\dot\gamma^*\to 0$ and has a peak around $\dot\gamma^*\approx 3$.
Then $\mu$ takes multiple values as the result of the S-shape in the flow curve.
The asymptotic form of $\mu$ in the limit $\dot\gamma^*\to \infty$ strongly depends on $e$ as
$\mu\to \sqrt{d/2}\lambda^*/\{\nu^*\sqrt{\nu^*-\lambda^*} \}$ for $e<1$
and $\mu\to d/\dot\gamma^*$ for $e=1$.
It is interesting that inelastic collisions ($e<1$) create a finite stress ratio, which might be one of important characteristics of macroscopic collections of granular particles.

Although the second normal stress difference $N_{p,2}$ is zero if we adopt the linearization, i. e. $\Lambda_{\alpha\beta}= \Lambda_{\alpha\beta}^L$, 
$N_{p,2}$ can be finite if we consider $\Lambda_{\alpha\beta}^{\rm NL}$ as shown in Eq.~\eqref{eq_N2}.
From the derivation in Appendix \ref{details_app} (the final result in Eq.~\eqref{N2}), we obtain
\begin{equation}
N_{p,2}=-\frac{C_3 n \sigma^3}{(\nu^*\sqrt{\theta}+2)\theta^{3/2}}P_{xy}^{*2}\sqrt{\frac{T_{\rm ex}}{m\sigma^2\zeta^2}}
\label{N2_main}
\end{equation}
for $d=3$, 
where $C_3=8(6-6601A)\sqrt{\pi}A/105$ and $A=(1+e)/2$.
We have also used Eqs.~\eqref{nu**}, \eqref{P_{xy}**} and \eqref{dgamma**}. 


\section{Simulation for hard-core suspensions}

\begin{figure}
 \begin{center}
  \includegraphics[width=\linewidth]{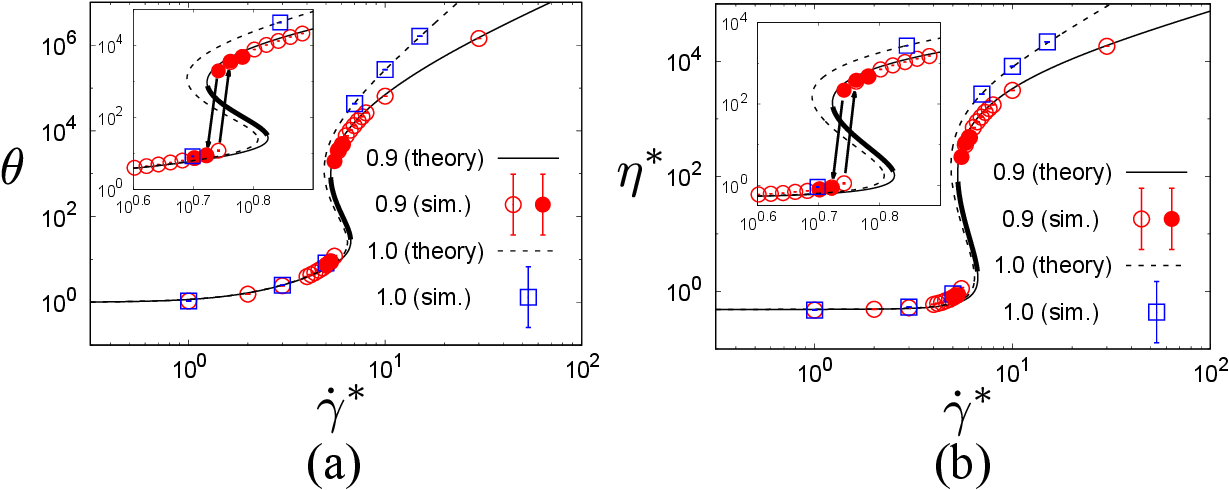}
 \end{center}
  \caption{
(a) Plots of $\theta$ against $\dot\gamma^*$ for $e=0.90$ (circles for simulation) and $e=1$ (open squares for simulation) for $d=3$ and $n\sigma^3=0.01$.
  (b) Plots of  $\eta^*$ against $\dot\gamma^*$ for $e=0.90$ (circles for simulation) and $e=1$ (open squares for simulation) for $d=3$ and $n\sigma^3=0.01$.
The theoretical curves are represented by the solid line for $e=0.9$ and the dashed line for $e=1$, respectively. The data are obtained by the EDLSHS. Near the DST, we gradually increase/decrease $\dot\gamma^*$ corresponding to the open and the solid circles. 
The thick lines stand for the linearly unstable regions for $e=0.9$.
The insets show enlarged ones around the discontinuous transition.
  }
\label{fig:fig1}
\end{figure}

\begin{figure}
 \begin{center}
  \includegraphics[width=0.5\linewidth]{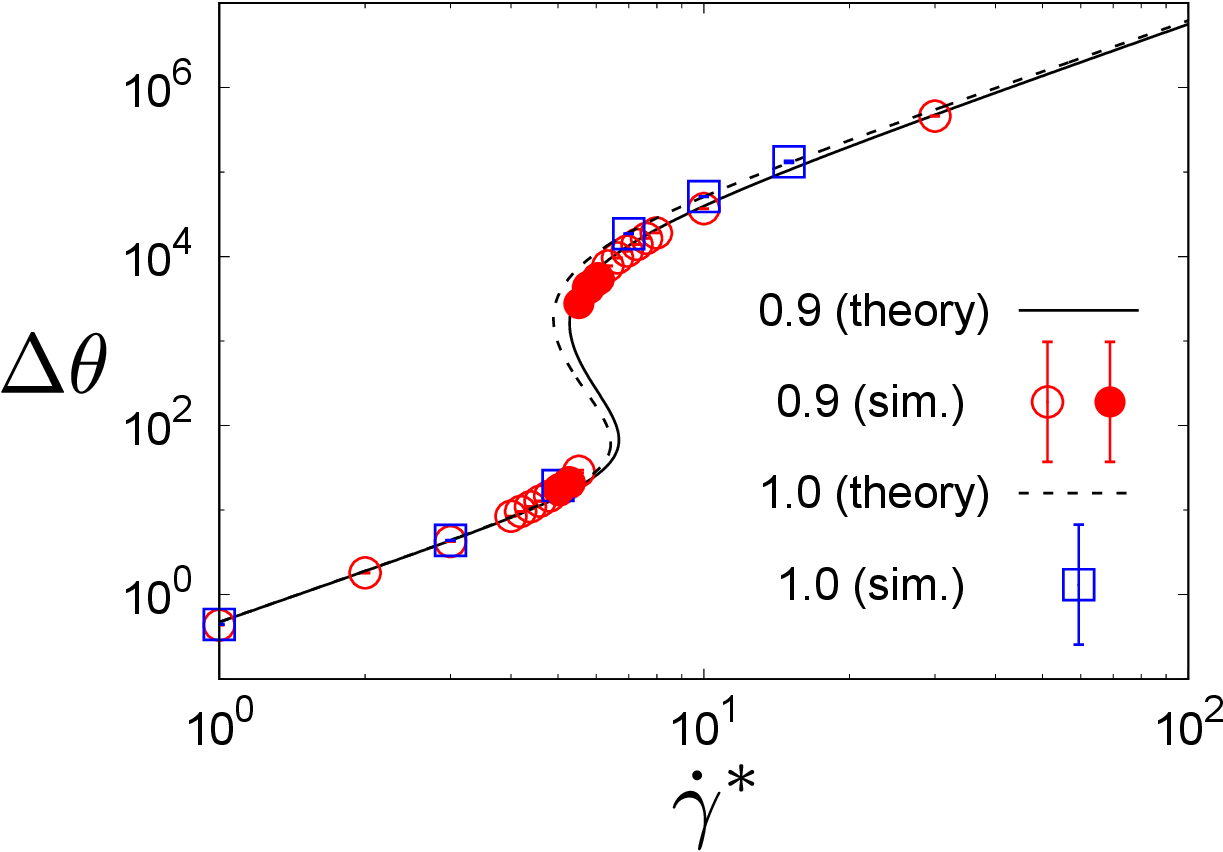}
 \end{center}
  \caption{
 Plots of  $\Delta \theta$ against $\dot\gamma^*$ for $e=0.90$ and $e=1$ with $d=3$ and $n\sigma^3=0.01$.
The legend and the parameters are the same as those in Fig.~\ref{fig:fig1}
.
  }
\label{fig:fig2}
\end{figure}

\begin{figure}
 \begin{center}
  \includegraphics[width=0.5\linewidth]{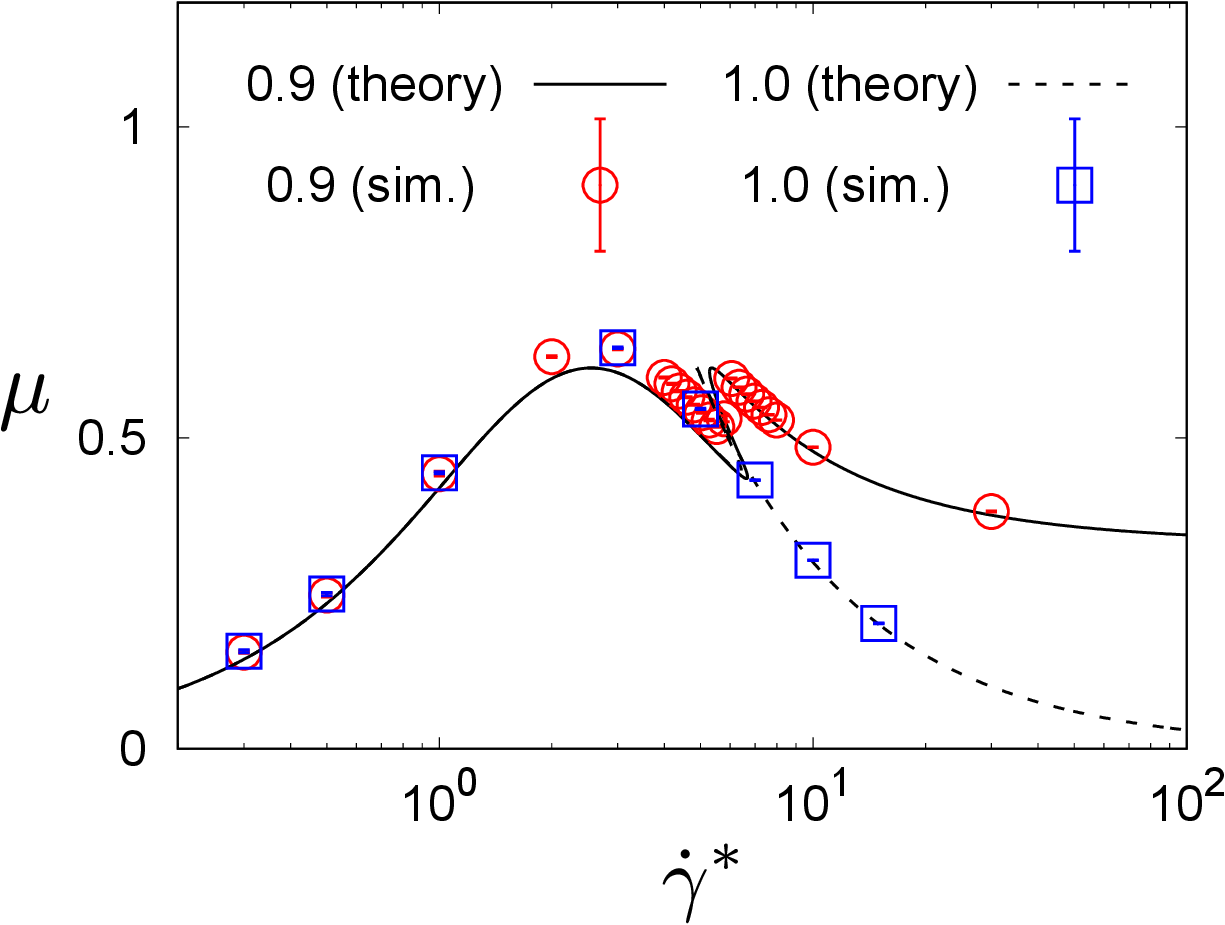}
 \end{center}
  \caption{
   Plots of  the stress ratio $\mu=-P_{xy}/P$ against $\dot\gamma^*$ for $e=0.90$ and $e=1$ in the case of $d=3$ and $n\sigma^3=0.01$.
The legend and the parameters are the same as those in Fig.~\ref{fig:fig1}.
  }
\label{fig:fig3}
\end{figure}


Let us check the quantitative validity of our theoretical results by using the three dimensional simulation under the Lees-Edwards boundary condition~\cite{Scala12,Bannerman11,Lees72}.
Because the Langevin equation \eqref{Langevin_eq} with Eq.~\eqref{noise} for dilute and uniform suspensions can be mapped onto the Boltzmann equation \eqref{boltzmann} with Eq.~\eqref{J(V|f)},
we simulate Eq.~\eqref{Langevin_eq} instead of numerically solving Eq.~\eqref{boltzmann}. 
Note that the results explained in this section are only those for hard-core dilute suspensions. 

It is difficult to implement the standard event-driven code for hard spheres because of the existence of the drag term in Eq.~\eqref{Langevin_eq}.
On the other hand, it is almost impossible to adopt soft-core models for the simulation of
 Eq.~\eqref{Langevin_eq} to reproduce the DST in this setup, because spheres are largely overlapped if the shear rate $\dot\gamma$ is large. 
Moreover, there is numerical difficulty to simulate the situation if $\eta^*$ discontinuously changes with the order of $10^3$ at a fixed $\dot\gamma^*$.
To overcome such difficulties, 
the event-driven Langevin simulation of hard spheres (EDLSHS)~\cite{Scala12}
is a powerful simulator for hard spheres.  
See Appendix \ref{EDLSHS} for the outline of the method of the EDLSHS.

In our simulation, we fix the number of particles $N=1000$ for $d=3$.
  In the vicinity of the DST for $\dot\gamma^* \in [0.400, 0.798]$, we gradually change the shear rate from $\dot\gamma_0^*=0.400 (0.798)$ to sequentially increasing (decreasing) values as $\dot\gamma^*=\dot\gamma_0^*$, $a\dot\gamma_0^*$, $a^2\dot\gamma_0^*$, $\cdots$, $a^{15} \dot\gamma_0^*=0.798(0.400)$ with the rate $a=10^{\pm 0.02}$.\

The main results of our simulation are presented in Figs.~\ref{fig:fig1}--\ref{fig:fig3}. 
All of the numerical results perfectly agree with the theoretical results without 
any fitting parameters.
We also find that the discontinuous jumps of $\eta^*$ take place at different $\dot\gamma^*$ depending on the protocol (if $\dot\gamma^*$ increases/decreases). 
This protocol dependence means that there is a hysteresis in the DST. 

Thus, we have confirmed the quantitative validity of our simple kinetic theory in terms of the Boltzmann equation to describe the DST.
We also confirm that the dilute suspension described by Eq.~\eqref{boltzmann} or Eq.~\eqref{Langevin_eq} can exhibit the DST.

\begin{figure}
 \begin{center}
  \includegraphics[width=0.5\linewidth]{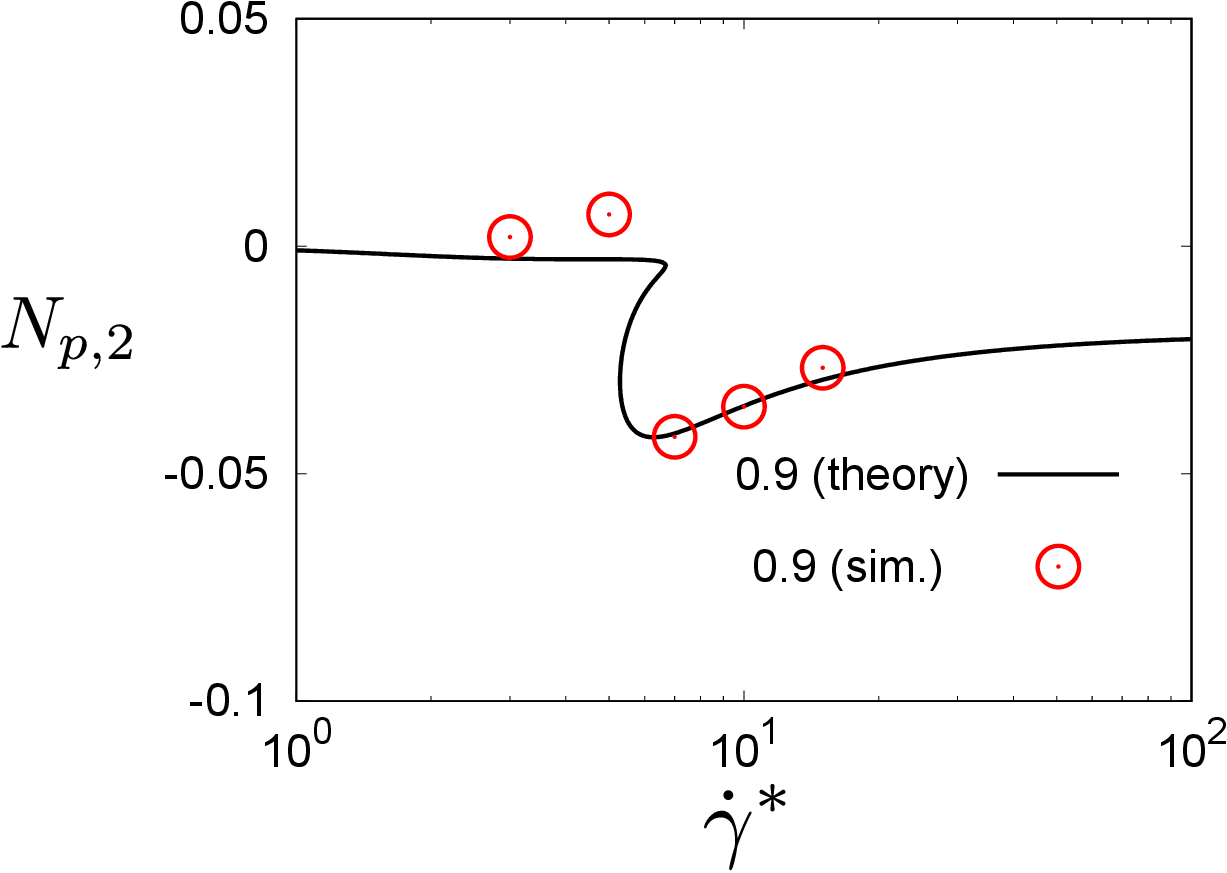}
 \end{center}
  \caption{
  The second normal difference $N_{p,2}$ from the simulation and the theory.
  Here, we adopt $\xi_{\rm ex}=1.2247\times 10^{-3}$ as a fitting parameter.
  }
  \label{fig:fig4}
\end{figure}

We plot $N_{p,2}$ obtained in Eq.~\eqref{N2_main} against the shear rate in Fig.~\ref{fig:fig4}.
Although we have adopted $\zeta\Delta t=0.1$ as explained in Appendix \ref{EDLSHS} and $\xi_{\rm ex}=1.0$, 
we simply adopt $\xi_{\rm ex}=1.2247\times10^{-3}$ as a fitting parameter.
Then, the theoretical result of $N_{p,2}$ agrees with the simulation results within this treatment.

\begin{figure}
 \begin{center}
  \includegraphics[width=150mm]{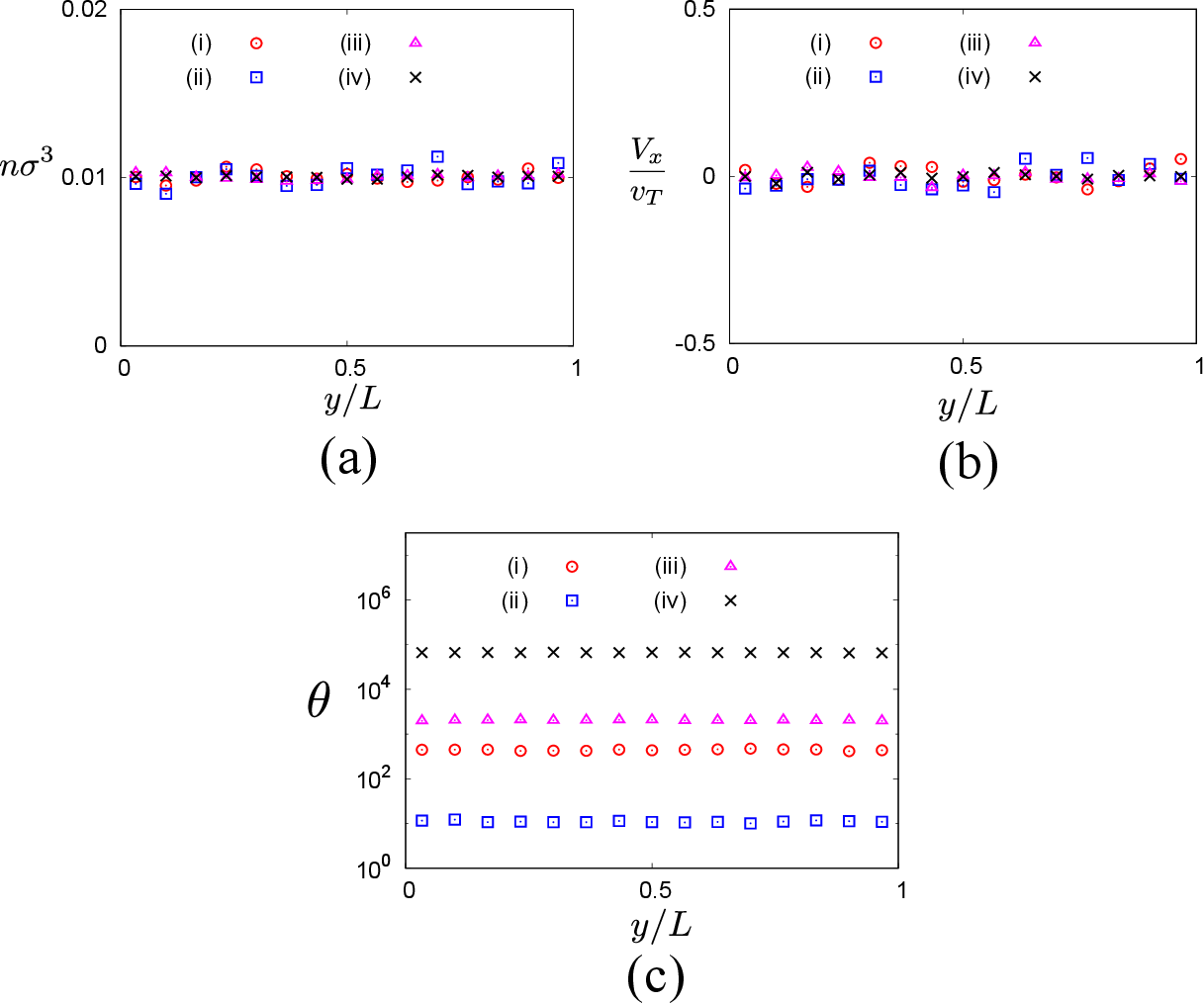}
 \end{center}
  \caption{
 (a)  Plots of the density ($n\sigma^3$) profiles in $y$-direction in the steady states for (i) $\dot\gamma^*=0.527$ (open circles), (ii) $0.552$ (open squares)  in the upper branch, (iii) $0.552$ (open triangles) in the lower branch and (iv) 1.0 (crosses) for $e=0.90$ in the case of $d=3$ and $n\sigma^3=0.01$ with the system size $L=46.4\sigma$.
In this figure, the density $n\sigma^3$ is averaged over $x$ and $z$ directions.
  (b)
 Plots of $V_x/v_T$ in $y$-direction for $e=0.90$, $d=3$ and $n\sigma^3=0.01$, where $v_T\equiv \sqrt{2T/m}$ for various shear rates. 
 The legend and the parameters are the same as those in (a). 
 (c)
  Plots of the temperatures in $y$-direction for $e=0.90$ in the case of $d=3$ and $n\sigma^3=0.01$ for various shear rates.
  The legend and the parameters are the same as those in (a).
  }
  \label{fig:fig5}
\end{figure}

Note that the spatial inhomogeneities for $\theta$, $\bm{V}$, and $n$ have not been observed for all $\dot\gamma^*$, at least, within our simulation~
(see Fig.~\ref{fig:fig5}) in steady states. 
These results are partially because
the thermal noise in Eq.~\eqref{Langevin_eq} stabilizes the homogeneous structure for small $\dot\gamma^*$, though such homogeneity might be violated for highly dissipative systems.  

We comment on two points before closing this section.
First, the relaxation time to reach the steady state becomes longer as the observed point approaches the critical point of the DST (see Appendix \ref{app_relaxation}).
Second, the time evolution of the domain growth can be observed if we start the simulation from an unstable point (see Appendix~\ref{app_domain}). 
This is analogous to the phase ordering process after the system is quenched into an unstable point~\cite{Bray94}.

\section{The role of an attractive interaction}

\begin{figure}
 \begin{center}
  \includegraphics[width=\linewidth]{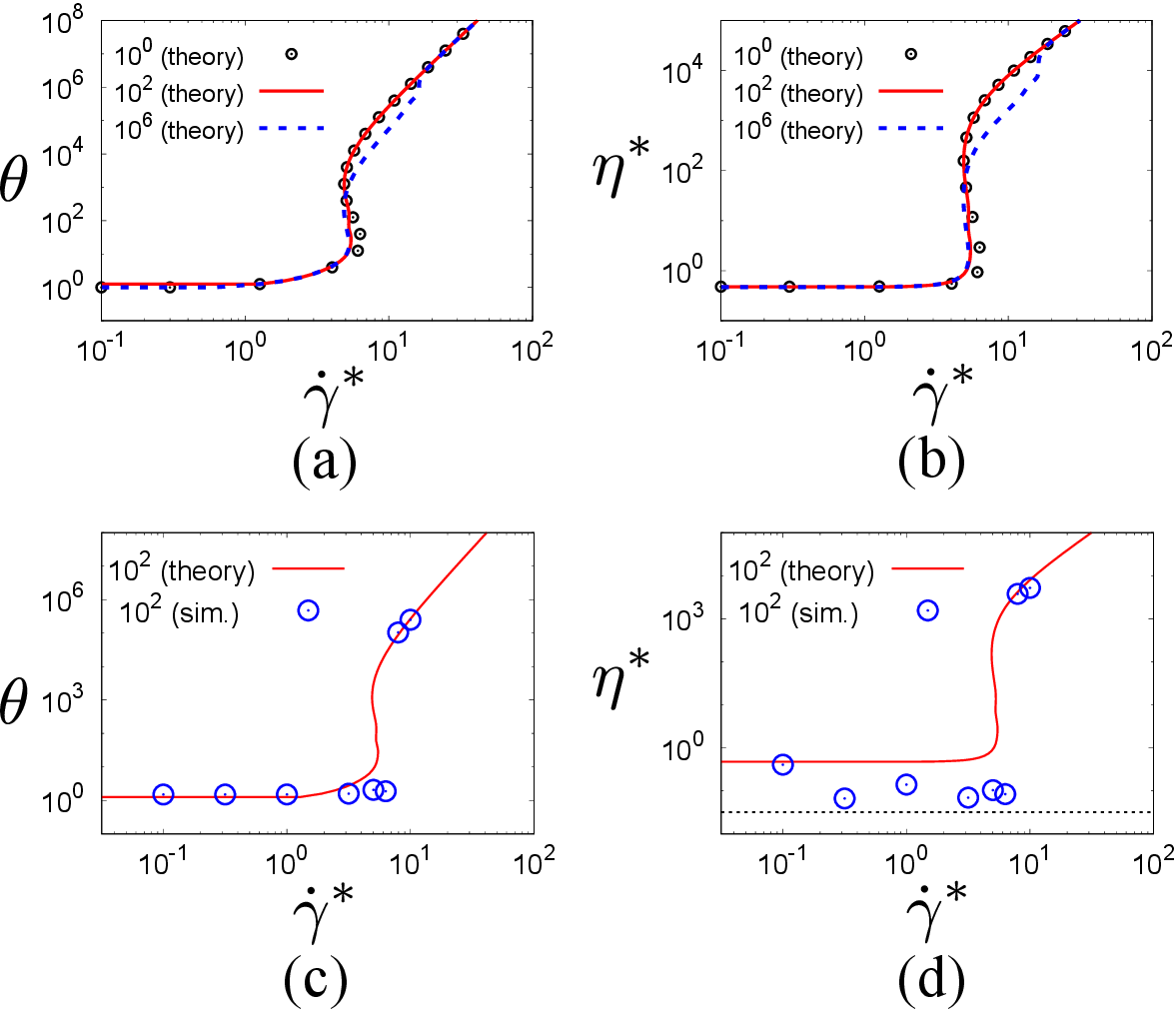}
 \end{center}
  \caption{
   Plots of (a) the temperature and (b) the shear viscosity against the shear rate for $\omega=1$ (black open circles), $10^2$ (red solid lines), and $10^6$ (blue dashed lines) with $n\sigma^3=0.01$, $\kappa=1.5$, and $e=1$.
   The lower panels ((c) and (d)) show the comparisons of the theory (red solid line) with the simulations (blue open circles) for $\omega=10^2$.
   The dotted line in (d) represents the effective shear viscosity \eqref{eq:vis_eff} for $\sigma_{\rm cl}=2.5\sigma$.
  }
  \label{fig:fig6}
\end{figure}
\begin{figure}
 \begin{center}
  \includegraphics[width=\linewidth]{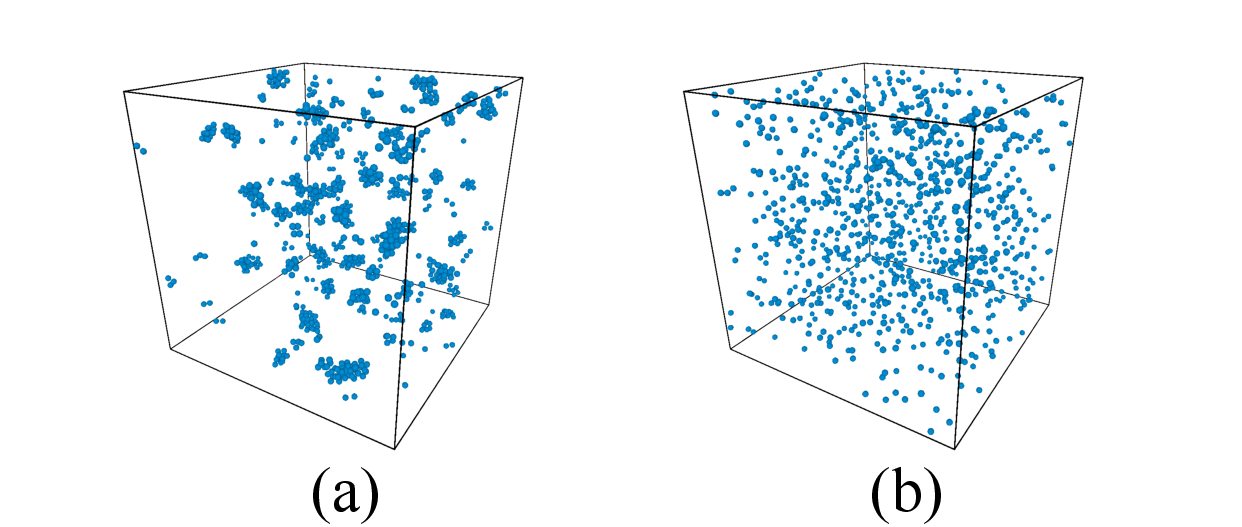}
 \end{center}
  \caption{
 Typical snapshots for cohesive systems for (a) $\dot\gamma^*=10^{-1.5}$ (low shear regime) and (b) $10^{-0.5}$ (high shear regime) with $\omega=10^2$ and $e=1$.
  }
  \label{fig:fig7}
\end{figure}


As mentioned in Introduction, a typical example of the dilute inertial suspension is the aerosol. 
Because there exists an attractive force between the aerosol particles,   
we need to clarify the role of an attractive interaction between grains in dilute suspensions.

For simplicity, we only analyze the cases of $d=3$ and $e=1$.
We assume that the interparticle potential is given by the square-well potential (its abbreviation is SW):
\begin{equation}\label{U_SW}
	U_{\rm SW}(r)=
	\begin{cases}
	\infty & (r\le \sigma)\\
	-\varepsilon & (\sigma< r\le \kappa \sigma)\\
	0 & (r>\kappa\sigma)
	\end{cases},
\end{equation}
where $\varepsilon$ and $\kappa$ are the well depth and well width ratio, respectively.

Now, the collision integral $J_{\rm SW}(\bm{V}|f)$ for the cohesive particles described by Eq.~\eqref{U_SW} corresponding to $J(\bm{V}|f)$ for the hard-core case becomes a little complicated,
but its explicit form (see, e.g. Eq. (12) in Ref.~\cite{Takada18}) is not important for our analysis.
The most important quantity for the rheology is the moment of the collision integral $\Lambda_{\alpha\beta}^{\rm SW}$ introduced in Eq.~\eqref{Lambda_maintext},
which is reduced to
\begin{equation}\label{Lambda_SW}
	\Lambda_{\alpha\beta}^{\rm SW}=\nu_{\rm SW}(P_{\alpha\beta}-nT \delta_{\alpha\beta}),
\end{equation}
where $\nu_{\rm SW}$ is given by \cite{Takada18}
\begin{align}
	\nu_{\rm SW}&= \frac{4}{5}n\sigma^2 \sqrt{\frac{\pi \varepsilon}{m}}\left(\frac{\varepsilon}{T}\right)^{7/2}
		\int_0^\infty dc_{12} \int_0^\infty d\tilde{b} \hspace{0.2em}\tilde{b}c_{12}^7
		\sin^2\frac{\chi}{2}\cos^2\frac{\chi}{2}\exp\left(-\frac{\varepsilon}{2T}c_{12}^2\right),
\end{align}
with
\begin{align}
	\chi&=
	\begin{cases}
		\pi - 2\sin^{-1}\frac{\tilde{b}}{\kappa} - 2\sin^{-1}\frac{\tilde{b}}{\mathfrak{N}} + 2\sin^{-1}\frac{\tilde{b}}{\mathfrak{N}\kappa} & (\tilde{b}\le \min(\kappa, \mathfrak{N}))\\
		2\sin^{-1}\frac{\tilde{b}}{\mathfrak{N}\kappa} - 2\sin^{-1}\frac{\tilde{b}}{\kappa} & (\min(\kappa, \mathfrak{N}) < \tilde{b}\le\kappa)\\
		0 & (\tilde{b}> \kappa) ,
	\end{cases}
\end{align}
where $\mathfrak{N}=\{1+2\varepsilon/(Tc_{12}^2)\}^{1/2}$ is the refractive index.
Here, the ratio of the potential well depth to the external temperature $\omega\equiv \varepsilon/T_{\rm ex}$ plays important roles.
It should be noted that $\nu_{\rm SW}$ depends on $\omega$ and $\theta$. 
We have also introduced the dimensionless relaxation rate $\nu_{\rm SW}^*=\nu_{\rm SW}/(\sqrt{\theta}\zeta)$.

After straightforward and parallel calculation as in Secs. 2 and 3 (see Appendix \ref{SQUARE_WELL}), we obtain the steady rheological relations: 
\begin{eqnarray}\label{eq:38}
	\frac{\Delta \theta_{\rm SW}}{\theta_{\rm SW}} &=& \frac{6(1-\theta_{\rm SW}^{-1})}{\nu_{\rm SW}^*\sqrt{\theta_{\rm SW}}+2},\\
\label{eq:39}
	P_{xy}^{{\rm SW}*} &=& -\frac{3(\theta_{\rm SW}-1)}{\dot\gamma^*},\\
\label{eq:40}
	\dot\gamma^* &=& (\nu_{\rm SW}^*\sqrt{\theta_{\rm SW}}+2)\sqrt{\frac{3(1-\theta_{\rm SW}^{-1})}{\nu_{\rm SW}^*\sqrt{\theta_{\rm SW}}+2\theta_{\rm SW}^{-1}}},\\
	\eta_{\rm SW}^* &=& \frac{\theta_{\rm SW}(\nu_{\rm SW}^*\sqrt{\theta_{\rm SW}}+2\theta_{\rm SW}^{-1})}{(\nu_{\rm SW}^*\sqrt{\theta_{\rm SW}}+2)^2} ,
\label{eq:41}
\end{eqnarray}
where the subscript SW stands for the corresponding variable under the influence of the square-well potential. 

As shown in Fig.~\ref{fig:fig6}
(theoretical curves for $\theta$ and $\eta^*$ are, respectively,  given by Eqs.~\eqref{eq:40} and \eqref{eq:41}), 
the flow curves are basically unchanged from those of hard-core grains for $\omega=\varepsilon/T_{\rm ex} \leq 10^2$, but we find the existence of another discontinuous jump of $\theta$ 
for $\omega \gg 1$.
The quenched-ignited transition takes place at $\dot\gamma^*\simeq 5$, while  the discontinuous jump caused by the attractive interaction takes place at
$\dot\gamma^*\simeq 20$ which corresponds to $\theta\simeq \omega$, at least, for $\omega\gg 1$. 
Unfortunately, our simulation for $\omega=10^2$ cannot capture the second discontinuous jumps for $\theta$ and $\eta^*$ (see Figs.~\ref{fig:fig6} (c) and (d)), but the agreement between the theory and the simulation for $\omega=10^2$ is reasonable, at least, above the first discontinuous jump.
We can also indicate that the steady flow curve in low shear rate is still stable in contrast to the case of sheared granular gases without the noise~\cite{Takada18}.

In the low shear regime, the viscosity obtained from the simulation deviates from that from the theory.
This is because clusters form in this regime as shown in Fig.~\ref{fig:fig7}.
Here, the mean diameter of clusters is approximately given by $\sigma_{\rm cl}\simeq 2.5\sigma$.
If we assume that all clusters are identical, the number density for these clusters are given by $n_{\rm cl}=n(\sigma/\sigma_{\rm cl})^3$.
With the aid of Eq.~\eqref{Newton_phys}, the effective viscosity in the low shear regime is given by
\begin{equation}\label{eq:vis_eff}
	\eta_{\rm eff}=\frac{n_{\rm cl}T_{\rm min}}{2\zeta}
	=\frac{nT_{\rm min}}{2\zeta}\left(\frac{\sigma}{\sigma_{\rm cl}}\right)^3.
\end{equation}
This estimation is reasonable in the low shear regime as shown in Fig.~\ref{fig:fig6} (d).

The results in this section can be summarized as follows: (i) The attractive interaction can cause a new discontinuous jump in the flow curve for $\omega\gg 1$. (ii) The hard-core model discussed in the previous sections gives qualitatively reasonable results for grains with the attractive interaction for $\omega \le 10^2$, where the clustering effects in the low shear regime can be absorbed by the mean cluster diameter $\sigma_{\rm cl}$.

Although we have already recognized the discontinuous quenched-ignited transition for hard-core particles discussed in the previous sections disappears as the density increases~\cite{Sangani96,Saha17,Hayakawa-Takada-Garzo},
the discontinuous jump caused by the attractive interaction is expected to survive even if we consider dense suspensions as observed in Ref.~\cite{Irani18}.
The further careful research on this problem will be necessary.

\section{Discussion and conclusion}

 Let us discuss our results. 
This section consists of the discussion and the conclusion.

\subsection{Discussion}

One take-home message here is that the DST can take place in dilute inertial suspensions without mutual frictions between particles, though we have ignored the hydrodynamic interaction.
Absence of the hydrodynamic interactions can be justified because we are only interested in suspensions in the dilute limit.
Indeed, some previous papers used the density dependent $\zeta$ which reduces to $\zeta\propto 1+3\sqrt{\varphi/2}\to 1$ in the dilute limit~\cite{Tsao95,Sangani96}.
This encourages experimentalists to try to find the DST for solid-gas suspensions such as aerosols in which particles are only influenced by the Stokesian drag and the collisional force. 
The easiest experimental setup might be a sheared granular gas under vibration, because we often use Eq.~\eqref{boltzmann} for a simplified model of such a system. 




The Langevin equation \eqref{Langevin_eq} employed in our study assumes that the gravity force is perfectly balanced with the drag
force by the air. 
This assumption seems to be valid for aerosols within our observing time scale, though attractive interaction between grains is not negligible. 
If the homogeneous state is unstable, one
would need to consider the time evolution of local structure as well as the consideration of the inhomogeneous drag.
The research in the unstable region would be an interesting subject in the near future.

Let us compare our results with those of the pioneer paper~\cite{Tsao95} which found the discontinuous transition of the kinetic temperature.
Our theoretical results $T\sim \dot\gamma^4/n^2$ and $\eta\sim \dot\gamma^2/n$ agree with theirs in the high shear regime of elastic suspensions ($e=1$).
Nevertheless, there are some differences between theirs and ours as explained below.
First, Ref.~\cite{Tsao95} only focuses on the rheology for elastic suspensions ($e=1$), but we include the results for granular suspensions ($e<1$) which have distinct behavior in the high shear regime as shown Figs.~\ref{fig:fig1} and \ref{fig:fig3}.
(Note that Sangani {\it et al.} derived Banoldian expressions in Eq.~\eqref{bagnold_dim} for $\dot\gamma^*\gg 1$, though they have not written their explicit forms~\cite{Sangani96}.) 
Second, their kinetic calculation is only used for the ignited state, whereas physical quantities in the quenched state are calculated separately.
Our analysis, however, can use a unified calculation for both the ignited state and the quenched state. 
Third,  we believe that
their calculation is not applicable to the behavior in the low shear limit, because there is neither the (Newtonian) viscosity nor the kinetic temperature in this limit.
Indeed, their calculation in this regime suggests $T\sim \varphi {\rm St}^3$ and $\eta\sim \varphi^2{\rm St}^2$ which approach zero in the limit ${\rm St}\to 0$~\cite{Tsao95},
which is the reason why non-existence of quenched viscosity in Fig.~\ref{fig:fig1} of Ref.~\cite{Tsao95}.
This structure is unchanged even in a recent paper~\cite{Saha17}.
We believe that they recognized the drawback of their analysis, because
they added the Newtonian viscosity to their calculated viscosity in low shear regime (see the argument to derive Eq.~(4.13) in Ref.~\cite{Tsao95}) by hand. 
Our model, however, can describe the Newtonian rheology in the low shear regime without any artificial trick.
Therefore, we believe that it is crucially important to introduce the temperature dependence of the drag and the thermal noise in Eq.~\eqref{Langevin_eq}. 
In other words, the previous model without the noise term is structurally unstable, because once we introduce very small $T_{\rm ex}$ the rheology for the low shear regime is changed completely.
It is obvious that the suspensions must be equilibrated in the absence of the external shear, which can be achieved only if we take into account the thermal noise. 
Therefore, we believe that our model is more appropriate than the previous model in the low shear regime. 


It is straightforward to extend the present analysis to moderately dense gases by using Enskog equation~\cite{Sangani96,Garzo13}.
Therefore, the extension of the analysis presented here is important.
After the completion of this work, we have already developed the analysis of a moderately dense suspension by the Enskog equation~\cite{Hayakawa-Takada-Garzo}.
As a result of the merge of two turning points for the flow curve of $\theta$ or $\eta^*$
we have verified that the DST disappears at a quite low density around $\varphi\approx 0.0176$, though the CST still survives above the critical density.
The appearance of the DST with the decrease of the density can be regarded as a saddle-node bifurcation as indicated by Refs.~\cite{Wyart14,BGK2016}.
The absence of the DST in finite density can be understood as follows.
As shown in Eqs. \eqref{bagnold_dim} and \eqref{high_shear_dim} the viscosity has the coefficient inversely proportional to the density in the high shear limit, which connect that on the Newtonian branch in Eq. \eqref{Newton_phys} around at $\dot\gamma^*=10$.
If the coefficient in the high shear regime is too large, the simple interpolation should be much larger than that on the Newtonian branch.
On the other hand, because the coefficient in the high shear regime decreases as the density increases, two branches can be connected continuously for finite density situation.
This result is qualitatively consistent with the previous theory for the temperature~\cite{Sangani96}.
Therefore, the mechanism of the DST presented here is not directly related to typical DSTs observed in dense suspensions.

We can indicate that Eq.~\eqref{Grad} is consistent with the  Green-Kubo formula~\cite{Evans08,Chong10,Hayakawa13,Suzuki15} as shown in Appendix \ref{G-K_formula}.
Nevertheless, the model only based on the Green-Kubo formula has isotropic stress.
Therefore, it is essential for the DST to adopt Grad's approximation \eqref{Grad}. 
An extended Grad's approximation is applicable to dense non-Brownian suspensions near the jamming point~\cite{Suzuki17}, in which we can successfully reproduce the divergent pressure viscosity and shear viscosity observed in experiments~\cite{Boyer11,Dagois-Bohy15}.

Our model does not include any mutual friction between grains, though many papers stressed important roles of the friction in DST for dense suspensions and dense dry granular flows~\cite{Otsuki11,Seto13,Mari14,Bi11,Pica11,Haussinger13,Grob14}.
In this sense, our analysis does not answer the mechanism of DST observed in typical experiments and simulations for dense suspensions.
  A recent paper has already partially reproduced a hysteresis in the flow curve, which is one of the characteristics of the DST by choosing the shear stress, the temperature (the pressure) and the spin temperature~\cite{Saitoh16}. 

One can replace the base distribution $f_{\rm eq}(\bm{V})$ by $f_{\rm ss}(\bm{V})$ which is the steady velocity distribution of the granular gas under the Fokker-Planck thermostat when we discuss highly dissipative granular gases.
This $f_{\rm ss}(\bm{V})$ can be represented by
\begin{equation}
f_{\rm ss}(\bm{V})=n v_T^{-d}(t)\tilde{f}(\bm{c}); \quad \bm{c}=\bm{V}/v_T(t) ,
\end{equation}
where $v_T=\sqrt{2T/m}$ and $\tilde{f}(\bm{c})$ satisfies
\begin{equation}\label{Sonine}
\tilde{f}(\bm{c})\approx \phi(c)(1+a_2S_2(c^2))
\end{equation}
with
\begin{equation}
\phi(c)=\pi^{-d}\exp(-c^2); \quad S_2(x)=\frac{1}{2}x^2-\frac{1}{2}(d+2)x+\frac{1}{8}d(d+2) .
\end{equation}
If we further expand $f_{\rm ss}(\bm{V})$ around $f_{\rm eq}(\bm{V})$, this is nothing but Grad's 14 moment method used by Ref.~\cite{Saha17}.
Nevertheless, $a_2$ in Eq.~\eqref{Sonine} is not given by Ref.~\cite{vanNoije98} as used in Ref.~\cite{Saha17} but is given by Ref.~\cite{Hayakawa03}.

The analysis of the attractive interaction in Sec. 5 is potentially important, because
(i) we illustrate the existence of a new mechanism of the discontinuous jump in the flow curve, (ii) this mechanism might survive even if we consider dense suspensions.
If our conjecture (ii) is right, the analysis of this model parallel to Ref.~\cite{Hayakawa-Takada-Garzo} may explain the DST in realistic suspensions.
This is one of interesting future directions of our study.

\subsection{Conclusion}

In conclusion, we have demonstrated the existence of a S-shape flow curve in dilute inertial suspensions based on the analysis of the inelastic Boltzmann equation.
This model exhibits the crossovers from the Newtonian viscosity to the Bagnoldian viscosity for $e<1$ and from the Newtonian to the viscosity proportional to $\dot\gamma^2$ for $e=1$.
The event-driven Langevin simulation for hard spheres reproduces the DST and
perfectly agrees with the theoretical results without any fitting parameters.
Therefore, we have confirmed the existence of the DST for dilute inertial suspensions.



\vspace*{0.4cm}

{\bf Acknowledgment}:
We thank Koshiro Suzuki, Takeshi Kawasaki, Michio Otsuki, Kuniyasu Saitoh and Meheboob Alam for their useful comments. 
We also express our sincere gratitude to Vicente Garz\'{o} for his collaboration to extend our work to finite density and discussion on the second normal difference.
This work is partially supported by the Grant-in-Aid of MEXT for Scientific Research (Grant No.\ 16H04025) and the YITP activities (Grant Nos.\ YITP-T-18-03 and YITP-W-18-17).
Numerical computation in this work was partially carried out at the Yukawa Institute Computer Facility.

\appendix



\begin{center}
{\bf Appendix}
\end{center}

This appendix contains some detailed information. 
In Appendix \ref{G-K_formula}, we explain the relationship between Grad's approximation and Green-Kubo formula within the BGK approximation.
In Appendix \ref{details_app} we give some detailed calculation for $\Lambda_{\alpha\beta}$ in Eq.~\eqref{Lambda} as well as some formulas of Gaussian integrals and angle integrals. 
This calculation is basically written in Refs.~\cite{Garzo02,Santos04} but we write their details as a self-contained form.
Appendix \ref{app:linear} is devoted to the detailed calculation of the linear stability analysis.
In Appendix \ref{EDLSHS}, we explain the outline of the algorithm of EDLSHS. 
In Appendix \ref{app_relaxation}, we briefly show the time evolution of the shear stress when we gradually change the shear rate to reach the steady state.
In Appendix \ref{app_domain}, we discuss the domain growth when we start from an unstable point, which is analogous to the spinodal decomposition after the system is quenched into an unstable point.  
In Appendix \ref{SQUARE_WELL}, we present the detailed analysis for the case of the attractive interaction.

\section{Relationship between Grad's approximation and Green-Kubo formula}
\label{G-K_formula}

The result in this paper strongly depends on the choice Eq.\eqref{Grad}.
In this section, we discuss the validity of the Grad expansion in Eq.~\eqref{Grad}.

For simplicity, let us consider
\begin{equation}\label{BGK}
\left(\frac{\partial}{\partial t}
-\dot\gamma
V_{y}\frac{\partial}{\partial V_{x}}
\right)
f(\bm{V},t) 
=
\zeta\frac{\partial}{\partial \bm{V}} \cdot \left(
\left\{ \bm{V}+ \frac{T_{\rm ex}}{m} \frac{\partial}{\partial \bm{V}} \right\} 
 f(\bm{V},t) \right)
+
\frac{f_{\rm eq}(\bm{V})-f(\bm{V},t)}{\tau} ,
\end{equation}
where the relaxation time $\tau$ satisfies
$\tau=(16/5)n\sigma^2\sqrt{\pi T/m}$ for $d=3$ and $e=1$~\cite{Santos98}.
BGK equation \eqref{BGK} is a well known simplified model to reduce to the linearized Boltzmann equation, which is compatible with Chapman-Enskog approximation with the proper choice of $\tau$ as in Ref.~\cite{Santos98}.
We restrict our interest to the case of $d=3$ and $e=1$ in this Appendix.

Within the framework of BGK equation \eqref{BGK} is reduced to
\begin{equation}\label{B1}
\frac{\delta f}{f_{\rm eq}}=-\dot\gamma^* \tau^* \frac{m V_x V_y}{T}+
\tau^*\left(1-\frac{T_{\rm ex}}{T} \right)\left(3-\frac{p^2}{mT} \right) ,
\end{equation}
where $\bm{p}=m\bm{V}$, $\delta f\equiv f-f_{\rm eq}$ and $\tau^*=\zeta \tau$.
To derive Eq. \eqref{B1} we have used the relations:
\begin{eqnarray}
\left( y \frac{\partial}{\partial x}-p_y \frac{\partial}{\partial p_x} \right) f_{\rm eq}
&=& \frac{m V_xV_y}{T}, \\
\left(\bm{p}+ m T_{\rm ex} \frac{\partial}{\partial \bm{p}} \right) f_{\rm eq}&=& 
\left(1-\frac{T_{\rm ex}}{T} \right) \bm{p} f_{\rm eq} .
\end{eqnarray}
It should be noted that there is no contribution of the second term on the right hand side of Eq. \eqref{BGK} for the macroscopic stress obtained by $P_{xy}=m \int d\bm{V}f(\bm{V}) V_x V_y$ because of its parity. Therefore, the first term on the right hand side of Eq. \eqref{B1} is only important for the consistency between Eqs. \eqref{Grad} and \eqref{B1}.

The macroscopic shear stress $\sigma_{xy}$ and the viscosity $\eta=-P_{xy}/\dot\gamma$ are determined by
\begin{equation}\label{Kubo}
P_{xy}=m \int d\bm{V} \delta f(\bm{V}) V_xV_y=-\frac{\dot\gamma^*\tau^* m^2}{T}\langle V_x^2V_y^2 \rangle_{\rm eq},
\end{equation}
where $\langle \cdot \rangle_{\rm eq}$ is the average in terms of the equilibrium distribution $f_{\rm eq}(\bm{V})$.
It should be noted that Eq. \eqref{Kubo} is identical to the Green-Kubo formula under an exponential relaxation~\cite{Suzuki15}.

On the other hand, if we adopt Eq. \eqref{Grad}, the normal stress difference disappears within the framework of the linearized BGK equation.
Indeed, substituting Eq.~\eqref{B1} into the expression of $P_{xx}$, we obtain
\begin{eqnarray}
P_{xx}&=&\int d\bm{p} f_{\rm eq}\frac{p_x^2}{m}\left\{1+\tau^*\left(1-\frac{T_{\rm ex}}{T} \right)\left(3-\frac{p^2}{mT}\right)\right\}
\nonumber\\
&=& \frac{1}{m}\int d\bm{p} p_x^2 f_{\rm eq}
+ 
\frac{\tau^*}{m}\left(1-\frac{T_{\rm ex}}{T}\right) 
\int d\bm{p} f_{\rm eq} p_x^2\left(3-\frac{p^2}{mT}\right) .
\label{D_sigma_xx^K}
\end{eqnarray}
Taking into account the relations $\int d\bm{p} (p_x^2/m) f_{\rm eq}=nT$, 
$\int d\bm{p} (p_x^4/m)f_{\rm eq}=3m n T$ and $\int d\bm{p} p_x^2p_y^2 f_{\rm eq}=n m~2 T^2$, the last term of Eq.~\eqref{D_sigma_xx^K} can be rewritten as
\begin{equation}
\int d\bm{p} f_{\rm eq} p_x^2\left(3-\frac{p^2}{mT} \right)=-2mn T .
\end{equation}
Thus, $\sigma_{xx}$ is reduced to
\begin{equation}
P_{xx}=nT \left\{1+2\tau^*\left(\frac{T_{\rm ex}}{T}-1\right) \right\} . 
\label{sigma_xx^K}
\end{equation}
It is obvious that the kinetic normal stress is isotropic, {\it i. e.} 
\begin{equation}
P_{xx}=P_{yy}=P_{zz} .
\label{iso_sigma^K}
\end{equation}

Now, let us consider the result in this appendix.
First, the consistency under the weakly sheared situation or nearly Newtonian situation is reasonable, because the system remains in a nearly equilibrium situation in which the linear response theory can be used, though there still exists a little disagreement between asymptotic forms in this appendix and the Green-Kubo formula.
Nevertheless, Grad's expansion \eqref{Grad} should include the nonlinear term which causes the normal stress difference $\Delta T\ne 0$, which is not included in the linearized BGK equation.  
We need further investigation on the microscopic validation of Eq. (\ref{Grad}).

\section{Detailed calculations of the kinetic theory}\label{details_app}

\subsection{$\int d\hat{\bm{\sigma}}\Theta(\bm{c}\cdot\hat{\bm{\sigma}})(\bm{c}\cdot\hat{\bm{\sigma}})^n$}

First, let us prove the following identity
\begin{equation}\label{Garzo_A5}
\int d\hat{\bm{\sigma}}\Theta(\bm{c}\cdot\hat{\bm{\sigma}})(\bm{c}\cdot\hat{\bm{\sigma}})^n=\beta_n c^n ,
\end{equation}
where 
\begin{equation}\label{beta}
\beta_n=\pi^{(d-1)/2}\frac{\Gamma(\frac{n+1}{2})}{\Gamma(\frac{n+d}{2})}
\end{equation}

It is straightforward to rewrite the left hand side (LHS) of Eq.~\eqref{Garzo_A5} as
\begin{eqnarray}
\int d\hat{\bm{\sigma}}\Theta(\bm{c}\cdot\hat{\bm{\sigma}})(\bm{c}\cdot\hat{\bm{\sigma}})^n 
&=& S_{d-1}c^n \int_0^{\pi/2} d\theta (\sin\theta)^{d-2} \cos^n \theta
= \frac{S_{d-1}c^n}{2}B\left(\frac{d-1}{2},\frac{n+1}{2} \right) 
\nonumber\\
&=& \pi^{(d-1)/2}\frac{\Gamma(\frac{n+1}{2})}{\Gamma(\frac{n+d}{2})} c^n ,
\end{eqnarray}
where we have introduced the Beta function $B(x,y)$, the Gamma function $\Gamma(x)$ and the area of unit hypersurface $S_d$:
\begin{equation}\label{B,G&S}
B(x,y)=\frac{\Gamma(x)\Gamma(y)}{\Gamma(x+y)}=2\int_0^{\pi/2}d\theta (\sin\theta)^{2x-1}(\cos\theta)^{2y-1},
\quad
\Gamma(x)=\int_0^\infty dt t^{x-1} e^{-t},
\quad
S_d=\frac{2\pi^{d/2}}{\Gamma(d/2)} .
\end{equation}
This is the end of proof of Eqs.~\eqref{Garzo_A5} and \eqref{beta}.

\subsection{$\int d\hat{\bm{\sigma}}\Theta(\bm{c}\cdot\hat{\bm{\sigma}})(\bm{c}\cdot\hat{\bm{\sigma}})^n \hat{\bm{\sigma}}$}

In this subsection, let us prove the identity:
\begin{equation}\label{Garzo_A6}
\int d\hat{\bm{\sigma}}\Theta(\bm{c}\cdot\hat{\bm{\sigma}})(\bm{c}\cdot\hat{\bm{\sigma}})^n \hat{\bm{\sigma}}
=\beta_{n+1}c^{n-1}\bm{c} .
\end{equation}

Let us introduce the angle $\theta_i$ with $1\le i \le d-1$ to characterize the unit hypersurface. In this case $\hat{\bm{\sigma}}$ can be expressed as
\begin{equation}
\hat{\bm{\sigma}}
=
\left(
\begin{array}{c}
\cos\theta_1 \\
\sin\theta_1 \cos\theta_2 \\
\sin\theta_1 \sin \theta_2 \cos\theta_3 \\
\cdots \\
\sin\theta_1 \cdots \sin\theta_{d-2} \cos\theta_{d-1} \\
\sin\theta_1 \cdots \sin\theta_{d-2} \sin\theta_{d-1} 
\end{array}
\right) ,
\end{equation}
where $\theta_{d-1}$ ranges over $[0,2\pi)$ and the other angles $\theta_i$ ($1\le i\le d-2$) ranges over $[0,\pi)$.
The angle $\theta_1$ is regarded as the angle between $\bm{c}$ and the normal unit vector $\hat{\bm{\sigma}}$.
The integral \eqref{Garzo_A6} is
\begin{equation}
\int d\hat{\bm{\sigma}}\hat{\bm{\sigma}}\Theta(\bm{c}\cdot\hat{\bm{\sigma}})=
\int_0^{\pi/2}d\theta_1\int_0^{\pi}d\theta_2 \cdots \int_0^{2\pi}d\theta_{d-1} (\sin \theta_1)^{d-2} (\sin\theta_2)^{d-3} \cdots \sin \theta_{d-2}
\hat{\bm{\sigma}} ,
\end{equation}
where the first component of $\hat{\bm{\sigma}}$ gives finite contribution but all the other components cancel because $i-$th component satisfies $\int_0^{\pi}d\theta_i \cos \theta_i\sin^{d-i-1}\theta_i=[\sin^{d-i} \theta_i]_{\theta_i=0}^{\pi}/(d-i)=0$.
Therefore, only the first component in LHS of Eq.~\eqref{Garzo_A6} which is parallel to $\bm{c}$ survives.
Therefore, the LHS of Eq.~\eqref{Garzo_A6} can be evaluated as
\begin{eqnarray}
\int d\hat{\bm{\sigma}}\Theta(\bm{c}\cdot\hat{\bm{\sigma}})(\bm{c}\cdot\hat{\bm{\sigma}})^n \hat{\bm{\sigma}}
&=& 
S_{d-1}c^n \int_0^{\pi/2}d\theta (\sin\theta)^{d-2}\cos^n \theta (\hat{\bm{c}}\cdot\hat{\bm{\sigma}})\hat{\bm{c}} \nonumber\\
&=&S_{d-1}c^n \int_0^{\pi/2}d\theta (\sin\theta)^{d-2}\cos^{n+1}\theta\hat{\bm{c}}
\nonumber\\
&=&\beta_{n+1}c^n \hat{\bm{c}}=\beta_{n+1}c^{n-1}\bm{c} .
\end{eqnarray}
This is the end of proof of Eq.~\eqref{Garzo_A6}.

\subsection{$\int d\hat{\bm{\sigma}}\Theta(\bm{c}\cdot\hat{\bm{\sigma}})(\bm{c}\cdot\hat{\bm{\sigma}})^n \hat{\sigma}_\alpha \hat{\sigma}_\beta$}

Let us prove the following identity
\begin{equation}
\mathscr{B}_{\alpha\beta}\equiv
\int d\hat{\bm{\sigma}}\Theta(\bm{c}\cdot\hat{\bm{\sigma}})(\bm{c}\cdot\hat{\bm{\sigma}})^n \hat{\sigma}_\alpha \hat{\sigma}_\beta
=\frac{\beta_n}{n+d}c^{n-2}(n c_\alpha c_\beta+c^2 \delta_{\alpha\beta}) .
\label{Garzo_A17}
\end{equation}

Let us assume the form
\begin{equation}
\hat{\sigma}_\alpha \hat{\sigma}_\beta
=a(\delta_{\alpha\beta}+b \hat{c}_\alpha \hat{c}_\beta)
\end{equation}
Because $\hat{\bm{\sigma}}$ is the unit vector and has the relation $\hat{\sigma}_\alpha\hat{\sigma}_\alpha=1$, the trace of the LHS of Eq.~\eqref{Garzo_A17}, $\mathscr{B}_{\alpha\alpha}$, is given by
\begin{equation}
\int d\hat{\bm{\sigma}}\Theta(\bm{c}\cdot\hat{\bm{\sigma}})(\bm{c}\cdot\hat{\bm{\sigma}})^n \hat{\sigma}_\alpha \hat{\sigma}_\alpha
=a (d+b) \int d\hat{\bm{\sigma}}\Theta(\bm{c}\cdot\hat{\bm{\sigma}})(\bm{c}\cdot\hat{\bm{\sigma}})^n
=a(d+b)\beta_nc^n=\beta_n c^n .
\end{equation}
Therefore we obtain
\begin{equation}\label{a=}
a=\frac{1}{d+b}
\end{equation}

On the other hand, $\hat{\sigma}_\alpha\mathscr{B}_{\alpha\beta}\hat{\sigma}_\beta$
becomes 
\begin{eqnarray}
\hat{\sigma}_\alpha\mathscr{B}_{\alpha\beta}\hat{\sigma}_\beta
&=& a \int d\hat{\bm{\sigma}}\Theta(\hat{\bm{\sigma}}\cdot\bm{c})(\hat{\bm{c}}\cdot\hat{\bm{\sigma}})^n(\hat{c}^2+b \hat{c}^2)=a(1+b)\beta_n c^n \nonumber\\
&=& \int d\hat{\bm{\sigma}}\Theta(\hat{\bm{\sigma}}\cdot\bm{c})
(\hat{\bm{\sigma}}\cdot\bm{c})^n(\hat{\bm{\sigma}}\cdot\hat{\bm{c}})^2
=\beta_{n+2}c^n .
\end{eqnarray}
Therefore, we obtain the relation
\begin{equation}\label{n+2=>n}
\beta_{n+2}=a(1+b)\beta_n=\beta_n \frac{n+1}{n+d}
\end{equation}
From Eqs,~\eqref{a=} and \eqref{n+2=>n} we reach Eq.~\eqref{Garzo_A17}.

\subsection{Evaluation of $\Lambda_{\alpha\beta}$}\label{derivation_Lambda}

In this subsection, we evaluate $\Lambda_{\alpha\beta}$ introduced in Eq.~(9). 
Substituting Eq.~(3)
into Eq.~(9)
we obtain
\begin{eqnarray}\label{Lambda_{ab}}
\Lambda_{\alpha\beta}
&=&-m \sigma^{d-1} \int d\bm{v}_1\int d\bm{v}_2\int d\hat{\bm{\sigma}}\Theta(\bm{v}_{12}\cdot\hat{\bm{\sigma}})
|\bm{v}_{12}\cdot\hat{\bm{\sigma}}|V_{1,\alpha} V_{1,\beta}
\left\{
\frac{f(\bm{V}_1^{**})f(\bm{V}_2^{**})}{e^2}-f(\bm{V}_1)f(\bm{V}_2)
\right\}
\nonumber\\
&=& 
-\frac{m \sigma^{d-1}}{2} \int d\bm{v}_1\int d\bm{v}_2\int d\hat{\bm{\sigma}}\Theta(\bm{v}_{12}\cdot\hat{\bm{\sigma}})
|\bm{v}_{12}\cdot\hat{\bm{\sigma}}|\nonumber\\
&&\times f(\bm{V}_1)f(\bm{V}_2)(V_{1,\alpha}^*V_{1,\beta}^*+V_{2,\alpha}^*V_{2,\beta}^*-V_{1,\alpha}V_{1,\beta}-V_{2,\alpha}V_{2,\beta}) ,
\end{eqnarray}
where we have introduced the post-collisional velocities $\bm{v}_1^*$ and $\bm{v}_2^*$ defined by
\begin{equation}\label{post-collision}
\bm{v}_1^*=\bm{v}_1-\frac{1+e}{2}(\bm{v}_{12}\cdot\hat{\bm{\sigma}})\hat{\bm{\sigma}},
\quad
\bm{v}_2^*=\bm{v}_2+\frac{1+e}{2}(\bm{v}_{12}\cdot\hat{\bm{\sigma}})\hat{\bm{\sigma}} .
\end{equation}
To obtain the final expression of Eq.~\eqref{Lambda_{ab}}, we have converted $(\bm{V}_i^{**},\bm{V}_i)$ into $(\bm{V}_i,\bm{V}_i^*)$ and used the Jacobian $d\bm{v}_1^*d\bm{v}_2^*=d\bm{v}_1d\bm{v}_2/e$ and the collision rule $(\bm{v}_{12}^*\cdot\hat{\bm{\sigma}})\hat{\bm{\sigma}}=-e(\bm{v}_{12}\cdot\hat{\bm{\sigma}})\hat{\bm{\sigma}}$.
We have also symmetrized the final expression.
With the aid of Eq.~\eqref{post-collision} we have the relation
\begin{eqnarray}\label{V_1+V_2}
&&V_{1,\alpha}^*V_{1,\beta}^*+V_{2,\alpha}^*V_{2,\beta}^*
-V_{1,\alpha}V_{1,\beta}-V_{2,\alpha}V_{2,\beta}\nonumber\\
&&=-\frac{1+e}{2}
(\bm{v}_{12}\cdot\hat{\bm{\sigma}})
(v_{12,\alpha}\hat{\sigma}_\beta+\hat{\sigma}_\alpha v_{12,\beta})
+\frac{(1+e)^2}{2}(\bm{v}_{12}\cdot\hat{\bm{\sigma}})^2\hat{\sigma}_\alpha\hat{\sigma}_\beta .
\end{eqnarray}
Substituting Eqs.~(\ref{Grad}) and \eqref{V_1+V_2} into Eq.~\eqref{Lambda_{ab}} with the linearization around $f_{\rm eq}(\bm{V})$ $\Lambda_{\alpha\beta}$ can be expressed as the summation of the two terms:
\begin{equation}\label{Lambda_{alp,be}}
\Lambda_{\alpha\beta}=\Lambda_{\alpha\beta}^L+\Lambda_{\alpha\beta}^{\rm NL} .
\end{equation}
Here, $\Lambda_{\alpha\beta}^L$ is the contributions which is proportional to or independent of $P_{\alpha\beta}/(nT)-\delta_{\alpha\beta}$ defined as
\begin{eqnarray}\label{Lambda_{alp,be}^L}
\Lambda_{\alpha\beta}^L
&=& -\frac{m  \sigma^{d-1} }{2} n^2\left(\frac{m}{2\pi T} \right)^d
\int d\bm{G} \int d\bm{v}_{12}\int d\hat{\bm{\sigma}} \Theta(\bm{v}_{12}\cdot\hat{\bm{\sigma}})|\bm{v}_{12}\cdot\hat{\bm{\sigma}}|
\exp\left(-\frac{m G^2}{T}\right)\exp\left(-\frac{m v_{12}^2}{4T} \right)
\nonumber\\
&&\times \left[1+\frac{m}{2T}\left(\frac{P_{\gamma\delta}}{nT}-\delta_{\gamma\delta} \right)\left(2G_\gamma G_\delta+\frac{1}{2}v_{12,\gamma}v_{12,\delta} \right)  \right]
\nonumber\\
&& \times
\left[-A(\bm{v}_{12}\cdot\hat{\bm{\sigma}})(v_{12,\alpha}\hat{\sigma}_\beta+\hat{\sigma}_\alpha v_{12,\beta})+2A^2(\bm{v}_{12}\cdot\hat{\bm{\sigma}})^2\hat{\sigma}_\alpha\hat{\sigma}_\beta \right]
\end{eqnarray}
where we have introduced $\bm{G}=(\bm{v}_1+\bm{v}_2)/2$ and $A=(1+e)/2$.
Equation \eqref{Lambda_{alp,be}} is further rewritten as
\begin{eqnarray}\label{Lambda_{33}}
\Lambda_{\alpha\beta}^L
&=&-\frac{  n^2v_T^3m  \sigma^{d-1} }{2\pi^d}
\int d\bm{C} \int d\bm{c}\int d\hat{\bm{\sigma}} \Theta(\bm{c}\cdot\hat{\bm{\sigma}})|\bm{c}\cdot\hat{\bm{\sigma}}|
\exp\left(-2C^2\right)\exp\left(-\frac{1}{2}c^2 \right)
\nonumber\\
&&\times \left[1+\left(P^*_{\gamma\delta}-\delta_{\gamma\delta} \right)\left(2C_\gamma C_\delta+\frac{1}{2}c_{\gamma}c_{\delta} \right)  \right]
\left[2A^2(\bm{c}\cdot\hat{\bm{\sigma}})^2\hat{\sigma}_\alpha\hat{\sigma}_\beta -A(\bm{c}\cdot\hat{\bm{\sigma}})(c_{\alpha}\hat{\sigma}_\beta+\hat{\sigma}_\alpha c_{\beta})\right] \nonumber\\
&=& \frac{  n^2v_T^3m  \sigma^{d-1} }{2\pi^d} A \tilde{\Lambda}_{\alpha\beta}^L
\end{eqnarray}
with
\begin{equation}\label{tilde_Lambda}
\tilde{\Lambda}_{\alpha\beta}^L\equiv 
\Lambda_{\alpha\beta}^{L, (1)}-2A\Lambda_{\alpha\beta}^{L, (2)}+
(\hat{P}_{\gamma\delta}-\delta_{\gamma\delta}) (\Lambda_{\alpha\beta\gamma\delta}^{L, (3)}-2A\Lambda_{\alpha\beta\gamma\delta}^{L,(4)}) ,
\end{equation}
where we have introduced $\hat{P}_{\alpha\beta}=P_{\alpha\beta}/(nT)$, $v_T=\sqrt{2T/m}$, $\bm{C}=\bm{G}/v_T$ and $\bm{c}=\bm{v}_{12}/v_T$ in the first expression.
We have also introduced $\Lambda_{\alpha\beta}^{(i)}$ ($i=1,2$) and $\Lambda_{\alpha\beta\gamma\delta}^{(j)}$ ($j=3,4$) in Eq.~\eqref{Lambda_{33}} as
\begin{eqnarray}\label{Lambda^L1}
\Lambda_{\alpha\beta}^{L, (1)}&=&\int d\bm{C}\int d\bm{c} \int d\hat{\bm{\sigma}}\Theta(\bm{c}\cdot\hat{\bm{\sigma}})e^{-2C^2} e^{-c^2/2} (\bm{c}\cdot\hat{\bm{\sigma}})^2(c_{\alpha}\hat{\sigma}_\beta+\hat{\sigma}_\alpha c_{\beta})
\\
\label{Lambda^L2}
\Lambda_{\alpha\beta}^{L, (2)}&=&\int d\bm{C}\int d\bm{c} \int d\hat{\bm{\sigma}}
\Theta(\bm{c}\cdot\hat{\bm{\sigma}})e^{-2C^2} e^{-c^2/2}(\bm{c}\cdot\hat{\bm{\sigma}})^3\hat{\sigma}_\alpha\hat{\sigma}_\beta
\\
\label{Lambda^L3}
\Lambda_{\alpha\beta\gamma\delta}^{L, (3)}
&=& \int d\bm{C} \int d\bm{c}\int d\hat{\bm{\sigma}}\Theta(\bm{c}\cdot\hat{\bm{\sigma}})e^{-2C^2} e^{-c^2/2}
\left(2C_\gamma C_\delta+\frac{1}{2}c_{\gamma}c_{\delta} \right) 
(\bm{c}\cdot\hat{\bm{\sigma}})^2
\nonumber\\
&& \times
(c_{\alpha}\hat{\sigma}_\beta+\hat{\sigma}_\alpha c_{\beta})
\\
\Lambda_{\alpha\beta\gamma\delta}^{L, (4)}
&=& \int d\bm{C} \int d\bm{c}\int d\hat{\bm{\sigma}}\Theta(\bm{c}\cdot\hat{\bm{\sigma}})e^{-2C^2} e^{-c^2/2}
\left(2C_\gamma C_\delta+\frac{1}{2}c_{\gamma}c_{\delta} \right) 
(\bm{c}\cdot\hat{\bm{\sigma}})^3\hat{\sigma}_\alpha \hat{\sigma}_\beta .
\label{Lambda^L4}
\end{eqnarray}


On the other hand, $\Lambda_{\alpha\beta}^{\rm NL}$ is the nonlinear contribution of $P_{\alpha\beta}/(nT)-\delta_{\alpha\beta}$, which can be expressed as
\begin{equation}\label{Lambda_{a,b}^{NL}}
\Lambda_{\alpha\beta}^{\rm NL}
=\frac{n^2v_T^3m\sigma^{d-1}}{2\pi^d}A
(\hat{P}_{\gamma\delta}-\delta_{\gamma\delta})(\hat{P}_{\xi\eta}-\delta_{\xi\eta})[\tilde{\Lambda}_{\alpha\beta\gamma\delta\xi\eta}^{{\rm NL}(1)}-2A\tilde{\Lambda}_{\alpha\beta\gamma\delta\xi\eta}^{{\rm NL}(2)}] ,
\end{equation}
where
\begin{eqnarray}
\label{Lambda^NL(1)}
\tilde{\Lambda}_{\alpha\beta\gamma\delta\xi\eta}^{{\rm NL} (1)}
&=&
\int d\bm{C} \int d\bm{c}\int d\hat{\bm{\sigma}}
\Theta(\bm{c}\cdot\hat{\bm{\sigma}})e^{-2C^2}e^{-c^2/2}
(\bm{c}\cdot\hat{\bm{\sigma}})^2
(c_\alpha\hat\sigma_\beta+c_\beta\hat\sigma_\alpha)
[C_\gamma C_\delta C_\xi C_\eta 
+\frac{1}{16}c_\gamma c_\delta c_\xi c_\eta
\nonumber\\
&&
-\frac{1}{4}(C_\gamma c_\delta +C_\delta c_\gamma)(C_\xi c_\eta + C_\eta c_\xi)
+\frac{1}{4}(C_\gamma C_\delta c_\xi c_\eta+ C_\xi C_\eta c_\gamma c_\delta)
]
\end{eqnarray}
and 
\begin{eqnarray}
\label{Lambda^NL(2)}
\tilde{\Lambda}_{\alpha\beta\gamma\delta\xi\eta}^{{\rm NL} (2)}
&=&\int d\bm{C} \int d\bm{c}\int d\hat{\bm{\sigma}}
\Theta(\bm{c}\cdot\hat{\bm{\sigma}})e^{-2C^2}e^{-c^2/2}
(\bm{c}\cdot\hat{\bm{\sigma}})^3 \hat{\sigma}_\alpha \hat{\sigma}_\beta
[C_\gamma C_\delta C_\xi C_\eta 
+\frac{1}{16}c_\gamma c_\delta c_\xi c_\eta
\nonumber\\
&&
-\frac{1}{4}(C_\gamma c_\delta +C_\delta c_\gamma)(C_\xi c_\eta + C_\eta c_\xi)
+\frac{1}{4}(C_\gamma C_\delta c_\xi c_\eta+ C_\xi C_\eta c_\gamma c_\delta)
] .
\end{eqnarray}

\subsubsection{Evaluation of $\Lambda_{\alpha\beta}^L$}

Let us evaluate $\Lambda_{\alpha\beta}^L$ introduced in Eq. \eqref{Lambda_{alp,be}^L} as well as Eqs. \eqref{Lambda_{33}} and \eqref{tilde_Lambda}.

The explicit expressions of $\Lambda_{\alpha\beta}^{L, (1)}$ in Eq.~\eqref{Lambda^L1} is given by
\begin{eqnarray}\label{Lambda^1}
\Lambda_{\alpha\beta}^{L,(1)}
&=& \left(\frac{\pi}{2}\right)^{d/2}\int d\bm{c} \int d\hat{\bm{\sigma}}
\Theta(\bm{c}\cdot\hat{\bm{\sigma}})
e^{-c^2/2} (\bm{c}\cdot\hat{\bm{\sigma}})^2(c_{\alpha}\hat{\sigma}_\beta+\hat{\sigma}_\alpha c_{\beta}) \nonumber\\
&=&2\left(\frac{\pi}{2}\right)^{d/2}\beta_3 \int d\bm{c} e^{-c^2/2} c c_\alpha c_\beta \nonumber\\
&=& 2 \delta_{\alpha\beta} \left(\frac{\pi}{2}\right)^{d/2} \frac{\pi^{(d-1)/2}}{d\Gamma(\frac{3+d}{2})} S_d \int_0^\infty dc c^{d+2} e^{-c^2/2} \nonumber\\
&=& \delta_{\alpha\beta} 2\left(\frac{\pi}{2}\right)^{d/2} \frac{\pi^{(d-1)/2}}{d\Gamma(\frac{d+3}{2})}\frac{2\pi^{d/2}}{\Gamma(\frac{d}{2})}2^{(d+1)/2}\Gamma\left(\frac{d+3}{2}\right)
 \nonumber\\
&=&4\sqrt{2} \frac{\pi^{(3d-1)/2}}{d\Gamma(\frac{d}{2})}\delta_{\alpha\beta} ,
\end{eqnarray}
where we have used Eq.~\eqref{Garzo_A6} in the expression in the third line, Eqs.~\eqref{beta} and \eqref{c_xc_y} in the fourth line and Eq.~\eqref{B,G&S} for the last expression. 
We also note the relation $\int_0^\infty dcc^{d+a}e^{-c^2/2}=2^{(d+a-1)/2}\Gamma((d+a+1)/2)$.

Similarly, $\Lambda_{\alpha\beta}^{L, (2)}$ in Eq, \eqref{Lambda^L2} is given by
\begin{eqnarray}\label{Lambda^2}
\Lambda_{\alpha\beta}^{L,(2)}
&=& \left(\frac{\pi}{2}\right)^{d/2}\int d\bm{c} \int d\hat{\bm{\sigma}}\Theta(\bm{c}\cdot\hat{\bm{\sigma}})
e^{-c^2/2} (\bm{c}\cdot\hat{\bm{\sigma}})^3\hat{\sigma}_\alpha\hat{\sigma}_\beta
\nonumber\\
&=& \left(\frac{\pi}{2}\right)^{d/2}\frac{\beta_3}{3+d}\int d\bm{c} e^{-c^2/2}c(3 c_\alpha c_\beta+c^2\delta_{\alpha\beta}) \nonumber\\
&=& \delta_{\alpha\beta} \frac{1}{d} \left(\frac{\pi}{2}\right)^{d/2} \frac{\pi^{(d-1)/2}}{\Gamma(\frac{3+d}{2})} S_d \int_0^\infty dc c^{d+2} e^{-c^2/2} \nonumber\\
&=& \delta_{\alpha\beta} \frac{1}{d} \left(\frac{\pi}{2}\right)^{d/2} \frac{\pi^{(d-1)/2}}{\Gamma(\frac{3+d}{2})} \cdot \frac{2\pi^{d/2}}{\Gamma(d/2)}\cdot 2^{(d+1)/2}\Gamma\left(\frac{d+3}{2}\right) \nonumber\\
&=& \frac{2\sqrt{2}}{d}\frac{\pi^{(3d-1)/2}}{\Gamma(\frac{d}{2})}\delta_{\alpha\beta} ,
\end{eqnarray}
where we have used Eqs.~\eqref{Garzo_A17} and \eqref{c_xc_y} for the third line.
Therefore, we obtain the relation
\begin{equation}\label{Lambda1+2}
\Lambda_{\alpha\beta}^{L,(1)}-2A\Lambda_{\alpha\beta}^{L,(2)}
=\frac{4\sqrt{2}\pi^{(3d-1)/2}}{d\Gamma(d/2)}(1-A)\delta_{\alpha\beta} .
\end{equation}
The expression of $\Lambda_{\alpha\beta\gamma\delta}^{(3)}$ in Eq. \eqref{Lambda^L3} consists of two parts
\begin{eqnarray}\label{Gamma^3}
\Lambda_{\alpha\beta\gamma\delta}^{L,(3)}
&=& 2\Lambda_{\alpha\beta\gamma\delta}^{(3,1)}+\frac{1}{2}\Lambda_{\alpha\beta\gamma\delta}^{(3,2)}
\end{eqnarray}
where $\Lambda_{\alpha\beta\gamma\delta}^{(3,1)}$ is given by
\begin{eqnarray}\label{Gamma^{3,1}}
\Lambda_{\alpha\beta\gamma\delta}^{(3,1)}
&=&\int d\bm{C} e^{-2C^2}C_\gamma C_\delta \int d\bm{c} e^{-c^2/2}\int d\hat{\bm{\sigma}}\Theta(\bm{c}\cdot\hat{\bm{\sigma}})
(\bm{c}\cdot\hat{\bm{\sigma}})^2
(c_\alpha\hat{\sigma}_\beta+\hat{\sigma}_\alpha c_\beta)
\nonumber\\
&=& \delta_{\gamma\delta} \frac{S_d}{d}
\int_0^\infty dC C^{d+1}e^{-2C^2}\times 2\beta_3\delta_{\alpha\beta} \frac{S_d}{d}\int_0^\infty dcc^{d+2}e^{-c^2/2}
\nonumber\\
&=& \frac{\delta_{\alpha\beta}\delta_{\gamma\delta}}{d^2} 
\frac{4\pi^d}{\Gamma(\frac{d}{2})^2} 
\frac{\pi^{(d-1)/2}}{\Gamma(\frac{3+d}{2})} 
\frac{1}{2^{d/2+2}}\Gamma(\frac{d}{2}+1)\times 2^{(d+1)/2}
\Gamma\left(\frac{d+3}{2}\right)
\nonumber\\
&=& \delta_{\alpha\beta}\delta_{\gamma\delta} \frac{\pi^{(3d-1)/2}}{d\sqrt{2}\Gamma(\frac{d}{2})} 
\end{eqnarray}
with the relation $\int_0^\infty dC C^{d+1}e^{-2C^2}=2^{-d/2-2}\Gamma(d/2+1)$,  and $\Lambda_{\alpha\beta\gamma\delta}^{(3,2)}$ is given by
\begin{eqnarray}\label{Gamma^{3,2}}
\Lambda_{\alpha\beta\gamma\delta}^{(3,2)}
&=& \int d\bm{C}  e^{-2C^2} \int d\bm{c} e^{-c^2/2}c_\gamma c_\delta \int d\hat{\bm{\sigma}}
\Theta(\bm{c}\cdot\hat{\bm{\sigma}}) (\bm{c}\cdot\hat{\bm{\sigma}})^2(c_{\alpha}\hat{\sigma}_\beta+\hat{\sigma}_\alpha c_{\beta}) \nonumber\\
&=& 2\beta_3 \left(\frac{\pi}{2} \right)^{d/2}\int d\bm{c} e^{-c^2/2} c c_\gamma c_\delta 
  c_\alpha c_\beta \nonumber\\
&=& 2\frac{\pi^{(d-1)/2}}{\Gamma(\frac{3+d}{2})}\left(\frac{\pi}{2} \right)^{d/2}
\frac{2\pi^{d/2}}{\Gamma(\frac{d}{2})}
\int_0^\infty dc c^{d+4} e^{-c^2/2}
\frac{(\delta_{\alpha\beta}\delta_{\gamma\delta}+\delta_{\alpha\gamma}\delta_{\beta\delta}
+\delta_{\alpha\delta}\delta_{\beta\gamma} )}{d(d+2)} 
\nonumber\\
&=&
2\frac{\pi^{(d-1)/2}}{\Gamma(\frac{3+d}{2})}\left(\frac{\pi}{2} \right)^{d/2}
\frac{2\pi^{d/2}}{\Gamma(\frac{d}{2})}
\cdot 2^{(d+3)/2}
\Gamma\left(\frac{d+5}{2}\right)
\frac{(\delta_{\alpha\beta}\delta_{\gamma\delta}+\delta_{\alpha\gamma}\delta_{\beta\delta}
+\delta_{\alpha\delta}\delta_{\beta\gamma} )}{d(d+2)} 
 \nonumber\\
&=& 4\cdot 2^{3/2}\frac{\pi^{(3d-1)/2}}{\Gamma(\frac{d}{2})} \left(\frac{d+3}{2}\right)
\frac{(\delta_{\alpha\beta}\delta_{\gamma\delta}+\delta_{\alpha\gamma}\delta_{\beta\delta}
+\delta_{\alpha\delta}\delta_{\beta\gamma} )}{d(d+2)} 
\nonumber\\
&=& 2^{5/2}\frac{(d+3)}{d(d+2)}\frac{\pi^{(3d-1)/2}}{\Gamma(\frac{d}{2})} 
(\delta_{\alpha\beta}\delta_{\gamma\delta}+\delta_{\alpha\gamma}\delta_{\beta\delta}
+\delta_{\alpha\delta}\delta_{\beta\gamma} )  ,
\end{eqnarray}
where we have used Eq.~\eqref{c_1c_2c_3c_4} in the third line.
Substituting Eqs.~\eqref{Gamma^{3,1}} and \eqref{Gamma^{3,2}} into Eq.~\eqref{Gamma^3} we obtain
\begin{equation}\label{Lambda_3}
\Lambda_{\alpha\beta\gamma\delta}^{L,(3)}=
\frac{\sqrt{2}\pi^{(3d-1)/2}}{d(d+2)\Gamma(\frac{d}{2})}
\left\{(2d+7)\delta_{\alpha\beta}\delta_{\gamma\delta}+
2(d+3)(\delta_{\alpha\gamma}\delta_{\beta\gamma}+\delta_
{\alpha\delta}\delta_{\beta\gamma}) \right\} .
\end{equation}

Similarly, $\Lambda_{\alpha\beta\gamma\delta}^{(4)}$ in Eq. \eqref{Lambda^L4} also consists of two parts 
\begin{eqnarray}\label{Lambda^4}
\Lambda_{\alpha\beta\gamma\delta}^{L,(4)}
&=& \int d\bm{C} \int d\bm{c}\int d\hat{\bm{\sigma}}
\Theta(\bm{c}\cdot\hat{\bm{\sigma}})e^{-2C^2} e^{-c^2/2}
\left(2C_\gamma C_\delta+\frac{1}{2}c_{\gamma}c_{\delta} \right) 
(\bm{c}\cdot\hat{\bm{\sigma}})^3\hat{\sigma}_\alpha \hat{\sigma}_\beta \nonumber\\
&=& 2\Lambda_{\alpha\beta\gamma\delta}^{(4,1)}+
\frac{\Lambda_{\alpha\beta\gamma\delta}^{(4,2)}}{2}.
\end{eqnarray}
where $\Lambda_{\alpha\beta\gamma\delta}^{(4,1)}$ is given by
\begin{eqnarray}\label{Lambda^{4,1}}
\Lambda_{\alpha\beta\gamma\delta}^{(4,1)}&=& 
\int d\bm{C} e^{-2C^2}C_\gamma C_\delta \int d\bm{c} e^{-c^2/2}\int d\hat{\bm{\sigma}}\Theta(\bm{c}\cdot\hat{\bm{\sigma}})
(\bm{c}\cdot\hat{\bm{\sigma}})^3\hat{\sigma}_\alpha \hat{\sigma}_\beta
\nonumber\\
&=&\frac{\delta_{\gamma\delta} S_d}{d}\int_0^\infty dC C^{d+1}e^{-2C^2}\times
\frac{\beta_3}{3+d}S_d \int_0^\infty dc e^{-c^2/2}c^d 
(3c_\alpha c_\beta+c^2\delta_{\alpha\beta})
\nonumber\\
&=& \frac{\delta_{\gamma\delta} S_d}{d} 2^{-d/2-2}\Gamma(\frac{d}{2}+1)
\times \frac{\beta_3}{d+3}\cdot \frac{d+3}{d}\delta_{\alpha\beta}\int_0^\infty dc e^{-c^2/2}c^{d+2}
\nonumber\\
&=& \delta_{\alpha\beta}\delta_{\gamma\delta}\frac{4\pi^d}{\Gamma(\frac{d}{2})^2}
\cdot \frac{\Gamma(\frac{d}{2}+1)}{2^{d/2+2}}
\cdot \frac{\pi^{(d-1)/2}}{d^2\Gamma(\frac{d+3}{2})}
\cdot 2^{(d+1)/2}
\Gamma\left(\frac{d+3}{2}\right)
 \nonumber\\
&=& \frac{\pi^{(3d-1)/2}}{\sqrt{2}d\Gamma(\frac{d}{2})} \delta_{\alpha\beta}\delta_{\gamma\delta} ,
\end{eqnarray}
and $\Lambda_{\alpha\beta\gamma\delta}^{(4,2)}$ is given by
\begin{eqnarray}\label{Lambda^{4,2}}
\Lambda_{\alpha\beta\gamma\delta}^{(4,2)} &=&
\int d\bm{C}  e^{-2C^2} \int d\bm{c} e^{-c^2/2}c_\gamma c_\delta \int d\hat{\bm{\sigma}}
\Theta(\bm{c}\cdot\hat{\bm{\sigma}}) (\bm{c}\cdot\hat{\bm{\sigma}})^3\hat{\sigma}_\alpha\hat{\sigma}_\beta
\nonumber\\
&=& \left(\frac{\pi}{2}\right)^{d/2}
\int d\bm{c} e^{-c^2/2} c_\gamma c_\delta 
\times \frac{\beta_3}{3+d}c(3c_\alpha c_\beta +c^2\delta_{\alpha\beta})
\nonumber\\
&=& \left(\frac{\pi}{2}\right)^{d/2}\frac{\beta_3S_d}{d+3}
\left\{\frac{3}{d(d+2)}
(\delta_{\alpha\beta}\delta_{\gamma\delta}+\delta_{\alpha\gamma}\delta_{\beta\gamma}+\delta_{\alpha\delta}\delta_{\beta\gamma})+\frac{1}{d}\delta_{\alpha\beta}\delta_{\gamma\delta}
 \right\}\int_0^\infty dce^{-c^2/2}c^{d+4} 
\nonumber\\
&=&
\left(\frac{\pi}{2}\right)^{d/2} 
\frac{2\pi^{d/2}}{d(d+2)\Gamma(\frac{d}{2})}
 \frac{1}{(3+d)\Gamma(\frac{d+3}{2})}
\cdot \{(d+5)\delta_{\alpha\beta}\delta_{\gamma\delta}+
3(\delta_{\alpha\gamma}\delta_{\beta\delta}+\delta_{\alpha\delta}\delta_{\beta\gamma})
\}\nonumber\\
&&\times 2^{(d+3)/2}
\Gamma\left(\frac{d+5}{2}\right)
\nonumber\\
&=&
\frac{2\sqrt{2}\pi^{(3d-1)/2}}{d(d+2)\Gamma(\frac{d}{2})}
 \{(d+5)\delta_{\alpha\beta}\delta_{\gamma\delta}+
3(\delta_{\alpha\gamma}\delta_{\beta\delta}+
\delta_{\alpha\delta}\delta_{\beta\gamma}) 
\}.
\end{eqnarray}
Substituting Eq.~\eqref{Lambda^{4,1}} and \eqref{Lambda^{4,2}} into Eq.~\eqref{Lambda^4} we obtain
\begin{equation}\label{Lambda_4}
\Lambda_{\alpha\beta\gamma\delta}^{L,(4)}=
\frac{\sqrt{2}\pi^{(3d-1)/2}}{d(d+2)\Gamma(\frac{d}{2})}
\left\{(2d+7)\delta_{\alpha\beta}\delta_{\gamma\delta}
+3(\delta_{\alpha\gamma}\delta_{\beta\delta}+\delta_{\alpha\delta}\delta_{\beta\gamma}) \right\} .
\end{equation}
With the aid of Eqs.~\eqref{Lambda_3} and \eqref{Lambda_4} and
taking into account the relation  $(\hat{P}_{\alpha\beta}-\delta_{\alpha\beta})\delta_{\alpha\beta}
=\hat{P}_{\alpha\alpha}-\delta_{\alpha\alpha}=0$,
we obtain
\begin{equation}\label{Lambda_3+4}
(\hat{P}_{\gamma\delta}-\delta_{\gamma\delta})(\Lambda_{\alpha\beta\gamma\delta}^{L,(3)}-
2A\Lambda_{\alpha\beta\gamma\delta}^{L,(4)})
=\frac{4\sqrt{2}\pi^{(3d-1)/2}}{d(d+2)\Gamma(d/2)}
(d+3-3A)(\hat{P}_{\alpha\beta}-\delta_{\alpha\beta}) .
\end{equation}
Substituting Eqs.~\eqref{Lambda1+2} and \eqref{Lambda_3+4} into Eq.~\eqref{tilde_Lambda} we obtain
\begin{equation}\label{res_tilde_Lambda}
\tilde{\Lambda}_{\alpha\beta}^L
= \frac{4\sqrt{2}\pi^{(3d-1)/2}}{d\Gamma(d/2)}
\left\{(1-A)\delta_{\alpha\beta}+\frac{1}{d+2}(d+3-3A)(\hat{P}_{\alpha\beta}-\delta_{\alpha\beta}) \right\} .
\end{equation}
Substituting Eq.~\eqref{res_tilde_Lambda} into Eq.~\eqref{Lambda_{33}} we finally obtain
\begin{eqnarray}
\label{eq:Lambda}
\Lambda_{\alpha\beta}^L&=& \frac{2\sqrt{2}\pi^{(d-1)/2}  n^2v_T^3m \sigma^{d-1} }{d\Gamma(d/2)}
\left\{A(1-A)\delta_{\alpha\beta}+\frac{1}{d+2}A(d+3-3A)(\hat{P}_{\alpha\beta}-\delta_{\alpha\beta}) \right\} \nonumber\\
&=&
\frac{\sqrt{2}\pi^{(d-1)/2}  n v_T \sigma^{d-1} }{d\Gamma(d/2)}
\nonumber\\
&&\times
\left\{(1-e^2)nT \delta_{\alpha\beta}+\frac{1+e}{d+2}(2d+3-3e)(P_{\alpha\beta}-nT\delta_{\alpha\beta}) \right\}
.
\end{eqnarray}

\subsubsection{Evaluation of $\Lambda_{\alpha\beta}^{\rm NL}$}

Now, let us evaluate ${\Lambda}_{\alpha\beta}^{\rm NL}$ introduced in Eq.~\eqref{Lambda_{a,b}^{NL}} as well as $\tilde{\Lambda}_{\alpha\beta\gamma\delta\xi\eta}^{\rm NL(1)}$ in Eqs.~\eqref{Lambda^NL(1)} and $\tilde{\Lambda}_{\alpha\beta\gamma\delta\xi\eta}^{\rm NL(2)}$ \eqref{Lambda^NL(2)}.

With the aid of Eqs.~\eqref{Gauss_NL4} and \eqref{Gauss_NL2} we can rewrite $\tilde{\Lambda}_{\alpha\beta\gamma\delta\xi\eta}^{\rm NL(1)}$ as
 \begin{eqnarray}
\tilde{\Lambda}_{\alpha\beta\gamma\delta\xi\eta}^{\rm NL(1)}
&=&
\left(\frac{\pi}{2}\right)^{d/2}
\int d\bm{c}\int d\hat{\bm{\sigma}}\Theta(\bm{c}\cdot\hat{\bm{\sigma}})e^{-c^2/2}
(\bm{c}\cdot\hat{\bm{\sigma}})^2(c_\alpha\hat{\sigma}_\beta+c_\beta \hat{\sigma}_\alpha)
\nonumber\\
&&
\times
[ 
\frac{
\delta_{\gamma\delta}\delta_{\xi\eta}+\delta_{\gamma\xi}\delta_{\delta\eta}
+\delta_{\gamma\eta}\delta_{\gamma\xi}
}{d+2}+\frac{1}{16}c_\gamma c_\delta c_\xi c_\eta
\nonumber\\
&& 
-\frac{1}{16}(\delta_{\gamma\xi}c_\delta c_\eta+ \delta_{\gamma\eta}c_\delta c_\xi
+\delta_{\delta\xi}c_\gamma c_\eta +\delta_{\delta\eta} c_\gamma c_\xi-\delta_{\gamma\delta}c_\xi c_\eta-\delta_{\xi\eta} c_\gamma c_\delta)
] 
\nonumber\\
&=&
2\left(\frac{\pi}{2}\right)^{d/2}\beta_3
\int d\bm{c} e^{-c^2/2} c c_\alpha c_\beta
[ 
\frac{
\delta_{\gamma\delta}\delta_{\xi\eta}+\delta_{\gamma\xi}\delta_{\delta\eta}
+\delta_{\gamma\eta}\delta_{\gamma\xi}
}{d+2}+\frac{1}{16}c_\gamma c_\delta c_\xi c_\eta
\nonumber\\
&& 
-\frac{1}{16}(\delta_{\gamma\xi}c_\delta c_\eta+ \delta_{\gamma\eta}c_\delta c_\xi
+\delta_{\delta\xi}c_\gamma c_\eta +\delta_{\delta\eta} c_\gamma c_\xi-\delta_{\gamma\delta}c_\xi c_\eta-\delta_{\xi\eta} c_\gamma c_\delta)
] 
\nonumber\\
&=&
\frac{2^{1-d/2}\pi^{d-1/2}}{\Gamma((3+d)/2)}
\{
\frac{
\delta_{\gamma\delta}\delta_{\xi\eta}+\delta_{\gamma\xi}\delta_{\delta\eta}
+\delta_{\gamma\eta}\delta_{\gamma\xi}
}{d+2}\int d\bm{c}e^{-c^2/2} c c_\alpha c_\beta
\nonumber\\
&&
+\frac{1}{16}\int d\bm{c}e^{-c^2/2}c c_\alpha c_\beta c_\gamma c_\delta c_\xi c_\eta
\nonumber\\
&&
-\frac{1}{16} \int d\bm{c}e^{-c^2/2}c c_\alpha c_\beta
(\delta_{\gamma\xi}c_\delta c_\eta+ \delta_{\gamma\eta}c_\delta c_\xi
+\delta_{\delta\xi}c_\gamma c_\eta +\delta_{\delta\eta} c_\gamma c_\xi-\delta_{\gamma\delta}c_\xi c_\eta-\delta_{\xi\eta} c_\gamma c_\delta)
\}
\nonumber\\
&=&
\frac{2^{5/2}\pi^{(3d-1)/2}}{d(d+2)\Gamma(d/2)}
\{
\delta_{\alpha\beta}
(\delta_{\gamma\delta}\delta_{\xi\eta}+\delta_{\gamma\xi}\delta_{\delta\eta}
+\delta_{\gamma\eta}\delta_{\gamma\xi})
\nonumber\\
&&
-\frac{d+3}{16}
[
2\delta_{\alpha\beta}(\delta_{\gamma\xi}\delta_{\delta\eta}+\delta_{\gamma\eta}\delta_{\delta\xi}-\delta_{\gamma\delta}\delta_{\xi\eta})
+\delta_{\alpha\delta}(\delta_{\beta\eta}\delta_{\gamma\xi}
+\delta_{\beta\xi}\delta_{\gamma\eta}-\delta_{\beta\gamma}\delta_{\xi\eta})
\nonumber\\
&&
+\delta_{\alpha\eta}(\delta_{\beta\delta}\delta_{\gamma\xi}+\delta_{\beta\gamma}\delta_{\delta\xi}-\delta_{\beta\xi}\delta_{\gamma\delta})
+\delta_{\alpha\gamma}(\delta_{\beta\eta}\delta_{\delta\xi}+\delta_{\beta\xi}\delta_{\delta\eta}-\delta_{\beta\delta}\delta_{\xi\eta})
\nonumber\\
&&
+\delta_{\alpha\xi}(\delta_{\beta\gamma}\delta_{\delta\eta}+\delta_{\beta\delta}\delta_{\gamma\eta}-\delta_{\beta\eta}\delta_{\gamma\delta})
]
\nonumber\\
&&+\frac{(d+3)(d+5)}{16(d+4)}
[
\delta_{\alpha\beta}(\delta_{\gamma\delta}\delta_{\xi\eta}+\delta_{\gamma\xi}\delta_{\delta\eta}+
\delta_{\gamma\eta}\delta_{\delta\xi})
+\delta_{\alpha\gamma}(\delta_{\beta\delta}\delta_{\xi\eta}+\delta_{\beta\xi}\delta_{\delta\eta}
+\delta_{\beta\eta}\delta_{\delta\xi})
\nonumber\\
&&
+\delta_{\alpha\delta}(\delta_{\beta\gamma}\delta_{\xi\eta}+\delta_{\beta\xi}\delta_{\gamma\eta}
+\delta_{\beta\eta}\delta_{\gamma\xi})
+\delta_{\alpha\xi}(\delta_{\beta\gamma}\delta_{\delta\eta}+\delta_{\beta\delta}\delta_{\gamma\eta}
+\delta_{\beta\eta}\delta_{\gamma\delta})
\nonumber\\
&&
+\delta_{\alpha\eta}(\delta_{\beta\gamma}\delta_{\delta\xi}+\delta_{\beta\delta}\delta_{\gamma\xi}
+\delta_{\beta\xi}\delta_{\gamma\delta})
] \},
\end{eqnarray}
where we have used Eqs.~\eqref{Garzo_A6}, \eqref{Gauss_2}, \eqref{Gauss_3} and \eqref{D4}.
Therefore, we obtain 
$\delta\tilde{\Lambda}_{\rm NL}^{(1)}\equiv \tilde{\Lambda}_{yy\gamma\delta\xi\eta}^{{\rm NL}(1)}
- \tilde{\Lambda}_{zz\gamma\delta\xi\eta}^{{\rm NL}(1)}$:
\begin{eqnarray}
\delta\tilde{\Lambda}_{\rm NL}^{(1)}&=&\frac{(2\pi)^{7/2}}{20}
[
\delta_{\gamma\xi}(\delta_{z\delta}\delta_{z\eta}-\delta_{y\delta}\delta_{y\eta})
+\delta_{\delta\xi}(\delta_{z\gamma}\delta_{z\eta}-\delta_{y\gamma}\delta_{y\eta})
+\delta_{\delta\eta}(\delta_{z\xi}\delta_{z\gamma}-\delta_{y\xi}\delta_{y\gamma})
\nonumber\\
&&
+\delta_{\gamma\eta}(\delta_{z\xi}\delta_{z\delta}-\delta_{y\xi}\delta_{y\delta})
-\delta_{\xi\eta}(\delta_{z\delta}\delta_{z\gamma}-\delta_{y\delta}\delta_{y\delta})
-\delta_{\gamma\delta}(\delta_{z\eta}\delta_{z\xi}-\delta_{y\eta}\delta_{y\xi})
\nonumber\\
&&
+\frac{8}{7}
\{
\delta_{\xi\eta}(\delta_{y\gamma}\delta_{y\delta}-\delta_{z\gamma}\delta_{z\delta})
+\delta_{\delta\eta}(\delta_{y\gamma}\delta_{y\xi}-\delta_{z\gamma}\delta_{z\xi})
+\delta_{\gamma\eta}(\delta_{y\delta}\delta_{y\xi}-\delta_{z\delta}\delta_{z\xi})
\nonumber\\
&&
+\delta_{\gamma\xi}(\delta_{y\eta}\delta_{y\delta}-\delta_{z\delta}\delta_{z\eta})
+\delta_{\gamma\delta}(\delta_{y\eta}\delta_{y\xi}-\delta_{z\eta}\delta_{z\xi})
+\delta_{\delta\xi}(\delta_{y\gamma}\delta_{y\eta}-\delta_{z\gamma}\delta_{z\eta})
\}
]
\nonumber\\
&=&
\frac{(2\pi)^{7/2}}{140}
[
15\{
\delta_{\xi\eta}(\delta_{y\gamma}\delta_{y\delta}-\delta_{z\gamma}\delta_{z\delta})
+\delta_{\gamma\delta}(\delta_{y\eta}\delta_{y\xi}-\delta_{z\eta}\delta_{z\xi})
\}
+\delta_{\delta\eta}(\delta_{y\gamma}\delta_{y\xi}-\delta_{z\gamma}\delta_{z\xi})
\nonumber\\
&&
+\delta_{\gamma\eta}(\delta_{y\delta}\delta_{y\xi}-\delta_{z\delta}\delta_{z\xi})
+\delta_{\gamma\xi}(\delta_{y\eta}\delta_{y\delta}-\delta_{z\delta}\delta_{z\eta})
+\delta_{\delta\xi}(\delta_{y\gamma}\delta_{y\eta}-\delta_{z\gamma}\delta_{z\eta})
]
\end{eqnarray}
for $d=3$.
Then, we have the relation
\begin{eqnarray}\label{N_2^1PP}
\delta\tilde{\Lambda}_{\rm NL}^{(1)}(\hat{P}_{\gamma\delta}-\delta_{\gamma\delta})(\hat{P}_{\xi\eta}-\delta_{\xi\eta})&=&
\frac{4\cdot2^{7/2}\pi^{7/2}}{140}(\hat{P}_{yx}^2+\hat{P}_{yz}^2+\hat{P}_{yy}^2-2\hat{P}_{yy}-\hat{P}_{zx}^2-\hat{P}_{zy}^2-\hat{P}_{zz}^2+2\hat{P}_{zz})
\nonumber\\
&\approx&
\frac{8\sqrt{2}\pi^{7/2}}{35} \hat{P}_{xy}^2 ,
\end{eqnarray}
where we have used $\hat{P}_{\alpha\alpha}=3$ and $P_{yz}=P_{zy}=P_{zx}\approx 0$.
\footnote{For $(\alpha,\beta)=(y, z)$ and $(z, x)$, Eq.~\eqref{stress_eq} becomes
\begin{eqnarray*}
	0 = -(\nu+2\zeta)P_{yz},\quad
	\dot\gamma P_{yz}= - (\nu+2\zeta)P_{zx},
\end{eqnarray*}
which yield $P_{yz}=P_{zy}=P_{zx}=0$.}


On the other hand, $\tilde{\Lambda}_{\alpha\beta\gamma\delta\xi\eta}^{\rm NL(2)}$ in Eq. \eqref{Lambda^NL(2)} can be evaluated as
\begin{eqnarray}
\tilde{\Lambda}_{\alpha\beta\gamma\delta\xi\eta}^{\rm NL(2)}
&=&
\left(\frac{\pi}{2}\right)^{d/2}
\int d\bm{c}\int d\hat{\bm{\sigma}}\Theta(\bm{c}\cdot\hat{\bm{\sigma}})e^{-c^2/2}
(\bm{c}\cdot\hat{\bm{\sigma}})^3\hat{\sigma}_\alpha\hat{\sigma}_\beta
\nonumber\\
&&
\times
[ 
\frac{
\delta_{\gamma\delta}\delta_{\xi\eta}+\delta_{\gamma\xi}\delta_{\delta\eta}
+\delta_{\gamma\eta}\delta_{\gamma\xi}
}{d+2}+\frac{1}{16}c_\gamma c_\delta c_\xi c_\eta
\nonumber\\
&& 
-\frac{1}{4}(\delta_{\gamma\xi}c_\delta c_\eta+ \delta_{\gamma\eta}c_\delta c_\xi
+\delta_{\delta\xi}c_\gamma c_\eta +\delta_{\delta\eta} c_\gamma c_\xi-\delta_{\gamma\delta}c_\xi c_\eta-\delta_{\xi\eta} c_\gamma c_\delta)
] 
\nonumber\\
&=&
\left(\frac{\pi}{2}\right)^{d/2}\frac{\beta_3}{3+d}
\int d\bm{c} e^{-c^2/2} c(3c_\alpha c_\beta+c^2\delta_{\alpha\beta})
[
\frac{
\delta_{\gamma\delta}\delta_{\xi\eta}+\delta_{\gamma\xi}\delta_{\delta\eta}
+\delta_{\gamma\eta}\delta_{\gamma\xi}
}{d+2}+\frac{1}{16}c_\gamma c_\delta c_\xi c_\eta
\nonumber\\
&& 
-\frac{1}{4}(\delta_{\gamma\xi}c_\delta c_\eta+ \delta_{\gamma\eta}c_\delta c_\xi
+\delta_{\delta\xi}c_\gamma c_\eta +\delta_{\delta\eta} c_\gamma c_\xi-\delta_{\gamma\delta}c_\xi c_\eta-\delta_{\xi\eta} c_\gamma c_\delta)
]  ,
\end{eqnarray}
where we have used Eq.~\eqref{Garzo_A17}.
With the aid of Eqs.~\eqref{beta}, \eqref{Gauss_d1}, \eqref{Gauss_2}, \eqref{Gauss_3} and \eqref{D4} we obtain
\begin{eqnarray}
\tilde{\Lambda}_{\alpha\beta\gamma\delta\xi\eta}^{\rm NL(2)}
&=&
\frac{\pi^{d-1/2}}{2^{d/2}(d+3)\Gamma((d+3)/2)}
\{
\frac{
\delta_{\alpha\beta}(
\delta_{\gamma\delta}\delta_{\xi\eta}+\delta_{\gamma\xi}\delta_{\delta\eta}
+\delta_{\gamma\eta}\delta_{\gamma\xi})
}{d+2}\frac{2^{(d+3)/2}\pi^{d/2}\Gamma((d+3)/2)}{\Gamma(d/2)}
\frac{d+3}{d}
\nonumber\\
&&
-\frac{3}{4}\frac{2^{(d+5)/2}\pi^{d/2} \Gamma((d+5)/2)}{\Gamma(d/2)}
[
\delta_{\gamma\xi}(\delta_{\alpha\beta}\delta_{\delta\eta}+
\delta_{\alpha\delta}\delta_{\beta\eta}+\delta_{\alpha\eta}\delta_{\beta\delta})
+\delta_{\gamma\eta}(\delta_{\alpha\beta}\delta_{\delta\xi}+
\delta_{\alpha\delta}\delta_{\beta\xi}+\delta_{\alpha\xi}\delta_{\beta\delta}
)
\nonumber\\
&&
+
\delta_{\delta\xi}(\delta_{\alpha\beta}\delta_{\gamma\eta}+
\delta_{\alpha\gamma}\delta_{\beta\eta}+\delta_{\alpha\eta}\delta_{\beta\gamma})
+\delta_{\delta\eta}(\delta_{\alpha\beta}\delta_{\gamma\xi}+
\delta_{\alpha\gamma}\delta_{\beta\xi}+\delta_{\alpha\xi}\delta_{\beta\gamma}
)
\nonumber\\
&&
-\delta_{\gamma\delta}(
\delta_{\alpha\beta}\delta_{\xi\eta}+\delta_{\alpha\xi}\delta_{\beta\eta}
+\delta_{\alpha\eta}\delta_{\beta\xi})
-\delta_{\xi\eta}(
\delta_{\alpha\beta}\delta_{\gamma\delta}+\delta_{\alpha\gamma}\delta_{\beta\delta}
+\delta_{\alpha\delta}\delta_{\beta\gamma}
)
]
\nonumber\\
&&
+\frac{2^{(d+3)/2}\pi^{d/2}\Gamma((d+5)/2)}{d\Gamma(d/2)}
\delta_{\alpha\beta}(\delta_{\gamma\xi}\delta_{\delta\eta}+\delta_{\gamma\eta}\delta_{\delta\xi}
-\delta_{\gamma\delta}\delta_{\xi\eta})
\}
\nonumber\\
&&
+\frac{\pi^{(3d-1)/2}}{4\sqrt{2}d(d+2)\Gamma(d/2)}
\{
\nonumber\\
&&
\frac{3(d+5)}{d+4}
[
\delta_{\alpha\beta}(\delta_{\gamma\delta}\delta_{\xi\eta}+\delta_{\gamma\xi}\delta_{\eta\delta}
+\delta_{\gamma\eta}\delta_{\delta\xi})
+\delta_{\alpha\gamma}(\delta_{\beta\delta}\delta_{\xi\eta}+\delta_{\beta\xi}\delta_{\delta\eta}
+\delta_{\beta\eta}\delta_{\delta\xi})
\nonumber\\
&&
+\delta_{\alpha\delta}(\delta_{\beta\gamma}\delta_{\xi\eta}+\delta_{\beta\xi}\delta_{\gamma\eta}
+\delta_{\beta\eta}\delta_{\gamma\xi})
+\delta_{\alpha\xi}(\delta_{\beta\gamma}\delta_{\delta\eta}+\delta_{\beta\delta}\delta_{\gamma\eta}
+\delta_{\beta\eta}\delta_{\gamma\delta})
\nonumber\\
&&
+\delta_{\alpha\eta}(\delta_{\beta\gamma}\delta_{\delta\xi}+\delta_{\beta\delta}\delta_{\gamma\xi}
+\delta_{\beta\xi}\delta_{\gamma\delta})
]
\nonumber\\
&&
+\delta_{\alpha\beta}(\delta_{\gamma\eta}\delta_{\delta\xi}+\delta_{\gamma\xi}\delta_{\delta\eta}+\delta_{\gamma\delta}\delta_{\eta\xi})
+\delta_{\alpha\beta}(\delta_{\gamma\delta}\delta_{\xi\eta}
+\delta_{\gamma\xi}\delta_{\delta\eta}+\delta_{\gamma\eta}\delta_{\delta\xi})
\}
\nonumber\\
&=& \frac{\sqrt{2}(d+4)\pi^{(3d-1)/2}}{d(d+2)\Gamma(d/2)}
\delta_{\alpha\beta}(\delta_{\gamma\delta}\delta_{\xi}+\delta_{\gamma\eta}\delta_{\delta\xi})
-\frac{\sqrt{2}\pi^{(3d-1)/2}}{(d+2)\Gamma(d/2)}\delta_{\alpha\beta}\delta_{\gamma\delta}\delta_{\xi\eta}
\nonumber\\
&&-\frac{3(d+3)\pi^{(3d-1)/2}}{\sqrt{2}\Gamma(d/2)}
\{
2\delta_{\alpha\beta}(\delta_{\gamma\xi}\delta_{\delta\eta}+\delta_{\gamma\eta}\delta_{\delta\xi}-\delta_{\gamma\delta}\delta_{\xi\eta})
+\delta_{\alpha\gamma}(\delta_{\beta\eta}\delta_{\delta\xi}+\delta_{\beta\xi}\delta_{\delta\eta}-\delta_{\beta\delta}\delta_{\xi\eta})
\nonumber\\
&&+2\delta_{\alpha\delta}(\delta_{\beta\eta}\delta_{\gamma\xi}+\delta_{\beta\xi}\delta_{\gamma\eta})-\delta_{\alpha\delta}\delta_{\beta\gamma}\delta_{\xi\eta}
+\delta_{\alpha\xi}(\delta_{\beta\delta}\delta_{\gamma\eta}+\delta_{\beta\gamma}\delta_{\delta\eta}-\delta_{\beta\eta}\delta_{\gamma\delta})
\nonumber\\
&&
+\delta_{\alpha\eta}(\delta_{\beta\delta}\delta_{\gamma\xi}+\delta_{\beta\gamma}\delta_{\delta\xi}-\delta_{\beta\xi}\delta_{\gamma\delta})
\}
\nonumber\\
&&
+\frac{\pi^{(3d-1)/2}}{4\sqrt{2}d(d+2)(d+4)\Gamma(d/2)}
\{
(5d+23)\delta_{\alpha\beta}(\delta_{\gamma\delta}\delta_{\xi\eta}+\delta_{\gamma\xi}\delta_{\delta\eta}+\delta_{\gamma\eta}\delta_{\delta\xi})
\nonumber\\
&&
+(3d+5)
[
\delta_{\alpha\gamma}(\delta_{\beta\delta}\delta_{\xi\eta}+\delta_{\beta\xi}\delta_{\delta\eta}
+\delta_{\beta\eta}\delta_{\delta\xi})
+\delta_{\alpha\delta}(\delta_{\beta\gamma}\delta_{\xi\eta}+\delta_{\beta\xi}\delta_{\gamma\eta}
+\delta_{\beta\eta}\delta_{\gamma\xi})
\nonumber\\
&&
+\delta_{\alpha\xi}(\delta_{\beta\gamma}\delta_{\delta\eta}+\delta_{\beta\delta}\delta_{\gamma\eta}
+\delta_{\beta\eta}\delta_{\gamma\delta})
+\delta_{\alpha\eta}(\delta_{\beta\gamma}\delta_{\delta\xi}+\delta_{\beta\delta}\delta_{\gamma\xi}
+\delta_{\beta\xi}\delta_{\gamma\delta})
]
\}.
\end{eqnarray}
Therefore, $\delta \tilde{\Lambda}_{\rm NL}^{(2)}\equiv \Lambda_{yy\gamma\delta\xi\eta}^{(2)}-\Lambda_{zz\gamma\delta\xi\eta}^{(2)}$ for $d=3$ is reduced to
\begin{eqnarray}
\delta \tilde{\Lambda}_{\rm NL}^{(2)}&=&
-18\sqrt{2}\pi^{7/2}
\{
2\delta_{y\gamma}(\delta_{y\eta}\delta_{\delta\xi}+\delta_{y\xi}\delta_{\delta\eta}-\delta_{y\delta}\delta_{\xi\eta})
-2\delta_{y\xi}\delta_{y\eta}\delta_{\gamma\delta}+3\delta_{y\delta}(\delta_{y\xi}\delta_{\gamma\eta}+\delta_{y\eta}\delta_{\gamma\xi})
\nonumber\\
&&
-2\delta_{z\gamma}(\delta_{z\eta}\delta_{\delta\xi}+\delta_{z\xi}\delta_{\delta\eta}-\delta_{z\delta}\delta_{\xi\eta})
+2\delta_{z\xi}\delta_{z\eta}\delta_{\gamma\delta}-3\delta_{z\delta}(\delta_{z\xi}\delta_{\gamma\eta}+\delta_{z\eta}\delta_{\gamma\xi})
\}
\nonumber\\
&&+\frac{2\pi^{7/2}}{15\sqrt{2}}
\{
\delta_{y\gamma}(\delta_{y\delta}\delta_{\xi\eta}+\delta_{y\xi}\delta_{\delta\eta}+\delta_{y\eta}\delta_{\delta\xi})+\delta_{y\delta}(\delta_{y\xi}\delta_{\gamma\eta}+\delta_{y\eta}\delta_{\gamma\xi})+\delta_{y\xi}\delta_{y\eta}\delta_{\gamma\delta}
\nonumber\\
&&
-
\delta_{z\gamma}(\delta_{z\delta}\delta_{\xi\eta}+\delta_{z\xi}\delta_{\delta\eta}+\delta_{z\eta}\delta_{\delta\xi})+\delta_{z\delta}(\delta_{y\xi}\delta_{\gamma\eta}+\delta_{z\eta}\delta_{\gamma\xi})+\delta_{z\xi}\delta_{z\eta}\delta_{\gamma\delta}
\} .
\end{eqnarray}
Thus, we obtain
\begin{eqnarray}
\delta \tilde{\Lambda}_{\rm NL}^{(2)}
(\hat{P}_{\gamma\delta}-\delta_{\gamma\delta})
(\hat{P}_{\xi\eta}-\delta_{\xi\eta})
&=&\left(\frac{4}{15}-126\right)
\sqrt{2}\pi^{7/2}
\{
\hat{P}_{yx}^2+(\hat{P}_{yy}-1)^2+\hat{P}_{yz}^2
\nonumber\\
&&
-\hat{P}_{zx}^2-\hat{P}_{zy}^2-(\hat{P}_{zz}-1)^2
\}
\nonumber\\
&\approx&
-\frac{1886\sqrt{2}\pi^{7/2}}{15}
\hat{P}_{xy}^2
 .
\label{N_2^2PP}
\end{eqnarray}

Substituting Eqs.~\eqref{N_2^1PP} and \eqref{N_2^2PP} into Eq.~\eqref{Lambda_{a,b}^{NL}} we obtain
\begin{eqnarray}\label{normal_Lambda}
\Lambda_{yy}^{\rm NL}-\Lambda_{zz}^{\rm NL}
&=&\frac{12-13202A}{105}
\sqrt{2\pi} n^2v_T^3m\sigma^2 A \hat{P}_{xy}^2
\end{eqnarray}
for $d=3$.
Substituting Eq.~\eqref{normal_Lambda} into Eq.~\eqref{eq_N2}
we obtain
\begin{eqnarray}
N_{p,2}&=&-\frac{1}{(\nu+2\zeta) nT} \frac{2(6-6601A)}{105}\sqrt{2\pi} n^2\left(\frac{2T}{m}\right)^{3/2}m\sigma^2 A \left(\frac{T_{\rm ex}}{T}\right)^2 P_{xy}^{*2}\nonumber\\
&=&-\frac{C_3 n\sigma^3}{(\nu^*\sqrt{\theta}+2)\theta^{3/2}}P_{xy}^{*2}\sqrt{\frac{T_{\rm ex}}{m\sigma^2\zeta^2}},
\label{N2}
\end{eqnarray}
where $C_3=8(6-6601A)\sqrt{\pi}A/105$ and we have used Eqs.~\eqref{nu**}, \eqref{P_{xy}**} and \eqref{dgamma**}. 

\subsection{Gaussian integrals}
Let us summarize Gaussian integrals used in the previous subsections.
\begin{eqnarray}\label{Gauss_d1}
\int d\bm{C} e^{-a C^2}&=&\left(\frac{\pi}{a}\right)^{d/2} ,
\\
\label{Gauss_2}
\int d\bm{c} e^{-c^2/2}c^a c_\alpha c_\beta
&=&\frac{\delta_{\alpha\beta}}{d} \int d\bm{c} e^{-c^2/2} c^{a+2} 
=\frac{\delta_{\alpha\beta}}{d}S_d \int_0^\infty dc e^{-c^2/2} c^{d+a+1}
\nonumber\\
&=&\frac{\delta_{\alpha\beta}}{d}\frac{2^{(d+a+2)/2}\pi^{d/2}}{\Gamma(d/2)}
\Gamma\left(\frac{d+a+2}{2}\right)
\\
\int d\bm{c} e^{-c^2/2}c^a c_\alpha c_\beta c_\gamma c_\delta
&=& \frac{S_d}{d(d+2)}
(\delta_{\alpha\beta}\delta_{\gamma\delta}+\delta_{\alpha\gamma}\delta_{\beta\delta}+
\delta_{\alpha\delta}\delta_{\beta\gamma})
\int_0^\infty  dc e^{-c^2/2}c^{d+a+3}
\nonumber\\
&=&
\frac{2^{(d+a+4)/2}\pi^{d/2}}{d(d+2)\Gamma(d/2)}
\Gamma\left(\frac{d+a+4}{2}\right)
(\delta_{\alpha\beta}\delta_{\gamma\delta}+\delta_{\alpha\gamma}\delta_{\beta\delta}+
\delta_{\alpha\delta}\delta_{\beta\gamma}) ,
\label{Gauss_3}
\end{eqnarray}
where we have used the following calculations:
First let us show
\begin{equation}\label{c_xc_y}
\int d\hat{\bm{\sigma}}\hat{c}_\alpha\hat{c}_\beta=S_d\frac{\delta_{\alpha\beta}}{d}.
\end{equation}
Indeed, the trace of the left hand side of this equation gives
$\int d\hat{\bm{\sigma}}\hat{c}_\alpha\hat{c}_\alpha=  \int d\hat{\bm{\sigma}}=S_d$,
while we can write $\int d\hat{\bm{\sigma}}\hat{c}_\alpha\hat{c}_\alpha=S_d K_0d$
if we set $\int d\hat{\bm{\sigma}}\hat{c}_\alpha\hat{c}_\beta=S_d K_0 \delta_{\alpha\beta}$. Therefore, we must have $K_0=1/d$.
Second, we show
\begin{equation}\label{c_1c_2c_3c_4}
\int d\hat{\bm{\sigma}}\hat{c}_\alpha\hat{c}_\beta\hat{c}_\gamma \hat{c}_\delta
=S_d \frac{1}{d(d+2)}
(\delta_{\alpha\beta}\delta_{\gamma\delta}+\delta_{\alpha\gamma}\delta_{\beta\delta}+
\delta_{\alpha\delta}\delta_{\beta\gamma}) .
\end{equation}
Indeed, let us assume
\begin{equation}
\int d\hat{\bm{\sigma}}\hat{c}_\alpha\hat{c}_\beta\hat{c}_\gamma \hat{c}_\delta
=S_d K_1(\delta_{\alpha\beta}\delta_{\gamma\delta}+\delta_{\alpha\gamma}\delta_{\beta\delta}+
\delta_{\alpha\delta}\delta_{\beta\gamma}) .
\end{equation}
 If we set $\alpha=\beta$, then using $\hat{c}_\alpha \hat{c}_\alpha=1$
we obtain
\begin{equation}
\int d\hat{\bm{\sigma}}\hat{c}_\gamma \hat{c}_\delta
=S_dK_1(d+2)\delta_{\gamma\delta}=S_dK_0\delta_{\gamma\delta}.
\end{equation}
Therefore, we obtain $K_1=1/(d(d+2))$.

Similarly, we obtain the relation
\begin{eqnarray}
\label{D4}
\int d\bm{c} e^{-c^2/2}c^a c_\alpha c_\beta c_\gamma c_\mu c_\nu c_\rho
&=&\frac{S_d}{d(d+2)(d+4)}
\nonumber\\
&&
[
\delta_{\alpha\beta}(\delta_{\gamma\mu}\delta_{\nu\rho}+\delta_{\gamma\nu}\delta_{\mu\rho}+\delta_{\gamma\rho}\delta_{\mu\nu})
+\delta_{\alpha\gamma}(\delta_{\beta\mu}\delta_{\nu\rho}+\delta_{\beta\nu}\delta_{\mu\rho}+\delta_{\beta\rho}\delta_{\mu\nu})
\nonumber\\
&&
+\delta_{\alpha\mu}(\delta_{\beta\gamma}\delta_{\nu\rho}+\delta_{\beta\nu}\delta_{\gamma\rho}+\delta_{\beta\rho}\delta_{\gamma\nu})
+\delta_{\alpha\nu}(\delta_{\beta\gamma}\delta_{\mu\rho}+\delta_{\beta\mu}\delta_{\gamma\rho}+\delta_{\beta\rho}\delta_{\gamma\mu})
\nonumber\\
&&
+\delta_{\alpha\rho}(\delta_{\beta\gamma}\delta_{\mu\nu}+\delta_{\beta\mu}\delta_{\gamma\nu}+\delta_{\beta\nu}\delta_{\gamma\mu})
]\int_0^\infty dc e^{-c^2/2}c^{d+a+5}
\nonumber\\
&=& \frac{2^{(d+a+6)/2}\pi^{d/2}\Gamma((d+a+6)/2)}{d(d+2)(d+4)\Gamma(d/2)}
\nonumber\\
&& \times
[
\delta_{\alpha\beta}(\delta_{\gamma\mu}\delta_{\nu\rho}+\delta_{\gamma\nu}\delta_{\mu\rho}+\delta_{\gamma\rho}\delta_{\mu\nu})
+\delta_{\alpha\gamma}(\delta_{\beta\mu}\delta_{\nu\rho}+\delta_{\beta\nu}\delta_{\mu\rho}+\delta_{\beta\rho}\delta_{\mu\nu})
\nonumber\\
&&
+\delta_{\alpha\mu}(\delta_{\beta\gamma}\delta_{\nu\rho}+\delta_{\beta\nu}\delta_{\gamma\rho}+\delta_{\beta\rho}\delta_{\gamma\nu})
+\delta_{\alpha\nu}(\delta_{\beta\gamma}\delta_{\mu\rho}+\delta_{\beta\mu}\delta_{\gamma\rho}+\delta_{\beta\rho}\delta_{\gamma\mu})
\nonumber\\
&&
+\delta_{\alpha\rho}(\delta_{\beta\gamma}\delta_{\mu\nu}+\delta_{\beta\mu}\delta_{\gamma\nu}+\delta_{\beta\nu}\delta_{\gamma\mu})
] ,
\end{eqnarray}
where we have used
\begin{eqnarray}\label{sig^6}
\int d\hat{\bm{\sigma}} \hat{\sigma}_\alpha\hat{\sigma}_\beta \hat{\sigma}_\gamma \hat{\sigma}_\mu\hat{\sigma}_\nu\hat{\sigma}_\rho
&=& \frac{S_d}{d(d+2)(d+4)}
[
\delta_{\alpha\beta}(\delta_{\gamma\mu}\delta_{\nu\rho}+\delta_{\gamma\nu}\delta_{\mu\rho}+\delta_{\gamma\rho}\delta_{\mu\nu})
+\delta_{\alpha\gamma}(\delta_{\beta\mu}\delta_{\nu\rho}+\delta_{\beta\nu}\delta_{\mu\rho}+\delta_{\beta\rho}\delta_{\mu\nu})
\nonumber\\
&&
+\delta_{\alpha\mu}(\delta_{\beta\gamma}\delta_{\nu\rho}+\delta_{\beta\nu}\delta_{\gamma\rho}+\delta_{\beta\rho}\delta_{\gamma\nu})
+\delta_{\alpha\nu}(\delta_{\beta\gamma}\delta_{\mu\rho}+\delta_{\beta\mu}\delta_{\gamma\rho}+\delta_{\beta\rho}\delta_{\gamma\mu})
\nonumber\\
&&
+\delta_{\alpha\rho}(\delta_{\beta\gamma}\delta_{\mu\nu}+\delta_{\beta\mu}\delta_{\gamma\nu}+\delta_{\beta\nu}\delta_{\gamma\mu})
] .
\end{eqnarray}

Similarly, we have the relations
\begin{eqnarray}\label{Gauss_NL4}
\int d\bm{C} e^{-2C^2}C_\alpha C_\beta C_\gamma C_\delta
&=& \frac{\pi^{d/2}}{2^{d/2}(d+2)} (\delta_{\alpha\beta}\delta_{\gamma\delta}+\delta_{\alpha\gamma}\delta_{\beta\delta}+\delta_{\alpha\delta}\delta_{\beta\gamma}) \\
\int d\bm{C}e^{-2C^2}C_\alpha C_\beta
&=&\frac{1}{4} \left(\frac{\pi}{2}\right)^{d/2}  \delta_{\alpha\beta}.
\label{Gauss_NL2}
\end{eqnarray}

\section{Linear stability analysis\label{app:linear}}
\begin{figure}
  \centering
  \includegraphics[width=85mm]{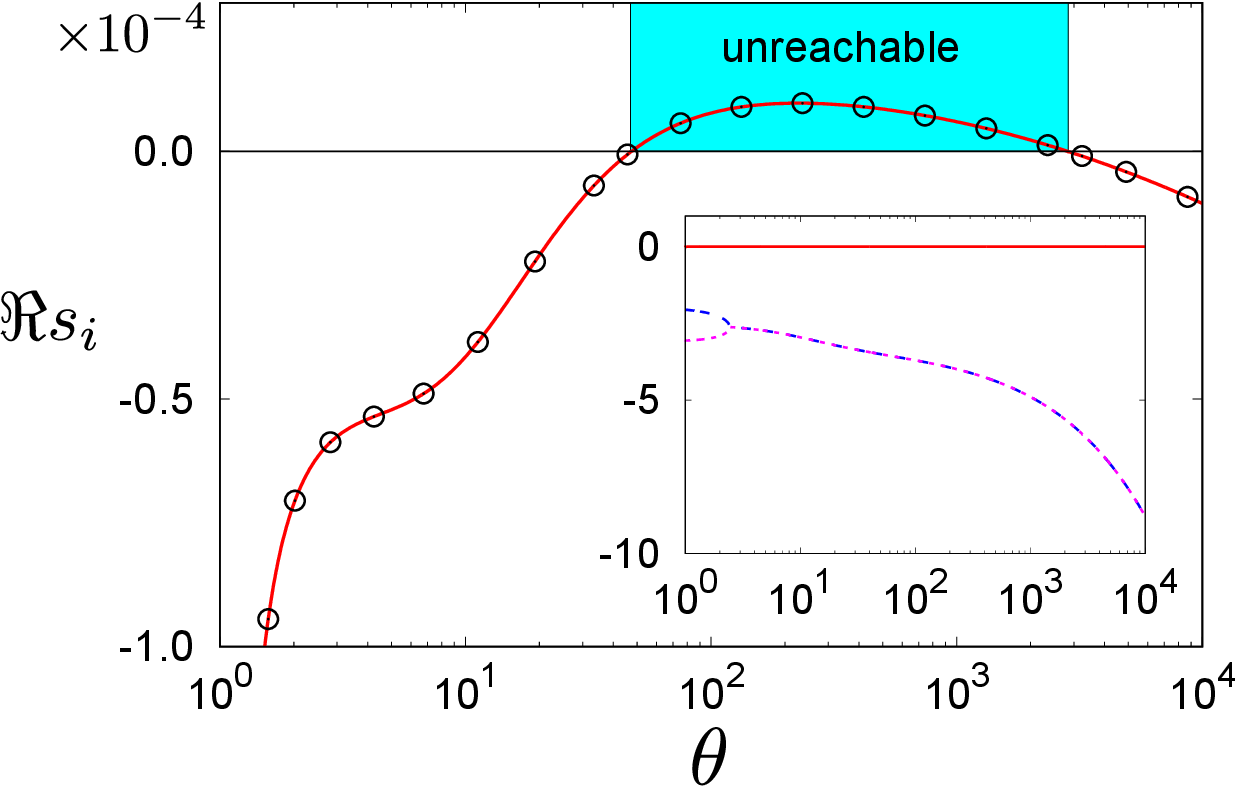}
  \caption{
  Plot of the real part of three eigenvalues $\Re s_i$ ($i=1, 2, 3$) against $\theta$ for $e=0.9$, where the lines and the open circles are, respectively, obtained from Eq.~\eqref{eq:eigen_eq} and \eqref{C8}. 
Here, the largest eigenvalue becomes positive in the intermediate $\theta$, while the other eigenvalues (see the inset) are always negative and are degenerated above a threshold $\theta$.
  }
  \label{fig:fig8}
\end{figure}

In this section, we analyze the linear stability of the steady state \eqref{DT/T**}--\eqref{eta**} obtained in the main text.
Let us rewrite the set of equations \eqref{d_tT}--\eqref{d_tP_{xy}} as
\begin{align}
 \frac{d \theta}{d t^*} &= -\frac{2}{d}\dot\gamma^* P_{xy}^* - \lambda^*\theta^{3/2}+2(1-\theta),\label{eq:eq1}\\
 \frac{d \Delta\theta}{d t^*}&= -2\dot\gamma^* P_{xy}^* - (\nu^*\sqrt{\theta}+2)\Delta \theta,\label{eq:eq2}\\
 \frac{d P_{xy}^*}{d t^*}&= -\frac{1}{d} \dot\gamma^* (d\theta-\Delta\theta)-(\nu^*\sqrt{\theta}+2) P_{xy}^*.\label{eq:eq3}
\end{align}
where we have introduced $t^*\equiv t\zeta$.
We can linearize Eqs.~\eqref{eq:eq1}--\eqref{eq:eq3} around the steady solution $(\theta_{\rm s}, \Delta \theta_{\rm s}, P_{xy,{\rm s}}^*)^{\rm T}$ in the presence of the shear rate $\dot\gamma^*$ as
\begin{equation}\label{eigeneq}
\frac{d}{d t^*} \Psi = M\Psi ,
\end{equation}
where $\Psi =(\delta\theta, \delta \Delta \theta ,\delta P_{xy}^*)^{\rm T}\equiv
(\theta-\theta_{\rm s}, \Delta\theta-\Delta \theta_{\rm s}, P_{xy}^*-P_{xy,{\rm s}}^*)$.
The explicit form of the $3\times 3$ matrix $M\equiv (M_{ij})$  is given by
\begin{equation}
M=
	\begin{pmatrix}
		\displaystyle -\left( \frac{2}{d}\dot\gamma^*_\theta P_{xy,{\rm s}}^*+\frac{3}{2}\lambda^*\sqrt{\theta_{\rm s}}+2\right) & 0 & \displaystyle -\frac{2}{d}\dot\gamma^*\\
		\displaystyle -\left(2\dot\gamma^*_\theta P_{xy,{\rm s}}^*+\frac{1}{2}\nu^* \theta_{\rm s}^{-1/2}\Delta \theta_{\rm s}\right) & -(\nu^*\sqrt{\theta_{\rm s}}+2) & -2\dot\gamma^*\\
		\displaystyle \frac{1}{d}\dot\gamma^*_\theta \Delta\theta_{\rm s} -(\dot\gamma^* + \dot{\gamma}^*_\theta \theta_{\rm s})-\frac{1}{2}\nu^*\theta_{\rm s}^{-1/2}P_{xy, {\rm s}}^* & \displaystyle \frac{1}{d}\dot\gamma^* & -(\nu^*\sqrt{\theta_{\rm s}}+2)
	\end{pmatrix} ,
\end{equation}
where we have introduced
$\dot\gamma_\theta\equiv (\partial \dot\gamma/\partial \theta)_{\rm s}$.
%
Introducing $\tilde{\Psi}(s)$ as the Laplace transform of $\Psi(t)$, the Laplace transform of Eq.\ (\ref{eigeneq}) is written as 
\begin{equation}
	\tilde{\Psi}(s)=(s \bm{1}-M)^{-1}{\Psi}(0)
\end{equation}
under the initial value ${\Psi}(0)$.
When the real part of the eigenvalue in the eigenequation \eqref{eigeneq} is positive, the steady solution is unstable under a perturbation \cite{Takada18}.
The eigenvalues are given by
\begin{align}
&\det(s \bm{1}-M)\nonumber\\
&=s^3 -(M_{11}+M_{22}+M_{33})s^2
	+(M_{11}M_{22}+M_{22}M_{33}+M_{33}M_{11}-M_{12}M_{21}-M_{23}M_{32}-M_{31}M_{13})s\nonumber\\
	&\hspace{1em}-M_{11}M_{22}M_{33}-M_{12}M_{23}M_{31}-M_{21}M_{32}M_{13}
	+M_{11}M_{23}M_{32}+M_{22}M_{31}M_{13}+M_{33}M_{12}M_{21}\nonumber\\
&=s^3 
	+ \left[\frac{2}{d}\dot\gamma_\theta^* P_{xy, {\rm s}}^*+\frac{3}{2}\lambda^* \sqrt{\theta_{\rm s}}+2(\nu^*\sqrt{\theta_{\rm s}}+3)\right]s^2\nonumber\\
&\hspace{1em}
	+\left[ (\nu^*\sqrt{\theta_{\rm s}}+2)^2 + 2(\nu^*\sqrt{\theta_{\rm s}}+2)\left(\frac{2}{d}\dot\gamma^*_\theta P_{xy, {\rm s}}^*+\frac{3}{2}\lambda^*\sqrt{\theta_{\rm s}}+2\right)\right.\nonumber\\
	&\hspace{2em}\left.-\frac{1}{d^2}\dot\gamma^*\dot\gamma^*_\theta \left(d\theta-\Delta \theta\right)-\frac{1}{d}\nu^*\dot\gamma^*\theta_{\rm s}^{-1/2}P_{xy, {\rm s}}^*\right]s\nonumber\\
&\hspace{1em}
	+(\nu^* \sqrt{\theta_{\rm s}}+2)^2\left(\frac{2}{d}\dot\gamma^*_\theta P_{xy, {\rm s}}^*+\frac{3}{2}\lambda^*\sqrt{\theta_{\rm s}}+2\right)\nonumber\\
	&\hspace{1em}
	+\frac{2}{d}\dot\gamma^*(\nu^*\sqrt{\theta_{\rm s}}+2)\left\{\frac{1}{d}\dot\gamma^*_\theta \Delta\theta_{\rm s} -(\dot\gamma^* + \dot{\gamma}^*_\theta \theta_{\rm s})-\frac{1}{2}\nu^*\theta_{\rm s}^{-1/2}P_{xy, {\rm s}}^*\right\}\nonumber\\
	&\hspace{1em}-\frac{1}{d^2}\nu^*\dot\gamma^{*2}\theta_{\rm s}^{-1/2}\Delta \theta_{\rm s}
	+\frac{3}{d}\lambda^*\dot\gamma^{*2}\sqrt{\theta_{\rm s}}+\frac{4}{d}
\dot\gamma^{*2}
=0 . \label{eq:eigen_eq}
\end{align}
Figure \ref{fig:fig8} plots the real parts of the eigenvalues (lines)  as the solutions of Eq.~\eqref{eq:eigen_eq} for $d=3$.
The linear stability is determined by the largest eigenvalue, which is approximately given by the linearized solution of Eq.~(\ref{eq:eigen_eq}) 
\begin{equation}\label{C8}	
s = \frac{N}{D},
\end{equation}
with
\begin{align}
N &=
(\nu^* \sqrt{\theta_{\rm s}}+2)^2\left(\frac{2}{d}\dot\gamma^*_\theta P_{xy, {\rm s}}^*+\frac{3}{2}\lambda^*\sqrt{\theta_{\rm s}}+2\right)\nonumber\\
	&\hspace{1em}
	+\frac{2}{d}\dot\gamma^*(\nu^*\sqrt{\theta_{\rm s}}+2)\left\{\frac{1}{d}\dot\gamma^*_\theta \Delta\theta_{\rm s} -(\dot\gamma^* + \dot{\gamma}^*_\theta \theta_{\rm s})-\frac{1}{2}\nu^*\theta_{\rm s}^{-1/2}P_{xy, {\rm s}}^*\right\}\nonumber\\
	&\hspace{1em}-\frac{1}{d^2}\nu^*\dot\gamma^{*2}\theta_{\rm s}^{-1/2}\Delta \theta_{\rm s}
	+\frac{3}{d}\lambda^*\dot\gamma^{*2}\sqrt{\theta_{\rm s}}+\frac{4}{d}\dot\gamma^{*2}_{\rm s}, 
 \\
D &
=-(\nu^*\sqrt{\theta_{\rm s}}+2)^2 
	- 2(\nu^*\sqrt{\theta_{\rm s}}+2)\left(\frac{2}{d}\dot\gamma^*_\theta P_{xy, {\rm s}}^*+\frac{3}{2}\lambda^*\sqrt{\theta_{\rm s}}+2\right)\nonumber\\
	&\hspace{1em}
	+\frac{1}{d^2}\dot\gamma^*\dot\gamma^*_\theta \left(d\theta-\Delta \theta\right)
	+\frac{1}{d}\nu^*\dot\gamma^*\theta_{\rm s}^{-1/2}P_{xy, {\rm s}}^*
.
\end{align}
This solution (open circles in Fig.~\ref{fig:fig8}) well reproduces the largest eigenvalue (the solid line) in Fig.~\ref{fig:fig8}.
In the intermediate regime $47\le \theta\le 2850$, the largest eigenvalue becomes the positive corresponding to the linearly unstable regime.
%
%
\section{Outline of EDLSHS}\label{EDLSHS}
In this section, we explain the outline of the event-driven Langevin simulation of hard spheres (EDLSHS)~\cite{Scala12} under a simple shear \cite{Evans08, Bannerman11} with the aid of the Lees-Edwards boundary condition \cite{Lees72}.
The time evolution of $i$-th particle at the position $\bm{r}_i$ and the peculiar momentum  of $i$-th particle are given by Eqs.~\eqref{Langevin_eq} and \eqref{noise}.
The velocity increment from time $t$ to $t+\Delta t$ in Eqs.~\eqref{Langevin_eq} and \eqref{noise} can be expressed as
\begin{equation}\label{random}
	\bm{v}_{i,\alpha}(t+\Delta t)= e^{-\zeta \Delta t}\bm{v}_{i,\alpha}(t)+\sqrt{\frac{T_{\rm ex}}{m}(1-e^{-2\zeta \Delta t})}\Gamma,
\end{equation}
where $\Gamma$ represents a zero mean random number whose variance is 1.
In this paper, we use $\Delta t = 0.1/\zeta$ \cite{Scala12}.
To consider the effect of particle collisions, we calculate the minimum time interval $\Delta \tau$ without the random force among the binary collisions of $i$-th and $j$-th particles $\Delta \tau_{ij}$ and the time for the $i$-th particle to reach the Lees-Edwards boundary $\Delta \tau_{i,{\rm wall}}$ \cite{Bannerman11}.
For $t<n\Delta t<t+\Delta \tau$  ($n$ is an integer) the positions of particles are updated without any collisions satisfying Eq.~\eqref{random}.
At $\Delta \tau = \Delta \tau_{ij}$, $i$-th and $j$-th particles collide and therefore their velocities change according to Eq.~(4), 
 while only the position of $i$-th particle is updated as $\bm{r}_i \mp \dot\gamma L \Delta t \to \bm{r}_i$ at $\Delta t=\Delta \tau_{i,{\rm wall}}$, where $L$ is the system size and the minus (plus) sign is selected if the velocity is positive (negative).

\section{Critical slowing down of the relaxation time}\label{app_relaxation}

\begin{figure}
 \begin{center}
  \includegraphics[width=140mm]{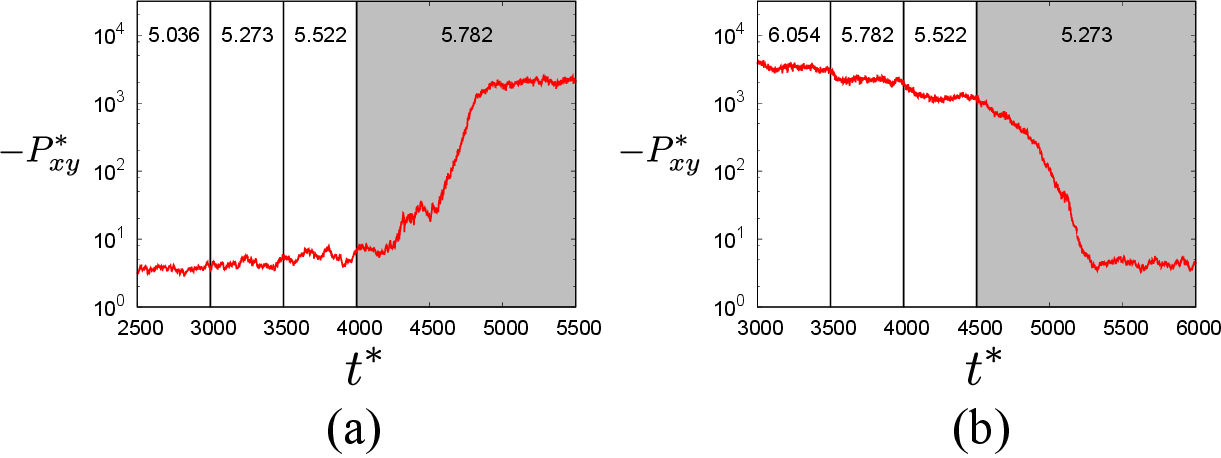}
 \end{center}
  \caption{
   Plots of the time evolution of the shear stress $P_{xy}^*$ when we gradually (a) increase and (b) decrease $\dot\gamma^*$ for $e=0.90$, $d=3$, and $n\sigma^3=0.01$.
Digits in the figure stand for the used $\dot\gamma^*$ in the simulation.
  }
  \label{fig:fig9}
\end{figure}

In this Appendix, we show the critical slowing down of the shear stress in the vicinity of the critical point of the DST when we gradually increase/decrease the shear rate to reach the steady state.
As expected (see Fig.~\ref{fig:fig9}), the relaxation time to reach the steady state becomes longer if the shear rate is closer to the discontinuous transition point.
Note that our data plotted in the main text are the results in the steady state.

%
\section{Domain growth in an unstable region}
\label{app_domain}
\begin{figure}
 \begin{center}
  \includegraphics[width=150mm]{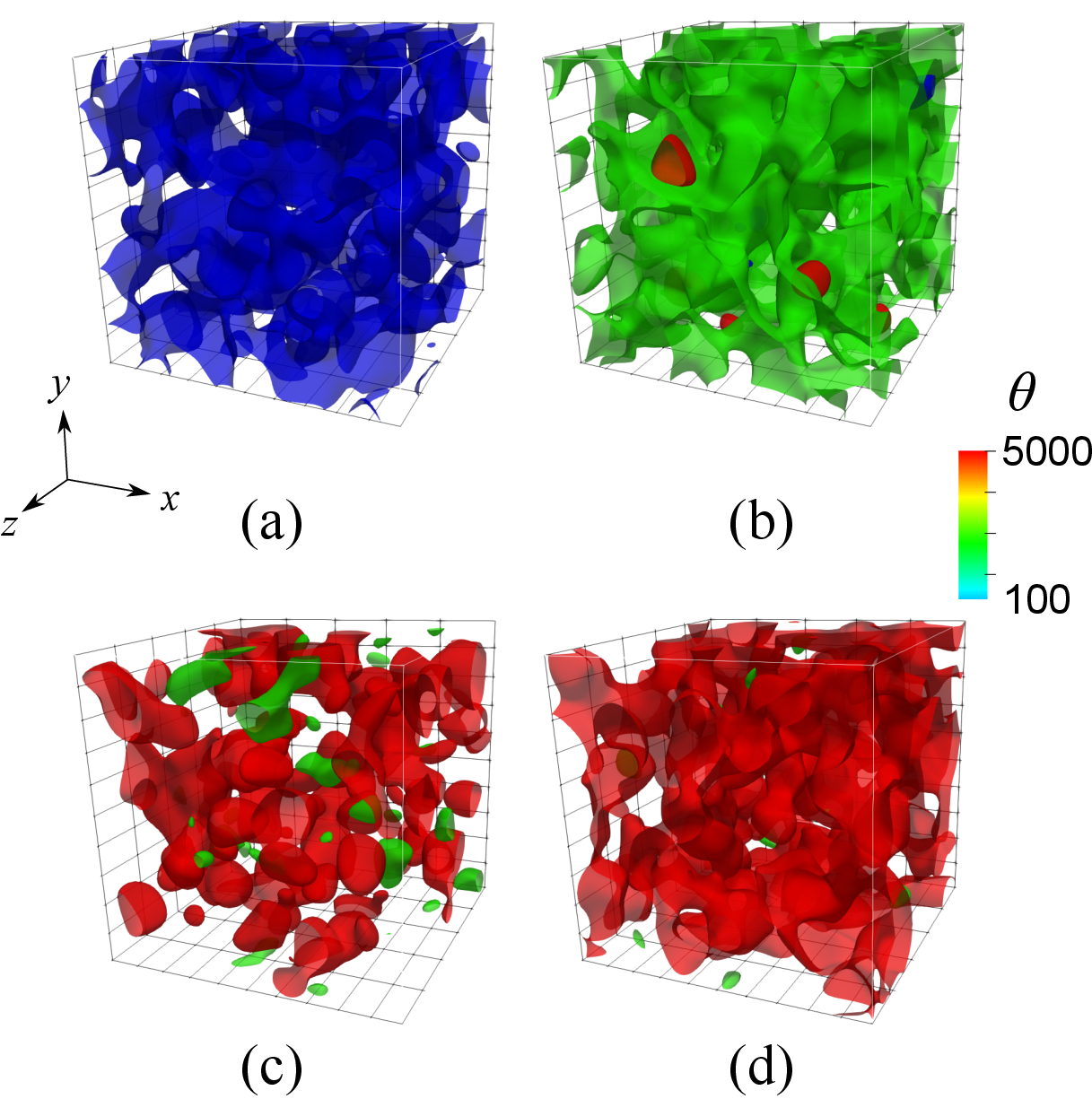}
 \end{center}
  \caption{
 Plots of the contours of the temperature field at (a) $t^*=0$, (b) $100$ , (c) $150$ , and (d) $500$  in the case of $d=3$ and $n\sigma^3=0.01$ with the system size $L=46.4\sigma$.
  }
  \label{fig:fig10}
\end{figure}

\begin{figure}
 \begin{center}
  \includegraphics[width=140mm]{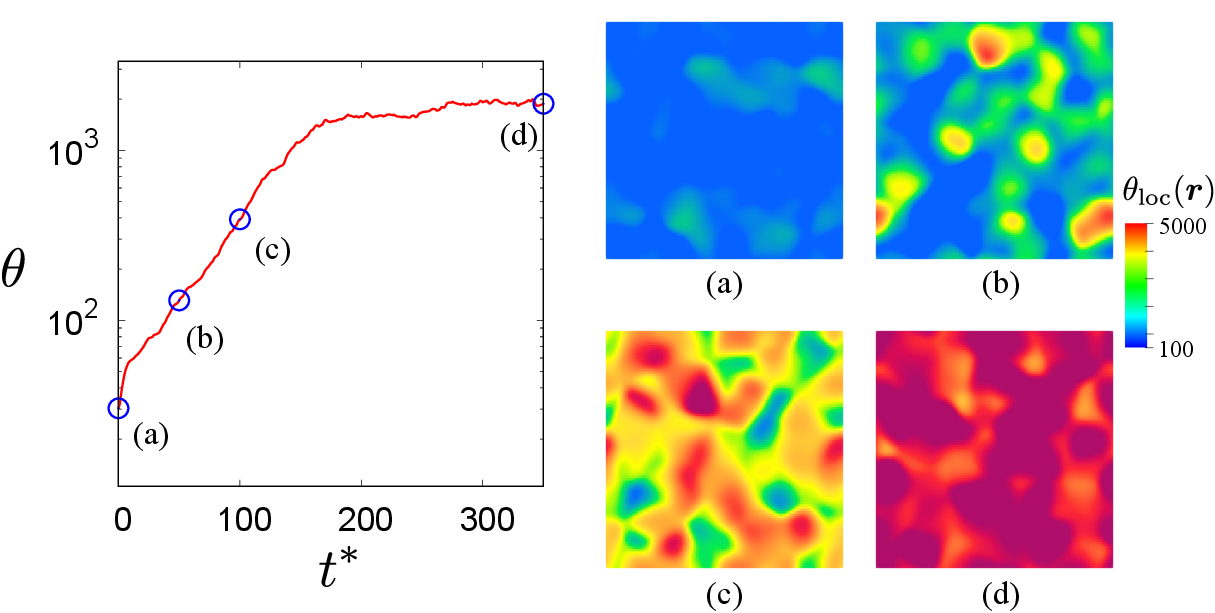}
 \end{center}
  \caption{
   (Left) Time evolution of $\theta$ for $\dot\gamma=4\times10^{0.18}$ when we start the simulation from the unstable regime $\theta=10^2$ shown in Fig.~\ref{fig:fig1}.
   (Right) Local temperature field $\theta_{\rm loc}(\bm{r})$ in $xy$ plane where $z$ dependence of $\theta(\bm{r})$ is averaged over $0\le z <L/100$ at (a) $t^*=0$, (b) $50$, (c) $100$, and (d) $350$.
  }
  \label{fig:fig11}
\end{figure}

In this Appendix, we discuss the time evolution of the temperature field $\theta$ after we start the simulation from an unstable point as transient before reaching a uniform state in the steady state.
Figures~\ref{fig:fig10} and \ref{fig:fig11} express the time evolution of the bi-continuous structure observed in the temperature field $\theta$, which is analogous to the phase ordering processes after the system is quenched into an unstable point~\cite{Bray94}.
Figure \ref{fig:fig11} (left) shows the time evolution of the temperature with an initial condition $\dot\gamma=4\times 10^{0.18}$, where $\theta$ forms a bi-continuous domain to approach the upper branch in the steady solution (see Fig.~\ref{fig:fig1}).
We plot the slices of the local temperature field $\theta_{\rm loc}(\bm{r})$ in Fig.~\ref{fig:fig11}(right).
We can observe that hot domains evolve in the cold background as time goes on, and then the hot regions dominate the system. 

%
\section{Detailed calculation for the rheology of particles under the square well potential}\label{SQUARE_WELL}
In this Appendix, let us show the detailed derivation of $\overleftrightarrow{\Lambda}^{\rm SW}$ in Eq.~\eqref{Lambda_SW}, when we have introduce the attractive interaction between grains as in Eq.~\eqref{U_SW}.
This appendix is basically the short summary of the needed equations for our analysis obtained in Ref.~\cite{Takada18}.

From the collision rule Eq.\ (\ref{collision_rule}), the following relationship is satisfied:
\begin{align}
\bm{c}_1^\prime\bm{c}_1^\prime + \bm{c}_2^\prime\bm{c}_2^\prime - \bm{c}_1\bm{c}_1 + \bm{c}_2\bm{c}_2
=-A(\bm{c}_{12}\cdot \hat{\bm{k}})(\bm{c}_{12} \hat{\bm{k}}+ \hat{\bm{k}}\bm{c}_{12})
	+2A^2 (\bm{c}_{12}\cdot \hat{\bm{k}})^2\hat{\bm{k}}\hat{\bm{k}}.\ ,\label{eq:c1c2_change}
\end{align}
where $A=(1+e)/2$.
Here, we have introduced the dimensionless velocity $\bm{c}_i \equiv \bm{v}_i / (2\varepsilon/m)^{1/2}$.
After some calculations (see Ref.\ \cite{Takada18}), $\overleftrightarrow{\Lambda}^{\rm SW}$ is expressed as
\begin{align}
\overleftrightarrow{\Lambda}^{\rm SW}
&= -\frac{1}{2}\pi^{-3}mn^2d^2 \left(\frac{\pi \varepsilon}{m}\right)^{3/2} \left(\frac{\varepsilon}{T}\right)^{5/2}
	\int d\bm{c}_{12}\int d\hat{\bm{k}} \tilde{S}(\chi,c_{12})c_{12} \exp\left(-\frac{1}{2}c_{12}^2\right)\nonumber\\
&\hspace{1em}\times \left[1+\frac{1}{2}(P_{\gamma\delta}^*-\delta_{\gamma\delta}) c_{12,\gamma}c_{12,\delta} \right]
	\left[-(\bm{c}_{12}\cdot \hat{\bm{k}})(\bm{c}_{12} \hat{\bm{k}}+ \hat{\bm{k}}\bm{c}_{12})
	+2 (\bm{c}_{12}\cdot \hat{\bm{k}})^2\hat{\bm{k}}\hat{\bm{k}}\right]\nonumber\\
&\equiv -\frac{1}{2}mn^2d^2 \left(\frac{\varepsilon}{\pi m}\right)^{3/2} \left(\frac{\varepsilon}{T}\right)^{5/2}
	\left(\overleftrightarrow{\Lambda_1^*} + \overleftrightarrow{\Lambda_2^*} 
	+ \overleftrightarrow{\Lambda_3^*} + \overleftrightarrow{\Lambda_4^*}\right),\label{eq:Lambda_sum}
\end{align}
where $\bm{c}_{12}=\bm{c}_1-\bm{c}_2$, $P^*_{\alpha\beta}=P_{\alpha\beta}/(nT)$
and $\tilde{S}(\chi,c_{12})$ is the collisional cross section~\cite{Takada18}.
Here, we have introduced $\overleftrightarrow{\Lambda_i^*}$ ($i=1,2,3,4$) as
\begin{align}
\overleftrightarrow{\Lambda_1^*}
&= -\int d\bm{c}_{12}\int d\hat{\bm{k}} \tilde{S}(\chi,c_{12})c_{12} \exp\left(-\frac{\varepsilon}{2T}c_{12}^2\right)
	(\bm{c}_{12}\cdot \hat{\bm{k}})(\bm{c}_{12} \hat{\bm{k}}+ \hat{\bm{k}}\bm{c}_{12}),\label{eq:Lambda1}\\
\overleftrightarrow{\Lambda_2^*}
&= 2\int d\bm{c}_{12}\int d\hat{\bm{k}} \tilde{S}(\chi,c_{12})c_{12} \exp\left(-\frac{\varepsilon}{2T}c_{12}^2\right)
	(\bm{c}_{12}\cdot \hat{\bm{k}})^2\hat{\bm{k}}\hat{\bm{k}},\label{eq:Lambda2}\\
\overleftrightarrow{\Lambda_3^*}
&= -\frac{1}{2} \int d\bm{c}_{12}\int d\hat{\bm{k}} \tilde{S}(\chi,c_{12})c_{12} \exp\left(-\frac{\varepsilon}{2T}c_{12}^2\right)
	(P_{\alpha\beta}^*-\delta_{\alpha\beta})c_{12,\alpha}c_{12,\beta} (\bm{c}_{12}\cdot \hat{\bm{k}})(\bm{c}_{12} \hat{\bm{k}}+ \hat{\bm{k}}\bm{c}_{12}),\label{eq:Lambda3}\\
\overleftrightarrow{\Lambda_4^*}
&= \int d\bm{c}_{12}\int d\hat{\bm{k}} \tilde{S}(\chi,c_{12})c_{12} \exp\left(-\frac{\varepsilon}{2T}c_{12}^2\right)
	(P_{\alpha\beta}^*-\delta_{\alpha\beta})c_{12,\alpha}c_{12,\beta} (\bm{c}_{12}\cdot \hat{\bm{k}})^2\hat{\bm{k}}\hat{\bm{k}},\label{eq:Lambda4}
\end{align}
respectively.

To evaluate $\overleftrightarrow{\Lambda_i^*}$ ($i=1,2,3,4$), we use the following relations:
\begin{align}
&\int d\hat{\bm{k}} \tilde{S}(\chi,c_{12}) c_{12}\hat{\bm{k}} 
= 2\pi \int_0^\infty d\tilde{b} \hspace{0.2em}\tilde{b} \sin^2\frac{\chi}{2} \bm{c}_{12},\label{eq:k1}\\
&\int d\hat{\bm{k}} \tilde{S}(\chi,c_{12}) c_{12}(\bm{c}_{12}\cdot \hat{\bm{k}})\hat{\bm{k}}\hat{\bm{k}} \nonumber\\
&=\pi \int_0^\infty d\tilde{b} \hspace{0.2em}\tilde{b} \sin^2\frac{\chi}{2} 
	\left[c_{12}^2 \cos^2\frac{\chi}{2}\overleftrightarrow{1}+\left(2\sin^2\frac{\chi}{2}-\cos^2\frac{\chi}{2}\right)\bm{c}_{12}\bm{c}_{12}\right] ,
\label{eq:k2}
\end{align}
where $\tilde{b}$ is the impact parameter~\cite{Takada18}.
Substituting Eq.\ (\ref{eq:k1}) into Eq.\ (\ref{eq:Lambda1}), $\overleftrightarrow{\Lambda_1^*}$ is rewritten as
\begin{align}
\overleftrightarrow{\Lambda_1^*}
&= -4\pi \int d\bm{c}_{12}\int_0^\infty d\tilde{b} 
	\tilde{b} c_{12} \sin^2\frac{\chi}{2}\exp\left(-\frac{\varepsilon}{2T}c_{12}^2\right) \bm{c}_{12}\bm{c}_{12}\nonumber\\
&= -\frac{16}{3}\pi^2 \int_0^\infty dc_{12}\int_0^\infty d\tilde{b} 
	\tilde{b} c_{12}^5
	\sin^2\frac{\chi}{2}\exp\left(-\frac{\varepsilon}{2T}c_{12}^2\right) \overleftrightarrow{1}.\label{eq:Lambda1_}
\end{align}
Similarly, in terms of Eqs.\ (\ref{eq:Lambda2}) and (\ref{eq:k2}), we rewrite $\overleftrightarrow{\Lambda_2^*}$ as
\begin{align}
\overleftrightarrow{\Lambda_2^*}
&= 2\pi \int d\bm{c}_{12}\int_0^\infty d\tilde{b} \tilde{b}c_{12} \sin^2\frac{\chi}{2}\exp\left(-\frac{\varepsilon}{2T}c_{12}^2\right)\nonumber\\
	&\hspace{1em}\times\left[c_{12}^2 \cos^2\frac{\chi}{2}\bm{1}+\left(2\sin^2\frac{\chi}{2}-\cos^2\frac{\chi}{2}\right)\bm{c}_{12}\bm{c}_{12}\right]\nonumber\\
&= \frac{16}{3}\pi^2 \int dc_{12}\int_0^\infty d\tilde{b} \tilde{b}c_{12}^5 \sin^2\frac{\chi}{2}\exp\left(-\frac{\varepsilon}{2T}c_{12}^2\right)\overleftrightarrow{1}.
\label{eq:Lambda2_}
\end{align}
Similarly, $\overleftrightarrow{\Lambda_3^*}$ and $\overleftrightarrow{\Lambda_4^*}$, respectively, reduce to
\begin{align}
\overleftrightarrow{\Lambda_3^*}
&= -\frac{16}{15}\pi^2 \int dc_{12}\int_0^\infty d\tilde{b} 
	\tilde{b}c_{12}^7 \sin^2\frac{\chi}{2}\exp\left(-\frac{\varepsilon}{2T}c_{12}^2\right)\left(\overleftrightarrow{P}^*- \overleftrightarrow{1}\right),\label{eq:Lambda3_}\\
\overleftrightarrow{\Lambda_4^*}
&= \frac{16}{15}\pi^2 \int dc_{12}\int_0^\infty d\tilde{b} \tilde{b}c_{12}^7 \sin^2\frac{\chi}{2}
	\left(1-\frac{3}{2}\cos^2\frac{\chi}{2}\right)\exp\left(-\frac{\varepsilon}{2T}c_{12}^2\right)\left(\overleftrightarrow{P}^*- \overleftrightarrow{1}\right).\label{eq:Lambda4_}
\end{align}
We substitute Eqs.\ (\ref{eq:Lambda1_})--(\ref{eq:Lambda4_}) into Eq.\ (\ref{eq:Lambda_sum}), we reach Eq.\ (\ref{Lambda_SW}).


\begin{thebibliography}{99}
\bibitem{comment_Tsao1} 
As will be shown, the viscosity $\eta$ in the low shear limit in Ref.~\cite{Tsao95} is proportional to $\varphi^2 \dot\gamma^2$, where $\varphi$ and $\dot\gamma$ are the volume fraction and the shear rate, respectively. Because their viscosity in the low shear limit disappears in the dilute limit ($\varphi\to 0$), Fig.~1 in their paper exists only on the ignited branch.
\bibitem{comment_Tsao2} More precisely, Ref~.\cite{Tsao95} recognized that the viscosity should recover the Newtonian in the low shear limit. Therefore, they assume the Newtonian viscosity Eq. (4.12) in their paper by hand. Then they combine their calculation with the Newtonian viscosity Eq.~(4.12) in their paper to obtain the Newtonian viscosity in Eq.~(4.13) in their paper. The description here is the result of their analysis without using their artificial combination of the Newtonian viscosity.
\end{thebibliography}
\end{document}